\documentclass[
    aps,
    prx,
    letterpaper,
    nobalancelastpage,
    twocolumn,
    superscriptaddress,
    nofootinbib,
    longbibliography
]{revtex4-2}
\usepackage[normalem]{ulem}
\usepackage{graphicx}
\usepackage{amsmath}
\usepackage{amssymb}
\usepackage{braket}
\usepackage[english]{babel}
\usepackage{color}
\usepackage{tcolorbox}
\usepackage[version=4]{mhchem}
\usepackage[hidelinks]{hyperref}
\usepackage{dsfont}
\usepackage{multirow}
\usepackage{booktabs}
\usepackage{array}
\usepackage{subfigure}
\usepackage{subcaption}

\usepackage{soul}

\usepackage{bm}
\newcommand{\blank}{\,&}

\newcommand{\He}{$^3$He$^*$}
\def\moire{moir\'{e}}
\def\Moire{Moir\'{e}}

\newcommand{\UIUC}{
    Department of Physics,
    The University of Illinois at Urbana-Champaign,
    Urbana, IL 61801, USA
}
\newcommand{\Purdue}{
    Department of Physics and Astronomy,
    Purdue University,
    West Lafayette, IN 47907, USA
}
\newcommand{\PurdueQ}{
    Purdue Quantum Science and Engineering Institute,
    Purdue University,
    West Lafayette, IN 47907, USA
}
\newcommand{\UC}{
    Department of Physics and James Frank Institute,
    University of Chicago,
    Chicago, IL 60637, USA
}
\newcommand{\PME}{
    Pritzker School of Molecular Engineering,
    University of Chicago, 
    Chicago, IL 60637, USA
}
\newcommand{\Rice}{
    Department of Physics and Astronomy, 
    Rice University, 
    Houston, TX 77005, USA
}
\newcommand{\RiceSCI}{
    Smalley-Curl Institute, 
    Rice University, 
    Houston, TX 77005, USA
}
\newcommand{\ANL}{
    Physics Division, 
    Argonne National Laboratory, 
    Lemont, IL 60439, USA
}

\begin{document}

\title{Quantum Science with Arrays of Metastable Helium-3 Atoms}
\author{Zheyuan Li}
\affiliation{\UIUC}
\author{Rupsa De}
\affiliation{\UIUC}
\author{Rishi Sivakumar}\email{Present address: HRL Laboratories, Malibu, CA 90265}
\affiliation{\UIUC}
\author{William Huie}
\affiliation{\UIUC}
\author{Hao-Tian Wei}
\affiliation{\Rice}
\affiliation{\RiceSCI}
\author{Justin D. Piel}
\affiliation{\Purdue}
\author{Chris H. Greene}\email{chgreene@purdue.edu}
\affiliation{\Purdue}
\affiliation{\PurdueQ}
\author{Kaden R. A. Hazzard}\email{kaden.hazzard@gmail.com}
\affiliation{\Rice}
\affiliation{\RiceSCI}
\author{Zoe Z. Yan}\email{zzyan@uchicago.edu}
\affiliation{\UC}
\author{Jacob P. Covey}\email{jcovey@uchicago.edu}
\affiliation{\UIUC}
\affiliation{\PME}
\affiliation{\UC}
\affiliation{\ANL}

\begin{abstract}
The motion of atoms in programmable optical tweezer arrays offers many new opportunities for neutral atom quantum science. These include inter- and intra-site atom motion for resource-efficient implementations of fermionic and bosonic modes, respectively, as well as tweezer transport for efficient compilation of arbitrary circuits. However, the exploitation of atomic motion for all three purposes and others is limited by the inertia of the atoms. We present a comprehensive architectural blueprint for the use of fermionic metastable helium-3 (\He) atoms -- the lightest trappable atomic species -- in programmable optical tweezer arrays. This includes a concrete analysis of atomic structure considerations as well as Rydberg-mediated interactions. We show that inter-tweezer hopping of \He ~atoms can be $\gtrsim3\times$ faster than previous demonstrations with lithium-6. We also demonstrate a new toolbox for encoding and manipulating qubits directly in the tweezer trap potential, uniquely enabled by the light mass of \He. Finally, we provide several examples of new opportunities for fermionic quantum simulation and computation that leverage the transport and inter-tweezer hopping of \He ~atom arrays. These tools present new methods to improve the resource efficiency of neutral atom quantum science that may also enable quantum simulations of lattice gauge theories and quantum chemistry outside the Born-Oppenheimer approximation.     
\end{abstract}
\maketitle

\section{Introduction and motivation}
 Arrays of neutral atoms in optical tweezers are rapidly emerging as a leading platform for quantum science applications including large-scale quantum computation~\cite{Saffman2010,Bluvstein2023}, many-body physics with programmable quantum matter~\cite{Browaeys2020,Young2022,Young2024}, and quantum-enhanced metrology with atomic array optical clocks~\cite{Finkelstein2024,Cao2024}. The emergence of these directions was largely enabled by the ability to gain control over the \textit{motional} properties of the atoms, as exemplified by the advent of reconfigurable tweezer arrays~\cite{Endres2016,Barredo2016,Kim2016,Bluvstein2022} and sideband cooling techniques to reach the motional ground state of the tweezer traps~\cite{Kaufman2012,Thompson2013,Cooper2018,Norcia2018b}.

\begin{figure}[t!]
    \centering
    \includegraphics[width=0.48\textwidth]{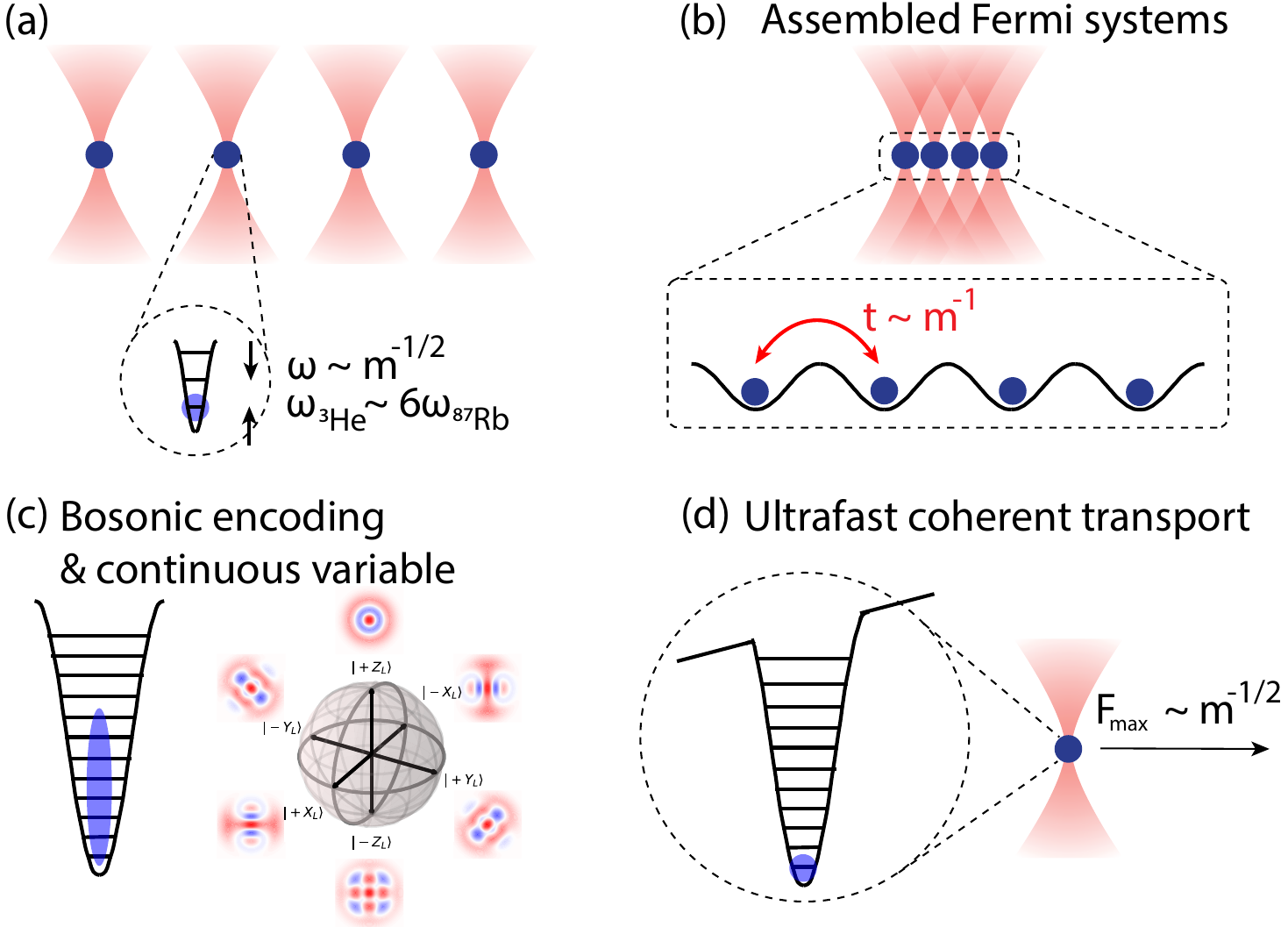}
    \caption{\textbf{Advantages of small mass for quantum science applications}. (a) The trap frequency of an atom in a tweezer scales as $\omega\sim m^{-1/2}$. (b) For fermionic hopping operations, the tunneling rate scales as $t\sim m^{-1}$ for the same trap depth in recoil units $E_R$. (c) A large trap frequency enables bosonic encodings with Raman sideband drives and/or direct trap modulation. (d) The trap frequency also sets the limit on tweezer acceleration during coherent transport, so light atoms enable faster transport.
        \label{Figure1}
    }
\end{figure}

Although these technologies have enabled countless new directions with neutral atom arrays, three avenues in particular have emerged that fully exploit this level of control over the atoms' position and momentum. First, coherent transport~\cite{Beugnon2007,Deist2022b,Bluvstein2022,Bluvstein2023} of atoms during quantum circuits has enabled fully programmable connectivity and `zoned' architectures that have opened the floodgates for logical quantum algorithms~\cite{Bluvstein2023,Reichardt2024,Bluvstein2025,Rines2025}. Second, the spatial wave function of the atom in the optical tweezer has emerged as an additional resource for storing and manipulating quantum information~\cite{Grochowski2023,Scholl2023b,Bohnmann2025}. Inspired by circuit QED, the harmonic oscillator (bosonic) mode can be exploited for bosonic error correction or enlarged code spaces~\cite{Crane2024,Liu2024a,Liu2024b}. Third, coherent tunneling of atoms \textit{between} optical tweezer traps opens the door to the direct encoding of fermionic quantum statistics at the hardware level~\cite{Murmann2015, bergschneider2019experimental, Spar2022}, obviating the overhead from the transformations that map fermionic operators onto qubit operations~\cite{Li2018,Gonzalez2023,schuckert2025fault,Ott2025,Gkritsis2025}. However, despite these opportunities, the extent to which they can be utilized is limited by the mass of the atom, which sets the timescale associated with motion. Minimizing the timescale of motion-based operations will be essential to overcome finite coherence times and to minimize the wall clock time of deep, error-corrected algorithms.

Here, we propose an architecture based on the lightest trappable atomic species, helium-3 (\He),
to advance quantum science of tweezer arrays along all three aforementioned directions. The speed at which an atom can be translated in an optical tweezer and the speed at which its harmonic oscillator state can be manipulated both depend on the trap frequency $\omega$, which scales with mass $m$ as $\omega\sim m^{-1/2}$. Hence, the use of helium-3 instead of rubidium-87 or strontium-88 increases this rate by $\approx6\times$, and the improvement is even larger relative to heavier species like cesium-133 and ytterbium-171. Additionally, the tunneling rate between optical traps scales inversely with $m$ when holding the trap depth fixed, offering an even more drastic improvement that will be indispensable in the simulation of fermionic matter. The light mass of \He ~also enables a new toolbox for encoding and manipulating qubits directly in the tweezer trap potential, exploiting the large anharmonicity from high trap frequency. These advantages of light atoms are summarized in Fig.~\ref{Figure1}. Crucially, \He~has an internal structure that is much better suited for optical Raman coupling (for Raman sideband cooling (RSC) and hyperfine qubit manipulation) than the light alkalis such as lithium-6. We present a platform for realizing \He~atoms in optical tweezers and for exploiting its unique properties for many quantum science directions.

\begin{figure*}[t!]
    \centering
    \includegraphics[width=0.8\textwidth]{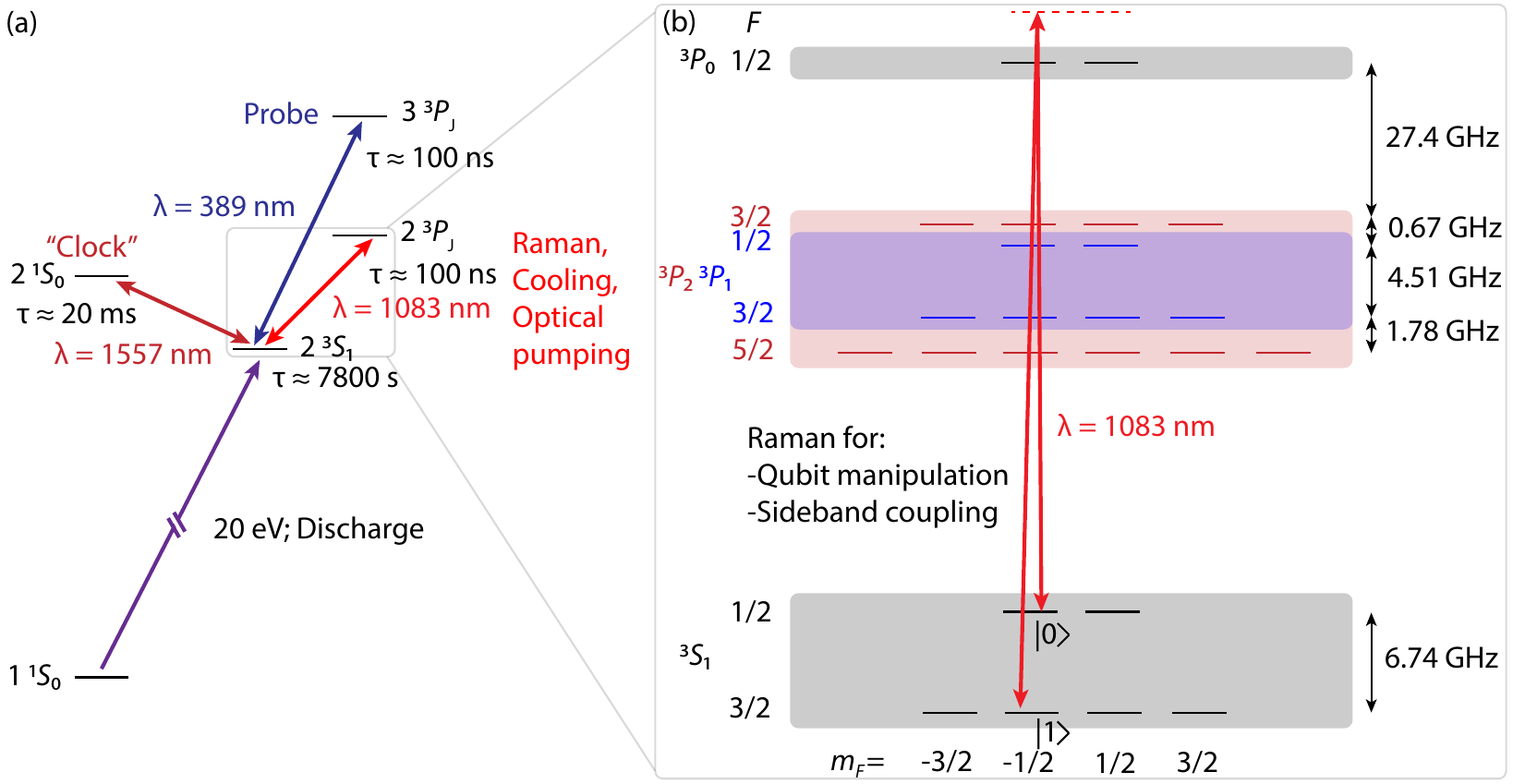}
    \caption{
        \textbf{Relevant level structure for $^3$He*.} (a) The level diagram including the absolute ground state $1s^2$ $^1S_0$; the $1s2s$ $^3S_1$ ``metastable ground" state; the $1s2s$ $^1S_0$ ``clock" state; the $1s2p$ $^3P_J$ manifold for cooling, optical pumping, and Raman coupling; and the $1s3p$ $^3$P$_J$ manifold that will be used for fluorescence readout. (b) A close-up of the $1s2s$ $^3S_1$ and $^3P_J$ transition with wavelength 1083 nm. This system defines the qubit, sideband cooling and motion-qubit coupling, and optical pumping. Note the large separation between $^3P_0$ and $^3P_1$/$^3P_2$ and the lack of hyperfine structure in the former. 
        \label{Figure2}
    }
\end{figure*}

\section{Detecting and cooling \He~atoms in optical tweezers}
\subsection{Introduction to $^3$He*}
Although it has rarely been in the spotlight, quantum science with ultracold metastable helium and other noble gases has been a vibrant and mature field for several decades. Bose-Einstein condensates of $^4$He*~\cite{Santos2001,Robert2001,Manning2014} and degenerate Fermi gases of $^3$He*~\cite{McNamara2006,Thomas2023} have been realized and offer unique applications for matter wave quantum optics~\cite{Schellekens2005,Manning2014} and precision measurements addressing the proton charge radius puzzle~\cite{vanRooij2011,Rengelink2018,LiMuli2025}. Additionally, magneto-optical traps (MOTs) of metastable noble gases have been used for trace isotope detection with implications for nuclear physics (via $^6$He and $^8$He~\cite{Lu2013,Muller2022}) and environmental science (using krypton isotopes~\cite{Chen1999}).

Noble gas atoms are characterized by high-energy, long-lived metastable states. The relevant level structure of He is shown in Fig.~\ref{Figure2}(a), and it is characterized by a ``metastable ground state" that is 20 eV above the true ground state but has a lifetime of $\approx7800$ seconds. $^3$He* is unique because it has two valence electrons, thus sharing the favorable level structure of alkaline earth(-like) atoms, and because it has a nuclear spin of 1/2. Hence, $^3$He* shares some features with ytterbium-171. However, since the ``metastable ground state" is an $s$-orbital, it also shares some features with heavy alkalis like rubidium and cesium. Namely, its strong S-P transition is towards the near-IR (1083 nm), the hyperfine ``ground state" splitting is large (6.7 GHz), and the excited-state hyperfine splittings are also large. These features are in stark contrast to light alkalis, thus presenting a unique combination of minimal mass and optimal level structure for laser-cooling.   

\begin{figure*}[t!]
    \centering
    \includegraphics[width=0.9\textwidth]{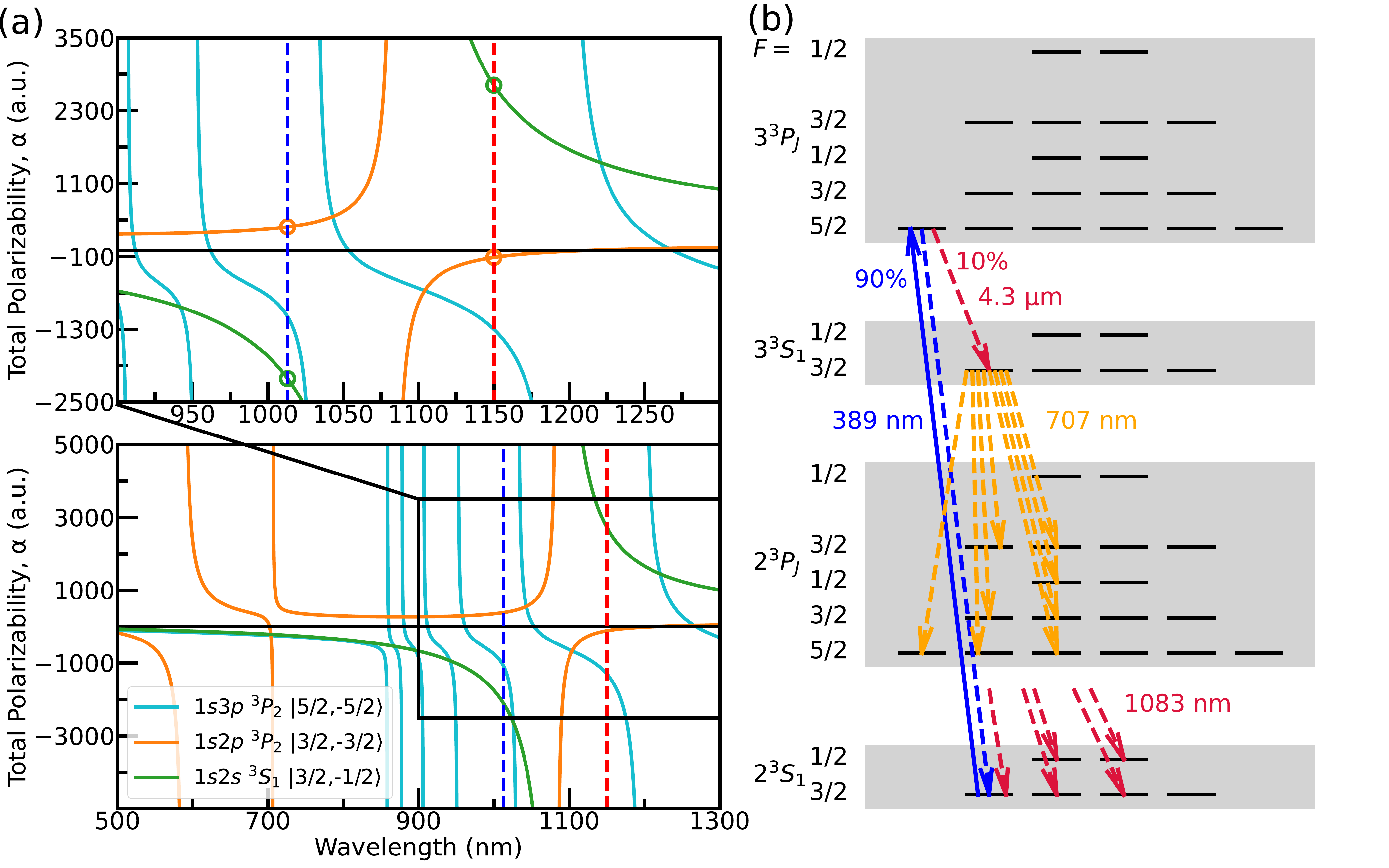}
    \caption{
        \textbf{The optical polarizability of key states.} (a) The polarizability of the ground state (green), upper state of optical pumping (orange), and excited state for fluorescence detection (light blue) as a function of wavelength, where the upper panel shows a close-up around the target wavelengths of 1013 and 1150 nm for blue- and red-detuned tweezer trapping, respectively, indicated by dashed lines. (b) The proposed scheme for fluorescence detection via the $1s3p$ $^3P_2$ level that is coupled to the $1s2s$ $^3S_1$ manifold with a favorable branching fraction and suitable wavelength (389 nm) for detection on a silicon camera. For detail on the polarizability calculation, see Appendix~\ref{Appendix, He polarizability}.  
        \label{Figure3}
    }
\end{figure*}

Qubit manipulation will be performed with stimulated Raman rotations via the 1083 nm transition as described below and benefits from the large fine and hyperfine structure splittings in the $^3P_J$ manifold. A similar paradigm will be used for Raman sideband cooling to the motional ground state in the tweezer trap~\cite{Kaufman2021,Jenkins2022}, which further benefits from the small recoil associated with the 1083-nm photon. The $1s3p$ $^3P_J$ manifold will be used as the probe transition, as discussed below, to benefit from its more favorable wavelength of 389 nm for silicon cameras. Magneto-optical trapping of He* using the $1s2s$ $^3S_1$ $\leftrightarrow$ $1s3p$ $^3P_J$ transition has been demonstrated~\cite{Koelemeij2003}.

He* Rydberg atoms have also been widely used for precision spectroscopy and hybrid systems~\cite{Hogan2018,Morgan2020}. Following these works and mirroring the development of high-fidelity Rydberg operations with alkali species~\cite{Levine2019,Radnaev2024}, we propose to use a two-photon excitation scheme via $1s2s$ $^3S_1$ $\leftrightarrow$ $1s3p$ $^3P_J$ (at 389 nm) and $1s3p$ $^3P_J$ $\leftrightarrow$ $1sns$ $^3S_1$ (at 785 nm). The Rydberg structure will be described below, and shares features with ytterbium-171 owing to its $^3S_1$ term symbol and $I=1/2$ nuclear spin.      

\subsection{Optical trapping}
Understanding the optical polarizability is crucial for trapping \He ~atoms in optical tweezers. Figure~\ref{Figure3}(a) shows the total polarizability of the three states that play an important role during readout and cooling: $1s2s$ $^3S_1$ $|F=3/2,m_F=-1/2\rangle$ (``g"; one of the qubit states as discussed below), $1s2p$ $^3P_2$ $|3/2,-3/2\rangle$ (``e"; for optical pumping during Raman sideband cooling; see below), and $1s3p$ $^3P_2$ $|5/2,-5/2\rangle$ (for the fluorescence detection; see Fig.~\ref{Figure3}(b) and the next subsection). We assume the tweezer is linearly polarized with electric field parallel to the bias magnetic field. The polarizability $\alpha(\omega)$ is used to characterize the spectral response of the atom (i.e., induced dipole moment) to the external electric field $\mathbf{d}=\alpha(\omega)\mathbf{E}$; therefore, the potential energy of such induced dipole can be expressed as a product of polarizability and intensity profile of the trap $U_\mathrm{dip} = -\alpha(\omega)I(\mathbf{r})/(2\epsilon_0c)$, where $\epsilon_0$ is the permittivity of free space and $c$ is the speed of light in vacuum.

Using the ac Stark shift formulation to describe the induced energy shift, the total polarizability of the atom can be expressed in terms of scalar $\alpha^{(0)}$ and tensor $\alpha^{(2)}$ components, given that the vector $\alpha^{(1)}$ light shift vanishes under linear polarization, as
\begin{equation}
\alpha(\omega) = \alpha^{(0)} + \alpha^{(2)}\left(\frac{3\cos^2\theta-1}{2}\right)\left[\frac{3m_F^2-F(F+1)}{F(2F-1)}\right]
\end{equation}
where $\theta=0$ is the angle between the tweezer polarization and the quantization axis determined by the applied magnetic field. The total polarizability results shown in Fig.~\ref{Figure3}(a) are calculated using energy levels and line strengths data from the NIST ASD database \cite{NIST_ASD}, and the detailed procedure is explained Appendix \ref{Appendix, He polarizability}. We note, however, that beyond-paraxial nature of tightly-focused optical tweezers introduces a small vector light shift even for nominally linear polarization, and that this vector light shift can be viewed as a synthetic magnetic field gradient~\cite{Thompson2013}. The effect of this light shift can be mitigated by, e.g., suppressing it with a perpendicular bias magnetic field~\cite{Thompson2013}. We do not expect significance differences between \He ~and other tweezer-trapped species.

Since the level structure in He* is similar to alkalis, the trapping options and considerations are similar. Specifically, two of our states of interest ($1s2s$ $^3S_1$ $|3/2,-1/2\rangle$ and $1s2p$ $^3P_2$ $|3/2,-3/2\rangle$) are coupled at 1083 nm and we thus see a pole in both curves in Fig.~\ref{Figure3}(a). With linearly polarized light, the hyperfine selection rule $\Delta m_F = 0$ dictates that there are more allowed transitions toward the $1s2s$ $^3S_1$ $|3/2,-1/2\rangle$ ground state than the $1s2p$ $^3P_2$ $|3/2,-3/2\rangle$ excited state; hence, the excited state exhibits a much narrower pole than that of the ground state. Due to its role in optical pumping during Raman sideband cooling, the polarizability of the ``e" state and its differential polarizability with respect to ``g" is intimately related to our ability to cool \He ~atoms to the motional ground state in optical tweezers. Finally, the polarizability of the $1s3p$ $^3P_2$ probe state, ``p", would matter if the tweezer traps were left on continuously during fluorescence detection.

We identify promising wavelengths for both blue- and red-detuned tweezer trapping. For red-detuned tweezer trapping, we identify 1150 nm as a suitable wavelength for which high-power Raman fiber amplifiers are commercially available. For a tweezer waist radius ($1/e^2$) of $\approx$1 $\mu$m, 1.23 mW is needed for a trap depth of $\approx$10 MHz ($\approx$0.5 mK). The polarizability ratio at 1150 nm is $\alpha_e/\alpha_g\approx-0.04$, meaning that the e-state is weakly ``anti-trapped". For comparison, alkalis (especially light alkalis) often have $\alpha_e/\alpha_g\approx-1$~\cite{Hutzler2017}. In fact, potassium has $\alpha_e/\alpha_g\approx-5$ at a trapping wavelength of 1064 nm~\cite{Cheuk2015}. For loading atoms into tweezers, cooling them to near their motional ground state, and performing fluorescence detection of them with high fidelity and high survival probability, this issue can be overcome (especially in the extremely mild case of \He), for example by performing gray molasses cooling~\cite{Angonga2022}. Note, however, that the ``p" state is more strongly anti-trapped at this wavelength than the ``e" state.

For blue-detuned trapping, we imagine using a digital hologram to create ``anti-tweezers" that are dark holes on a bright background (i.e. the ``negative" of a standard tweezer array)~\cite{Trisnadi2022}. We identify 1013 nm as a suitable wavelength for which $\sim$100-W ytterbium-doped fiber amplifiers are commercially available. The magnitude of the ``g"-state polarizability is similar at 1013 and 1150 nm due to their proximity to the 1083-nm pole in between. The ``e"/``g" polarizability ratio at 1013 nm is $\alpha_e/\alpha_g\approx-0.18$, meaning that the ``e" state again has the opposite sign as the ``g" state (not surprising given their shared transition at 1083-nm). The magnitude of the polarizability ratio is slightly higher in this case than 1150 nm, but it is still relatively small such that both trapping wavelengths offer similar prospects for cooling. Of course, one natural advantage of blue-detuned traps is the reduced off-resonant scattering or trap-induced photo-ionization since the atoms sit in the dark rather than in the light. Additionally, the polarizability of ``p" is more favorable at 1013 than 1150 nm. Specifically, it is the same sign as that of ``g" on the blue-detuned side, and in fact we predict a ``magic" wavelength of $\approx$1025.6 nm where the ``g"- and ``p"-state polarizabilities match.

Finally, we note that the tensor light shifts give rise to a finite differential polarizability between the hyperfine Zeeman states in the metastable ground manifold, which may impact Raman sideband cooling and hyperfine qubit coherence. We calculate that the differential polarizability is 0.015\% between the qubit states (see Fig.~\ref{Figure2}) at 1150 nm. Assuming a trap depth of 10 MHz and 1\% intensity fluctuations which give rise to 100-kHz light shift fluctuations, the differential light shift fluctuations on the qubit would be 15 Hz. Accordingly, we might expect $T_2^*\approx10$ ms, which is typical of rubidium and cesium qubits as well and can straightforwardly be extended by dynamical decoupling~\cite{Bluvstein2022,Singh2022}. The tensor light shift could be suppressed by detuning further to the red of the 1083-nm transition. In addition, we calculate that the differential polarizability at 1150 nm is 0.031\% between the ``g" state $1s2s$ $^3S_1$ $|3/2,-1/2\rangle$ and the other metastable ground state $1s2s$ $^3S_1$ $|3/2,-3/2\rangle$ involved in Raman sideband cooling (see below and Fig.~\ref{Figure4}).

\subsection{Fluorescence readout}
To perform fluorescence detection of \He ~atoms in optical tweezers, we propose to use a bichromatic imaging technique that is widely used with both alkaline earth(-like) species~\cite{Cooper2018,Covey2019,Jenkins2022,Senoo2025,Falconi2025} and alkali species~\cite{Angonga2022,Menon2024} in tweezers. Namely, we will cool the atoms via the 1083-nm $1s2s$ $^3S_1$$\leftrightarrow$$1s2p$ $^3P_J$ transition and simultaneously probe the atoms with the 389-nm $1s2s$ $^3S_1$$\leftrightarrow$$1s3p$ $^3P_J$ transition. This approach leverages the vastly superior quantum efficiency of silicon detectors at 389 versus 1083 nm and it offers flexibility to send 1083-nm cooling light along the imaging path. The latter may be helpful to perform three-dimensional RSC during readout. The detection scheme will be based on quickly chopping back and forth between a probe block (during which the tweezer could be off) and a RSC block~\cite{Jenkins2022}, such that the atom remains cold while scattering 389-nm photons at the desired scattering rate \textit{on average}. As shown in Fig.~\ref{Figure3}(b), the branching fraction of direct decay back to the ground state is $\approx$90\%, and the remaining $\approx$10\% cascades via the $1s3s$ $^3S_1$ state. The branching fraction is calculated using Einstein coefficient $A_{ki}$ of the corresponding decay path ($9.47\times10^6\mathrm{\ s^{-1}}$ for $1s3p$ $^3P_J$$\rightarrow$$1s2s$ $^3S_1$ and $1.07\times10^6\mathrm{\ s^{-1}}$ for $1s3p$ $^3P_J$$\rightarrow$$1s3s$ $^3S_1$), obtained from NIST ASD database~\cite{NIST_ASD}. This cascade is completely closed and sufficiently fast (the lifetime of $1s3s$ $^3S_1$ is $\approx$40 ns) to not present any appreciable losses or heating during the decay while the tweezer light is chopped off. The branching ratio for direct decay is significantly higher than the case of alkalis~\cite{Duarte2011,McKay2011} partially because of the relatively small optical frequency of the $1s3p$ $^3P_J$ $\rightarrow$ $1s3s$ $^3S_1$ transition and partially because of the absence of a lower-lying D state.

Figure~\ref{Figure3} also shows the polarizability of the $1s3p$ $^3P_2$ $|5/2,-5/2\rangle$ state, which is more strongly anti-trapped than the excited state of OP at 1150 nm. In case of any deleterious effects of the optical tweezer light on this state, we plan to synchronously strobe the tweezer trap such that the tweezer is off during the probe pulses but on during the RSC pulses~\cite{Hutzler2017}. It is straightforward to chop these laser intensities in 100 ns using acousto-optical deflectors. Chopping the tweezer light off also obviates any potential photoionization associated with occupying this state in the presence of the tweezer light. However, we note that the 1150-nm photon from the tweezer is well below the ionization threshold from the $1s3p$ $^3P_J$ manifold (the Rydberg excitation wavelength from this state is $\approx785$ nm). We also note that this issue would be mitigated by using blue-detuned ``anti-tweezers" in which the atom sits in the dark, as discussed above. 

Two-photon ionization during magneto-optical trapping on the 389-nm transition has been observed but has a low rate~\cite{Koelemeij2003}. Specifically, the two-photon photoionization loss rate was measured as a function of detuning $\Delta$ and intensity $I$ (with saturation parameter $s=I/I_\mathrm{sat}$, where $I_\mathrm{sat}$ is the saturation intensity), and scales as $\alpha_\text{2ph}\sim s^2/\Delta^2$ under the far-detuned regime $\Delta^2\gg(s+1)(\Gamma/2)^2$~\cite{Koelemeij2003}. It was observed that $\alpha_\text{2ph}\approx0.07$ s$^{-1}$ for $s=40$ and $\Delta=-35$ MHz. For fluorescence detection, we expect to employ a scattering rate of $\approx2\pi\times50$ kHz~\cite{Cooper2018,Covey2019}, which is well below the linewidth of $\Gamma\approx2\pi\times1.5$ MHz. For a detuning of $\Delta=-5$ MHz, this scattering rate $R_\mathrm{sc}$ could be obtained with a saturation parameter of $s=3.25$, since
\begin{equation}
R_\mathrm{sc} = \Gamma\left[\frac{s/2}{1+\left(2\Delta/\Gamma\right)^2+s}\right].
\end{equation}
Rescaling the measured loss rates to these conditions suggests that $\alpha_\text{2ph}\approx0.023$ s$^{-1}$ in our case, which corresponds to a lifetime of $\gtrsim44$ seconds under measurement conditions.

{We perform detailed simulations of the proposed chopped imaging protocol in Appendix~\ref{Appendix, Imaging}, finding a wide parameter regime in which scattering $\gtrsim1,000$ blue photons (sufficient for high-fidelity detection) while retaining the atom with probability exceeding 0.999 is realistic.

It should be noted that slightly different probe parameters were utilized in this chopped imaging simulation; therefore, the loss due to two-photon ionization has been reconsidered under such regime in the Appendix. Nevertheless, we found loss rate similar to $\alpha_\mathrm{2ph}$ of the far-detuned case discussed above.

\subsection{Cooling to the motional ground state}
Cooling to the motional ground state is the last remaining challenge after demonstrating high-fidelity detection and the techniques that it enables such as atom rearrangement. We consider both Raman sideband cooling and a measurement-based cooling scheme that can ``herald" the presence of thermal errors.

\begin{figure}[t!]
    \centering
    \includegraphics[width=0.48\textwidth]{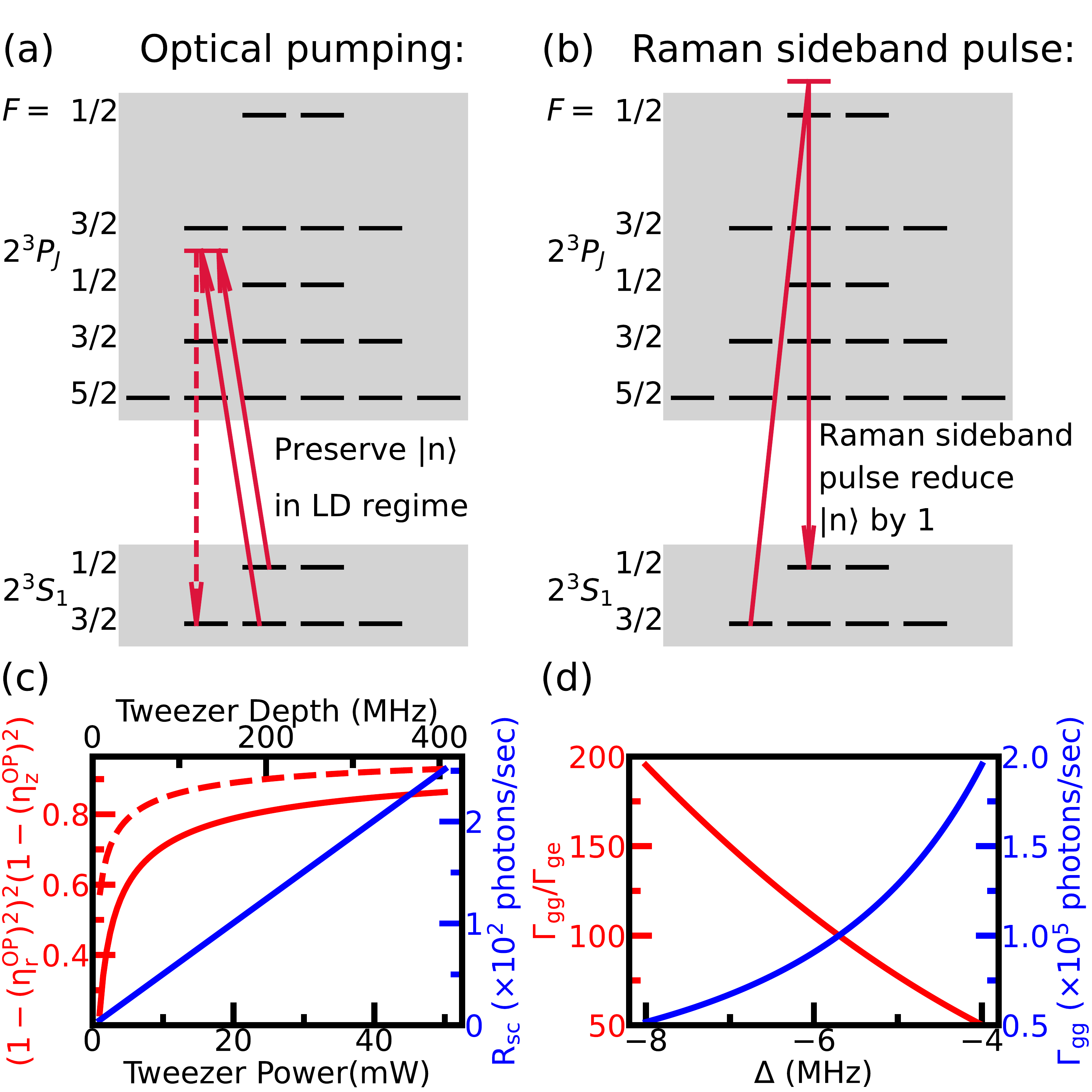}
    \caption{
        \textbf{The schematic approach for Raman sideband cooling.} (a) Optical pumping (OP) that should preserve the motional state $|n\rangle$. (b) Raman transitions between two ground states (one of which is the dark state of OP) that reduce $\langle n\rangle$. OP and the Raman sideband pulses are applied in alternation. (c) Interplay between probability of occupying motional ground state (red) and off-resonant scattering rate from trap $R_\mathrm{sc}$ (blue), both as functions of tweezer power/depth (assuming a tweezer waist of $w_0\approx1\mu $m). The red solid line represents $(1-(\eta_r^\text{OP})^2)^2(1-(\eta_z^\text{OP})^2)$ while the red dashed line represents $(1-(\eta_r^\text{OP})^2)^3$. Although the off-resonant scattering rate increases with increase in trap depth/power, as needed to achieve higher ground state probability, this trap scattering rate is still slow compared to the expected cooling rate. (d) Interplay between “good” to “bad” decay ratio $\Gamma_\mathrm{gg}/\Gamma_\mathrm{ge}$ and OP scattering rate $\Gamma_\mathrm{gg}$, both as functions of OP detuning $\Delta$. To avoid the anti-trapped excited state, finite OP detuning is utilized to realize a dressed state regime, such that the spontaneous decay happens predominantly between trapped dressed states (“good” decays), while still maintaining relatively fast OP scattering rate.
        \label{Figure4}
    }
\end{figure}

\begin{figure*}[t!]
    \centering
    \includegraphics[width=0.85\textwidth]{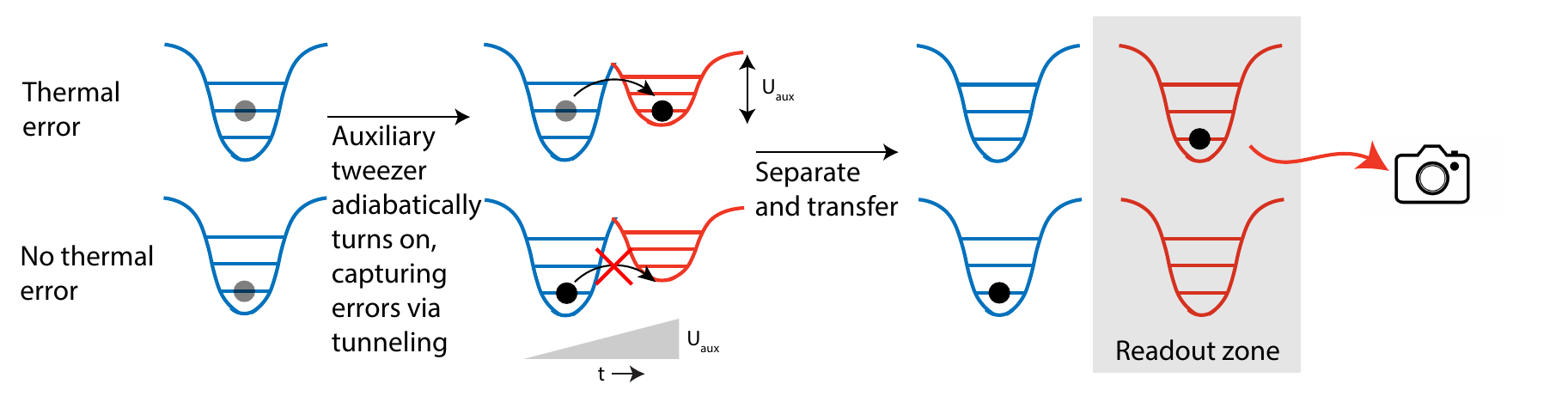}
    \caption{
        \textbf{Measurement-based cooling with a tunnel-coupled auxiliary tweezer.} An auxiliary tweezer  (red) captures a thermal error from the science tweezer (blue) by adiabatically ramping through resonance (see text).  Then, the auxiliary tweezer is separated and moved to a readout zone, where a fluorescence image reveals which science tweezer hosted an error. 
        \label{Figure5}
    }
\end{figure*}

\subsubsection{Raman sideband cooling}
We anticipate that the motional ground state can be initialized with high probability via direct RSC on the 1083-nm transition due to the large hyperfine splittings, large trap frequency, and small recoil of the 1083-nm photon. Figure~\ref{Figure4} shows excitation schemes that will be used for RSC. The OP beam will pump atoms to the $^3S_1$ $|3/2,-3/2\rangle$ ground state via the $^3P_2$ $|3/2,-3/2\rangle$ state. A sideband with frequency matching the ground state hyperfine splitting of 6.7 GHz will repump the $F=3/2$ ground state manifold. The Raman sideband pulse that removes motional quanta will be applied via a Raman transition between $|3/2,-3/2\rangle$ and $|1/2,-1/2\rangle$ via the $^3P_0$ manifold. The goal of OP is to pump it back while remaining in the same motional state. Several Raman sideband beams will be aligned onto the atoms from different directions such that motional quanta can be selectively and iteratively removed from each axis~\cite{Kaufman2012,Thompson2013,Jenkins2022}.

The ability to effectively cool to near the motional ground state via RSC is described by the Lamb-Dicke parameter associated with the optical pumping step $\eta_\text{OP}$. From the outset, we make a distinction between the radial and the axial directions of our tweezer trap. The radial Lamb-Dicke parameter is given by $\eta_r^\text{OP}\equiv r_0k$, where $r_0=\sqrt{\hbar/(2m\omega_r)}$ is the oscillator length for a particle of mass $m$ in the radial direction of the trap with angular frequency $\omega_r$, and $k=2\pi/\lambda$ is the wave number associated with the optical pumping light for which $\lambda=1083$ nm. The axial Lamb-Dicke parameter is given by $\eta_z^\text{OP}\equiv z_0k$, where $z_0=\sqrt{\hbar/(2m\omega_z)}$ is based on the axial angular trap frequency $\omega_z$. The radial and axial trap frequencies are given in terms of the optical tweezer $1/e^2$ waist radius $w_0$ and the Rayleigh length $z_R=k w_0^2/2$, respectively, combined with the maximum trap depth $U_0$ as $\omega_r = \sqrt{4U_0/(mw_0^2)}$ and $\omega_z = \sqrt{2U_0/(mz_R^2)}$, respectively.

At the beginning of RSC, the atom is in a mixed thermal state with a temperature corresponding to the average motional quanta $\bar{n}\gtrsim1$. During the OP step, the probability of increasing the motional quanta upon decay scales as $(\eta^\text{OP})^2$ for a given direction. However, for the initial condition $\bar{n}\gtrsim1$, we must account for the effect on the dipole matrix elements associated with decaying on a motional sideband, which serve to increase the probability of adding motional quanta upon decay. Namely, in this regime, the probability of increasing the motional quanta upon decay scales as $(\eta_\text{eff}^\text{OP})^2=(2\bar{n}+1)(\eta^\text{OP})^2$~\cite{Kaufman2012}. This additional $n$-dependence makes it more challenging to perform efficient sideband cooling at high initial temperatures. However, we expect that myriad sub-Doppler cooling schemes will allow us to initialize the system with $\bar{n}\lesssim1$ before RSC owing to the unique features of \He. Namely, the alkali-like hyperfine metastable ground state structure enables polarization gradient cooling~\cite{Kaufman2012,Manetsch2024} and gray molasses cooling~\cite{Angonga2022} while the relatively narrow optical transition and high trap frequencies enable cooling schemes devised for alkaline earth(-like) atoms in the optical sideband-resolved regime such as Sisyphus cooling~\cite{Taieb1994,Cooper2018,Covey2019}.

We use $\eta^\text{OP}$ to make a simple, conservative estimate of our ability to initialize atoms in the three-dimensional ground state. As $\bar{n}$ approaches zero, the probability of occupying the motional ground state in a given direction can be roughly estimated by $1-(\eta^\text{OP})^2$, although we expect this to be a lower bound due to the stochastic nature of spontaneous emission and the fact that the motional ground state is dark with respect to optical pumping. Since we require our atoms to be in the three-dimensional ground state for certain applications (e.g. to obtain identical fermions), a useful figure of merit (FoM) is $(1-(\eta_r^\text{OP})^2)^2(1-(\eta_z^\text{OP})^2)$ which roughly estimates the probability of occupying the ground state in all three dimensions. This FoM is shown in Fig.~\ref{Figure4}(c) as a function of tweezer power/depth. On the one hand, \He ~suffers from the small $m$ since $\omega_{r,z}\sim m^{-1/2}$ and thus $\eta_{r,z}^\text{OP}\sim m^{-1/4}$ for a given trap. On the other hand, large optical powers are available at the chosen blue- and red-detuned tweezer wavelengths. Hence, the major limitation when increasing the depth is the off-resonant scattering rate from the trap. However, as shown in Fig.~\ref{Figure4}(c), the scattering rate from 1150-nm tweezers (assuming a tweezer waist of $w_0\approx1\mu $m) is slow compared to the expected cooling rate for depths exceeding $\approx$500 MHz ($\approx$25 mK). We also note that an axial 1D lattice could be added to mitigate the challenges associated with cooling the axial direction~\cite{Young2020}, the red dashed line in Fig.~\ref{Figure4}(c) provides an estimate of the ground state probability in such isotropic tweezer configuration. 

So far, we have neglected the differential polarizability associated with the OP transition. As described above and shown in Fig.~\ref{Figure3}, the polarizability ratio at 1013 nm and 1150 nm is $\alpha_e/\alpha_g\approx-0.18$ and $\approx-0.04$, respectively, meaning that the e-state is weakly ``anti-trapped". Therefore, we must account for the different potential when assessing the probability to increase motional quanta associated with OP. Unlike heavy alkalis for which the linewidth $\Gamma\gg\omega$, \He ~is in a regime where $\Gamma\approx\omega$ which exaggerates the role of the excited state. Accordingly, we propose to follow the RSC approach taken by potassium quantum gas microscopes~\cite{Cheuk2015} and optical tweezer arrays~\cite{Lorenz2021} for which $\alpha_e/\alpha_g\approx-5$ at a trapping wavelength of 1064 nm~\cite{Cheuk2015}. These works use finite detunings $\Delta$ for their OP pulses to realize dressed states that obviate the challenge associated with anti-trapped excited states. Figure~\ref{Figure4}(d) shows the ratio of ``good" to ``bad" decays (i.e. the ratio of decays to the trapped dressed ground state versus decays to the anti-trapped dressed ground state) as a function of $\Delta$. We also plot the scattering rate during OP (a proxy for OP timescale) on a second vertical axis to show the interplay of the two. We find favorable conditions in which the ratio of good to bad decays can remain above 200 while OP scatters $\approx4\times10^4$ photons/sec. We leave an assessment of the optimal regime for future study.

\subsubsection{Measurement-based cooling}
To encode quantum information in atomic motion, we will need to deterministically initialize atoms in the motional ground state. 
In the event that RSC proves insufficient, we must add an additional step that can detect thermal errors. Such a technique, combined with reconfigurable atom arrays~\cite{Endres2016,Barredo2016} would allow us to remove thermal errors and reconfigure the tweezer array to obtain a zero-defect \textit{and} zero-temperature array~\cite{Scholl2023b}.

There are two approaches to such a measurement-based cooling technique. The first, enabled by the unique level structure of alkaline earth(-like) atoms involves shelving the atom in the metastable state via an optical clock sideband pulse that is conditional on the motional state~\cite{Scholl2023b}. Although this approach is very powerful and promising for compatible atomic species, we note that it only detects thermal errors along the direction of the clock laser beam. Extensions will be required in order to detect thermal errors in all three directions.

We focus on a second approach that is particularly well suited for light atomic species. 
This proposal follows the pioneering works demonstrating tunnel-coupled fermionic atoms in optical lattice experiments~\cite{esslinger2010fermi}, later extended to optical tweezers with deterministic single-atom preparation~\cite{Murmann2015, bergschneider2019experimental, Spar2022, Yan2022}. 
Namely, we intend to map a thermal error onto occupancy of an auxiliary tweezer (see Fig.~\ref{Figure5}). 
The predominant thermal errors (occupancies above $n=0$) will be in the axial, weak-trapping tweezer axis due to the poorer axial Lamb-Dicke parameter compared to the radial axes.
The auxiliary tweezer will be generated by a different objective and/or a different spatial light modulator (SLM) relative to the science tweezer, allowing a differential displacement $\Delta z$ in their centers of mass.
With independent control over its depth, we will tune the lowest energy state $U_{\rm aux}$ of the auxiliary tweezer's $n'_z=0$ level to be slightly deeper than the second-lowest energy state of the science tweezer ($n_z=1$ and $|U|<|U_{\rm aux}|$).  
Then, the auxiliary tweezer power is reduced adiabatically, such that its depth sweeps through resonance with $|U|$.
Any atom originally in $n_z=1$ will tunnel over to the auxiliary tweezer with high probability as long as the adiabatic condition $J_{n,n'}^2\gg \frac{d(U-U_{\rm aux})}{dt}$ is met. 
Here, $J_{n,n'}=\langle \psi_n |\hat{H} |\psi_{\rm aux, n'}\rangle$ is the resonant tunneling matrix element for a wavepacket in the science tweezer $\ket{\psi_n}$ to tunnel into the auxiliary tweezer supporting the wavepacket $\ket{\psi_{\rm aux,n'}}$.
If the centers-of-mass are approximately $\sim3z_R$ away, and for axial trap frequencies $\omega\sim 2\pi {\times}10\,$kHz, these tunneling rates can reach ${\sim}h{\times} 1\,$kHz for \He, as estimated by numerical calculations of intertweezer tunneling (see Sec.~\ref{sec:tunneling}).

The advantage of this method for measurement-based cooling is that thermal errors of $n_z>1$ are also simultaneously transported into the auxiliary tweezer, with $n_z$ mapping onto the auxiliary $n'_z=n_z+1$ level, due to the near-equidistant harmonic oscillator spacings $\omega_z\approx\omega_z'$.  
Crucially, the science tweezer's ground state occupancy is unaffected as it never enters resonance with any of the auxiliary tweezer levels.
This scheme, based off of successful Landau-Zener-type transitions between ground-state-occupied Li-6 tweezer arrays that reached over 99\% 3D transfer fidelities~\cite{Spar2022}, should be even more robust with lighter species. 
After the error is transported to the auxiliary sites, these are moved away to a readout zone where high-fidelity fluorescence images will reveal which science sites had thermal errors.
This error detection method could be extended to three spatial dimensions by having separate auxiliary tweezers displaced along the tweezer radial directions, as well.

\section{Coherent operations}
With the ability to initialize and detect a defect-free array of \He ~atoms in their motional ground state, we now turn to coherent manipulation of quantum information encoded in the atoms. We focus on three encoding schemes: a hyperfine qubit (spin encoding), the site-specific occupation of a delocalized atom (fermionic encoding), and the occupation of different motional states within an optical tweezer (bosonic encoding). We focus on the use of Rydberg-mediated entangling gates for each encoding scheme. Although Rydberg-mediated interactions for intra-tweezer motion-encoded qubits are nascent and outside the scope of this work, we note that ``phonon SWAP" gates mediated by Rydberg dressing have been proposed~\cite{Belyansky2019}.  

\subsection{Encoding and manipulating hyperfine qubits}

\subsubsection{Magnetic field-insensitive encoding}
Figure~\ref{Figure2}(b) shows the relevant level structure of $^3$He* including the hyperfine structure. As a fermion with nuclear spin \( I = 1/2 \), it naturally does not have magnetic-field-insensitive $m_F=0$ states, in contrast to bosonic isotopes like $^4$He. To identify a suitable qubit, we performed Breit-Rabi calculations to determine the hyperfine splitting as a function of external magnetic field, ranging from the weak-field regime through the Paschen-Back regime. The Land\'e g-factors for angular momentum and the zero-field hyperfine splitting of the $1s2s$ $^3S_1$ state, used in the Breit-Rabi calculations, are taken from~\cite{PhysRevA.32.2712}. These calculations revealed a `magic' magnetic field of 803.5 G [see Fig.~\ref{Figure6}], where the differential Zeeman shift between the states $F=3/2$, $m_F=-1/2$  and $F=1/2$, $m_F=-1/2$ vanishes, yielding a field-insensitive pair of qubit states. Our result is in good agreement with those from White \textit{et al.} in 1959~\cite{White1959}, who found that the energy difference between these two hyperfine levels reaches a minimum at approximately 802.5 G. For additional context, this field is similar to the location of the Feshbach resonance in $^6$Li. It is also similar to magnetic fields that have been used for strontium-88 atom arrays when driving the optical clock transition rapidly~\cite{Madjarov2020}.

\begin{figure}[t!]
    \centering
    \includegraphics[width=0.4\textwidth]{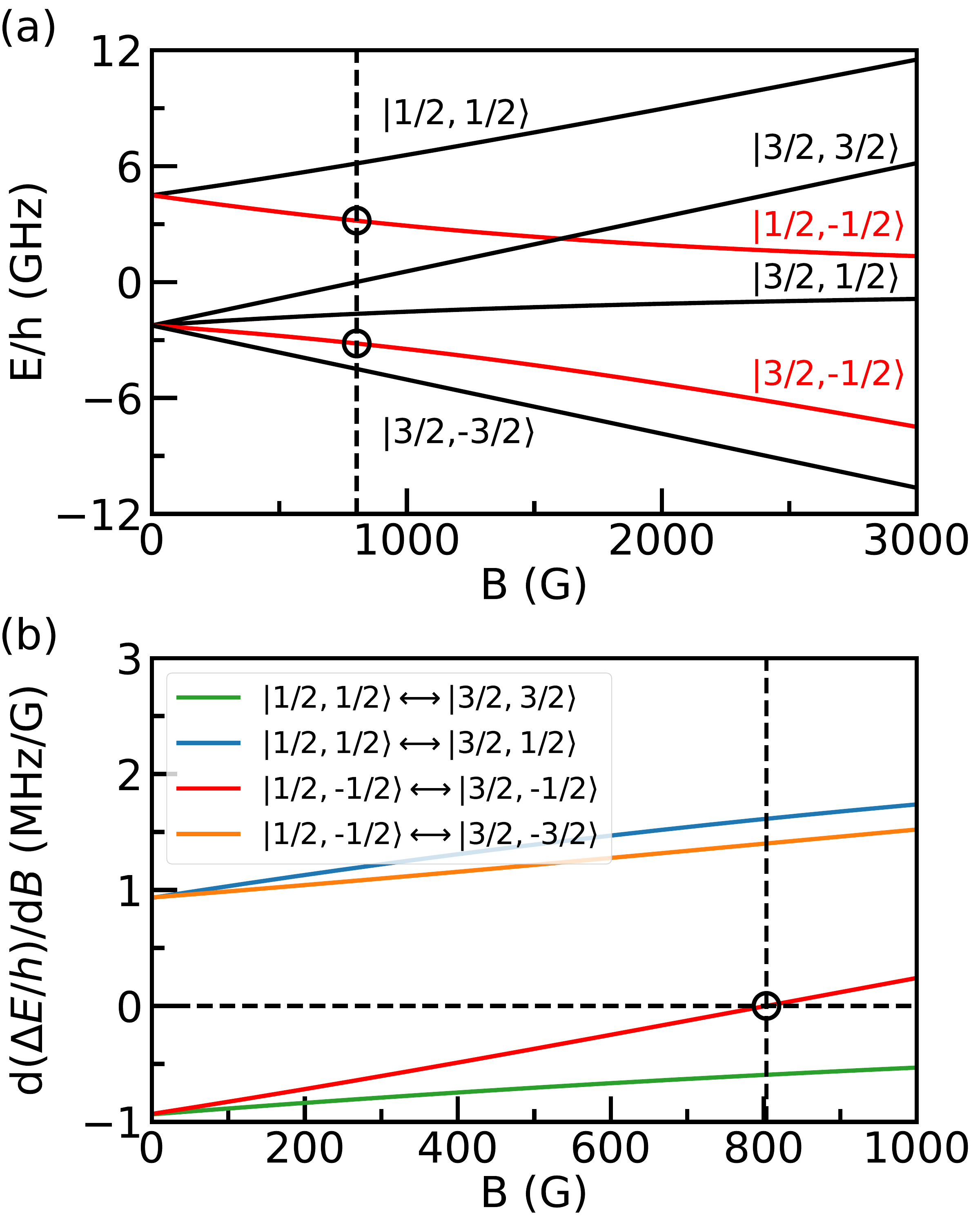}
    \caption{
        \textbf{Zeeman structure and `magic' magnetic field for $^3$He*.} (a) Breit-Rabi simulations of the hyperfine ground state Zeeman maps. The `magic field' of 803.5 G is shown for the qubit states: $|3/2,-1/2\rangle$ and $|1/2,-1/2\rangle$. (b) The differential Zeeman shift for each pair of states, where the qubit differential crosses zero at the `magic field'.
        \label{Figure6}
    }
\end{figure}

\begin{figure*}[t!]
    \centering
    \includegraphics[width=0.85\textwidth]{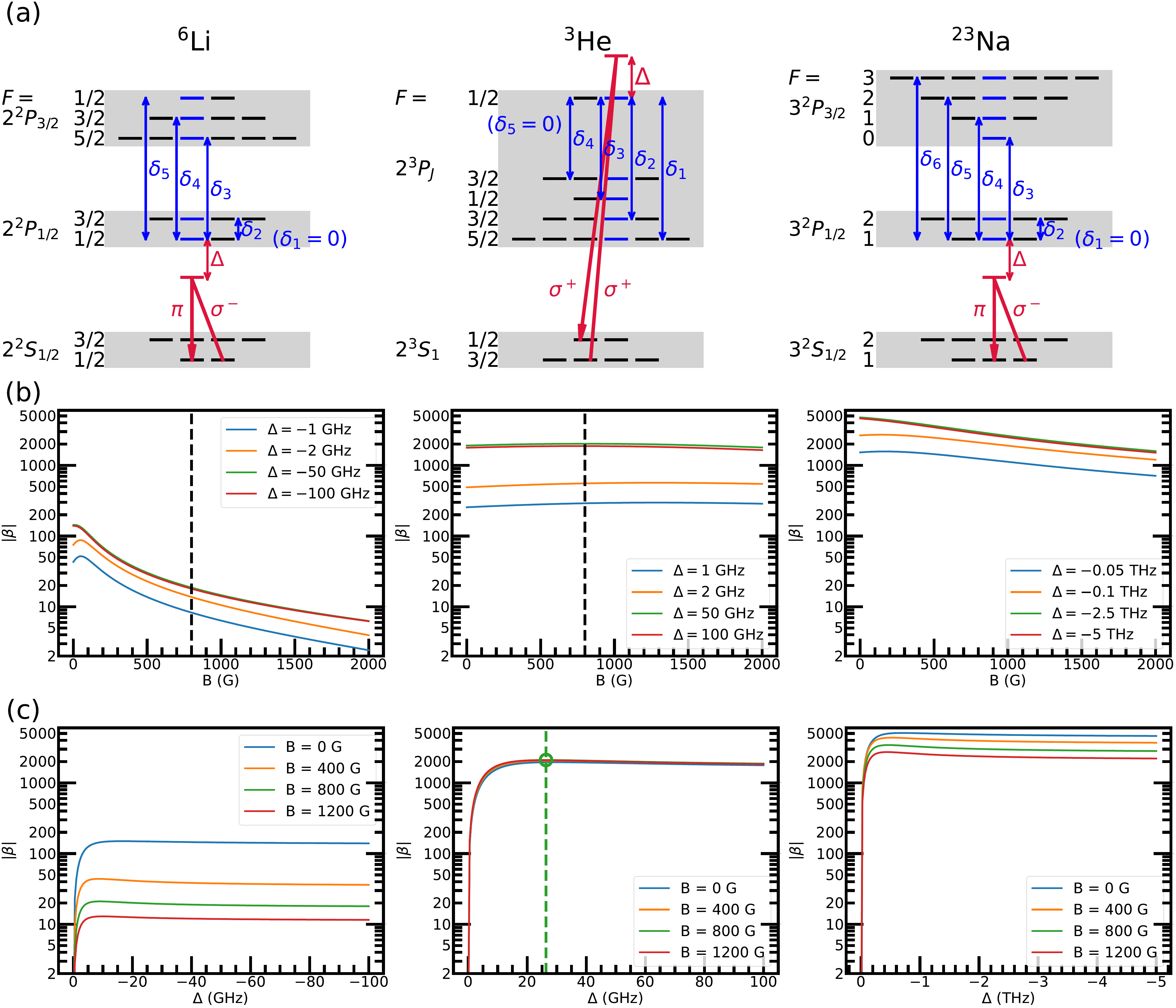}
    \caption{
        \textbf{Comparison of $|\beta|$ ratio between $^6$Li, \He, and $^{23}$Na.} (a) The hyperfine level structure of states involved in the stimulated Raman transition. (b) $|\beta|$ ratio as a function of magnetic field $B$ for different detuning $\Delta$, where the black dashed line is at $B=800\mathrm{\ G}$. (c) $|\beta|$ ratio as a function of $\Delta$ for different $B$, where the green dashed line and circle indicate the maximum $|\beta|$ value of \He~at $B=800\mathrm{\ G}$. For detail on the $|\beta|$ ratio calculation, see Appendix~\ref{Appendix, Beta Ratio and Raman Fidelity Limit}.
        \label{Raman_Fidelity_Limit_Figure}
    }
\end{figure*}

\subsubsection{Hyperfine structure-limited Raman gate fidelity}
Qubit manipulation will be performed with stimulated Raman rotations via the 1083-nm transition as shown in Fig.~\ref{Figure2}(b), and benefits from the large fine and hyperfine structure splittings in the $^3P_J$ manifold. A similar paradigm will be used for RSC to the motional ground state in the tweezer trap as discussed above [see Fig.~\ref{Figure4}(b)], which further benefits from the small recoil associated with the 1083-nm photon.

A specific consequence of the level structure stems from the universal figure of merit $\beta$ for Raman couplings: the ratio of the (effective) Raman Rabi frequency, $\Omega_R$, to the inelastic-scattering rate, $\Gamma_\mathrm{ine}$, directly quantifies the fundamental fidelity limit for single-qubit gates~\cite{Wei2013}, $\mathcal{F}\approx1-1/|\beta|$, which is intended to be a general estimate independent of the exact gate operation. This metric affects RSC, qubit manipulation, and even the possibility of engineering Raman-induced tunneling~\cite{Jaksch2003,Lin2011,Cheuk2012,Wang2012,Wei2013,Aidelsburger2013,Miyake2013}. For detail on the procedure used to calculate $|\beta|$ ratio, see Appendix~\ref{Appendix, Beta Ratio and Raman Fidelity Limit}.

As shown in Fig.~\ref{Raman_Fidelity_Limit_Figure}(b), compared to heavier alkali-metal atoms like $\mathrm{^{23}Na}$, $\mathrm{^6Li}$ exhibits a faster decrease in $|\beta|$ ratio with respect to increasing magnetic field, which could be attributed to the much weaker coupling between its electronic and nuclear spin~\cite{Wei2013}. Nevertheless, such magnetic-field suppression of $|\beta|$ ratio is not observed for \He. In fact, \He~ experiences the least suppression in comparison to $\mathrm{^6Li}$ and $\mathrm{^{23}Na}$.

On the other hand, as shown in Fig.~\ref{Raman_Fidelity_Limit_Figure}(c), with respect to increasing laser detuning $\Delta$, the $|\beta|$ ratio undergoes rapid increase for small $\Delta$, reaches a maximum, and then approaches an asymptotic value at large $\Delta$. Such a nonmonotonic change in $|\beta|$ gives rise to a maximum value at an optimal detuning around $\Delta\approx E_\mathrm{fs}$, where $E_\mathrm{fs}$ is the fine-structure splitting in energy of the coupled excited state manifolds~\cite{Wei2013}. At a magnetic field of $B=800\mathrm{\ G}$, approximately where the Feshbach resonance of $\mathrm{^6Li}$ ($B=834\mathrm{\ G}$) and the aforementioned “magic” magnetic field (magnetic-field insensitive qubit) of \He~ ($B=803.5\mathrm{\ G}$) occur, the corresponding maximum $|\beta|$ ratio of \He~ is $\sim100\times$ larger in comparison to that of $\mathrm{^6Li}$ (Table~\ref{tab:table1}), thus leading to a higher fidelity ($F=0.9995$) than that of $\mathrm{^6Li}$ ($F=0.9525$). Moreover, at $B=800\mathrm{\ G}$, the maximum $|\beta|$ ratio of \He~ ($|\beta| = 2.106\times10^3$) is on similar order of magnitude to that of $\mathrm{^{23}Na}$ ($|\beta| = 3.428\times10^3$), hence giving rise to fidelity that is comparable to that of $\mathrm{^{23}Na}$ ($F=0.9997$).

Together, these two points suggest that \He~is a better candidate than $\mathrm{^6Li}$ (the only other light fermionic species) for our scientific applications and others including laser-assisted spin-orbit coupling and Raman-induced tunneling~\cite{Jaksch2003,Lin2011,Cheuk2012,Wang2012,Wei2013,Aidelsburger2013,Miyake2013}.

\begin{table}[t!]
\caption{\label{tab:table1}
\textbf{Comparison of the figure of merit for Raman coupling $|\beta|$ and the fidelity $F$} between $\mathrm{^6Li}$, \He, and $\mathrm{^{23}Na}$ at various magnetic fields (B) and detunings ($\Delta$).
}
\begin{ruledtabular}
\begin{tabular}{l|r|r|l|l}
$\mathrm{Atomic\ Species}$&$B\mathrm{\ (G)}$&$\Delta\mathrm{\ (GHz)}\qquad$&$|\beta|$&$F$\\
\colrule
$\mathrm{^6Li}$&$0$&$-1.642\times10^1$&$1.497\times10^2$&$0.9933$\\
$\mathrm{^3He^*}$&$0$&$2.905\times10^1$&$1.959\times10^3$&$0.9995$\\
$\mathrm{^{23}Na}$&$0$&$-7.014\times10^2$&$5.081\times10^3$&$0.9998$\\
\colrule
$\mathrm{^6Li}$&$800$&$-9.008\qquad\ \ $&$2.104\times10^1$&$0.9525$\\
$\mathrm{^3He^*}$&$800$&$2.635\times10^1$&$2.106\times10^3$&$0.9995$\\
$\mathrm{^{23}Na}$&$800$&$-4.609\times10^2$&$3.428\times10^3$&$0.9997$\\
\colrule
$\mathrm{^6Li}$&$0$&$-10^5$&$1.341\times10^2$&$0.9925$\\
$\mathrm{^3He^*}$&$0$&$10^5$&$1.543\times10^3$&$0.9994$\\
$\mathrm{^{23}Na}$&$0$&$-10^5$&$4.411\times10^3$&$0.9998$\\
\colrule
$\mathrm{^6Li}$&$800$&$-10^5$&$1.719\times10^1$&$0.9418$\\
$\mathrm{^3He^*}$&$800$&$10^5$&$1.620\times10^3$&$0.9994$\\
$\mathrm{^{23}Na}$&$800$&$-10^5$&$2.665\times10^3$&$0.9996$\\
\end{tabular}
\end{ruledtabular}
\end{table}

\subsection{Rydberg-mediated entangling gates}
As noted above, we intend to use Rydberg excitation in the $^3S_1$ $F=3/2$ Rydberg series as one technique to perform entangling gates between \He~ atoms. Our excitation pathway [shown in Fig.~\ref{FigureRydberg}(a)] is via the $1s2s$ $^3S_1$ $\leftrightarrow$ $1s3p$ $^3P_J$ transition (at 389 nm) and the $1s3p$ $^3P_J$ $\leftrightarrow$ $1sns$ $^3S_1$ $F=3/2$ transition (at 785 nm). This scheme is essentially the same as that used for alkali species, and we therefore do not anticipate any unique challenges for \He~ in the context of optical excitation to Rydberg states. We note, however, that the light mass of \He~ will exaggerate motional sensitivity. For instance, in typical release-recapture Rydberg excitation schemes, the Doppler shifts will be more pronounced at a given temperature. This could be viewed as a feature in the context of motion-encoded quantum information, but is generally undesirable for spin-based two-qubit gates. Accordingly, it may be preferable to perform Rydberg-mediated gates with the tweezer traps on instead of blinking them off, as is the standard practice in the Rydberg atom array community. To this end, we note that blue-detuned trapping in ``anti-tweezers" offers an advantage in terms of realizing a near-``magic" condition in which the Rydberg state is trapped with nearly the same polarizability as the ``metastable ground" state since the Rydberg electron experiences ponderomotive repulsion from optical potentials~\cite{Isenhower2009,Barredo2020}.  

\begin{figure}[t!]
    \includegraphics[width=0.5\textwidth]{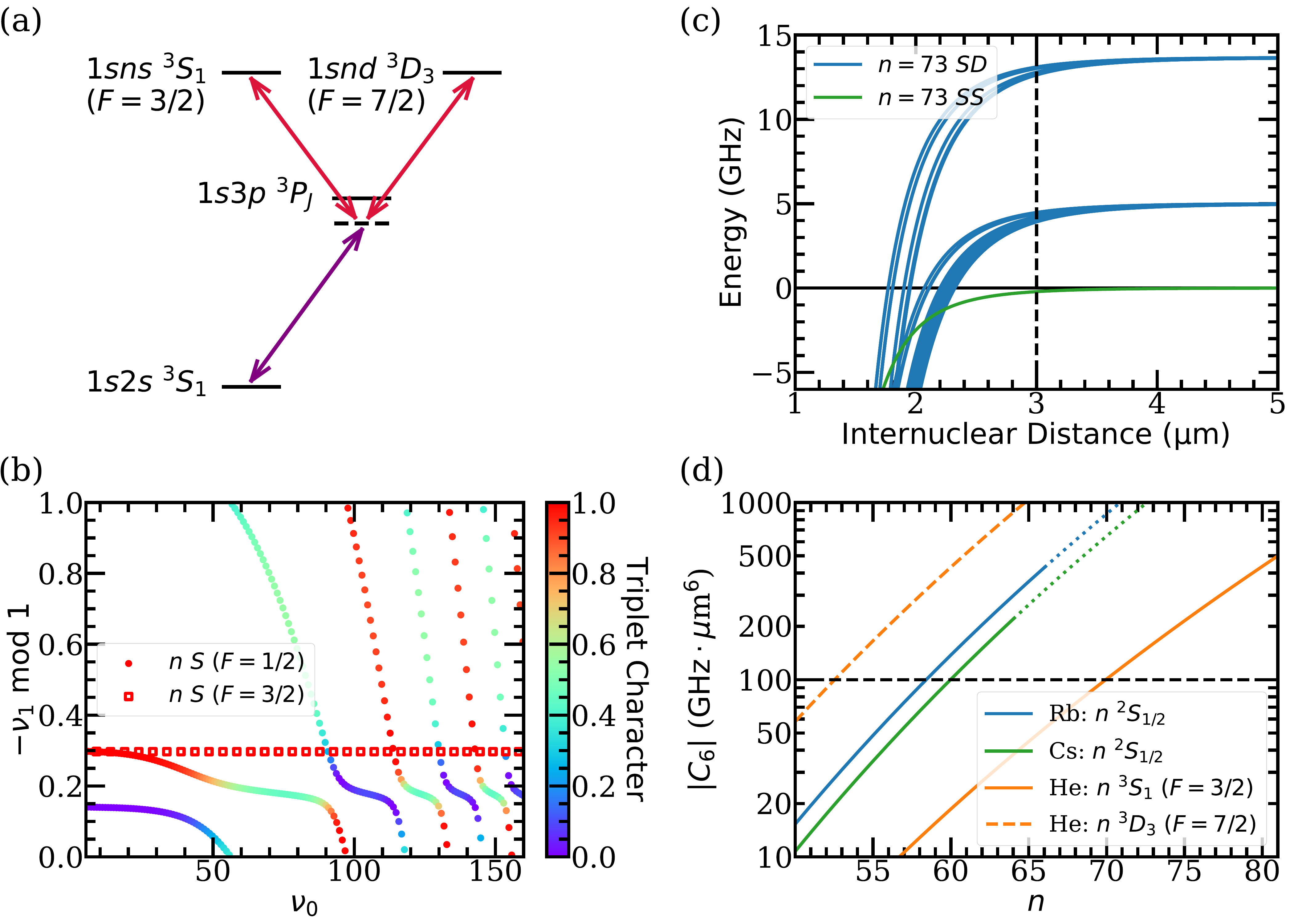}
    \caption{
        \textbf{Rydberg interactions for \He.}
        (a) The two-photon excitation pathway from the $1s2s$ $^3S_1$ ``metastable ground state" to the $1sns$ S-series or $1snd$ D-series Rydberg manifolds via the $1s3p$ $^3P_J$ manifold. (b) The Lu-Fano plots show the bound-state quantum defects of the two spin configurations within the S-series ($F=3/2$ and $F=1/2$). For $F=1/2$, the total electron spin $S$ is not a good quantum number and the series fluctuates between singlet and triplet character. We find that the $F=3/2$ states are spectrally well isolated for our target of $n\approx73$. (c) The spectroscopic energies of the S-S ($M=3$) and S-D ($M=0,1,2,3,4$) Rydberg pair states as a function of interatomic separation for $n\approx73$. This shows clean, perturbative ($C_6$) behavior and spectral isolation of the S-S pair state for separations of $d\gtrsim2.5$ $\mu$m. (d) We plot the $C_6$ coefficients corresponding to the perturbative pair interactions for the S- and D-series of \He ~as well as for the S-series of cesium and rubidium for comparison versus $n$. The target $C_6$ value is shown with the horizontal dashed black line. The $C_6$ coefficients for scenarios above the critical $n$ value, where the van der Waals regime is no longer valid at $\sim3\mathrm{\ \mu m}$, are shown as dotted lines. For detail on the quantum defect and $C_6$ coefficient calculation, see Appendices~\ref{Appendix, FT-MQDT} and~\ref{Appendix, Rydberg}.
        }
        \label{FigureRydberg}
\end{figure}

\subsubsection{Hyperfine structure of the $^3$S$_1$ Rydberg series}

As is the case for all atoms and positive ions, the energy levels of the Rydberg states of $^3$He are characterized by quantum defects that are slowly varying functions of energy. The mathematical expression for the energy levels of a Rydberg series going to ionization threshold $I_i$ is written as follows: 
\begin{equation}
    E = I_{i} - \frac{\mathrm{Ry}(m_{\rm red})}{\nu_{i}^2}=\begin{cases} I_{1}-\frac{\mathrm{Ry}(m_{\rm red})}{\nu_{1}^2}\\\\I_{0}-\frac{\mathrm{Ry}(m_{\rm red})}{\nu_{0}^2} \end{cases}
    \label{eq:mqdtEn}
\end{equation}
where $\nu$ is the effective quantum number, related to the quantum defect $\mu$ and principal quantum number $n$ of a given atomic state by $\nu=n-\mu$ and where $\mu$ varies smoothly and slowly with energy. The reduced-mass-modified Rydberg constant $\mathrm{Ry}(m_{\rm red})=\frac{M}{M+m_e}\mathrm{Ry}_\infty$ is proportional to the reduced mass of the electron-ion system, $\frac{\mathrm{Ry}(\mathrm{^3He})}{\mathrm{Ry}_\infty}\approx0.9998181118795$. For $^3$He there are two hyperfine thresholds, a lower hyperfine level $f_c=1$ and an upper hyperfine level $f_c=0$, split by $I_0-I_1 = 8.66564986577$\,GHz \cite{Schneider2022-gs}.  The Rydberg series of $1sns$ states of $^3$He therefore involve two channels for some symmetries.  For instance, the $1sns$ states with total angular momentum $F= \frac{1}{2}$ have one channel that consists of an $f_c=1$ ionic hyperfine state coupled to the spin-1/2 Rydberg $s$-electron, and a second channel that consists of 
an $f_c=0$ ionic hyperfine state coupled to the spin $\frac{1}{2}$ Rydberg $s$-electron.  These two channels do interact, and that interaction can be expressed as a coupling between the two different total electron spin channels, $^3$S and $^1$S.  The total electron spin is an approximately good quantum number for low $n$ Rydberg states, where the exchange splitting is much larger than the hyperfine splitting of the ion. This contrasts the $1sns$ states with $F= \frac{3}{2}$, which have only one channel that consists of an $f_c=1$ ionic hyperfine state coupled to the spin-1/2 Rydberg $s$-electron. Because this is a single, nondegenerate channel, it implies that the total electron spin is still a good quantum number for this symmetry Rydberg series, see Fig.~\ref{FigureRydberg}(b). Note that this story is largely the same as in ytterbium-171 which is also helium-like and has a spin-1/2 nucleus~\cite{Chen2022,Ma2022,Peper2024}.

\subsubsection{$C_6$ coefficients}

Our goal is to find a Rydberg state that is relatively isolated from nearby perturbers in order to effectively work in the Rydberg blockade regime for implementing high-fidelity controlled-Z gates~\cite{Levine2019,Graham2019}. This requires both sufficient spectral isolation and sufficient Rydberg-Rydberg interaction strength to be deeply in the blockade regime~\cite{Saffman2010}. As noted above and shown in Fig.~\ref{FigureRydberg}(a), we focus on a two-photon excitation scheme from the $1s2s$ $^3S_1$ metastable ground state to the triplet Rydberg series via the $1s3p$ $^3P_J$ intermediate state. Dipole selection rules thus allow us to couple to the $1sns$ $^3S_1$ and $1snd$ $^3D_3$ Rydberg series. We employ frame transformation-multichannel quantum defect theory (FT-MQDT, discussed in Appendix~\ref{Appendix, FT-MQDT}) to study the hyperfine structures and pair interactions of these Rydberg series, and we perturbatively calculate the $C_6$ coefficients for the $ns+ns$ and $ns+nd$ pairs as well as for a stretched state pair of $nd+nd$. See Appendix~\ref{Appendix, Rydberg} for details.

The Lu-Fano plot in Fig.~\ref{FigureRydberg}(b) shows that the $nS\ F=1/2$ series is not ideal to work with due to its hyperfine channel interactions in the energy range of consideration, $n=50-80$. This is not the case for the $nS\ F=3/2$ series, seen on the same figure, which is well separated from the $nS\ F=1/2$ states and which has no interchannel interactions since the series converges only to one ionic hyperfine threshold level. Therefore, our study of the relevant long-range atom-atom interactions concentrates on perturbative calculation of the $C_6$ coefficients for the both atoms in the same $ns\ F=3/2$ state. To ensure that the desired internuclear distance of $\gtrsim$ 2.5$\mu m$ was free of nearby levels that could affect the Rydberg blockade, the $C_6$ coefficients for an $ns\ F=3/2$ state interacting with an $nd$ state atom have also been calculated. The result of this study can be seen in Fig.~\ref{FigureRydberg}(c) where the $ns\ F=3/2 + nd$ curves come down to mark the beginning of the ``spaghetti'' zone, as probed from the $1s3p\ ^3P$ state. A numerical diagonalization confirms that the $ns\ F=3/2$ are free of undesirable $ns+nd$ perturbing states at the internuclear distance of $\gtrsim$ 2.5 $\mu m$. 

In fact for the present scenario, excitation of $nd+nd$ states can also be considered as an alternative to the $ns\ F=3/2$ equal state pairs discussed above. Hence, the $C_6$ values for the $nd+nd$ state interactions have been explored as well. The $nd$ states are quite different than the previously discussed $ns+ns$ state interactions; owing to the smaller $nd$ quantum defects their energy levels lie very close to the degenerate manifold of $^3$He, most importantly close to the $nf$ levels that have nearly zero quantum defects. While their position in the spectrum of $^3$He is not as close to the degenerate manifold as the $f$ and $g$ states, it is close enough that not taking its effect into account would only give an inadequate approximation. A proper analysis of the $nd$ equal state pairs in $^3$He would take the full degenerate manifold of all $n\ell$ levels with $\ell>2$ into account. But in the present exploratory calculation, only the $C_6$ for the stretched pair state $nd\ F=7/2\ +\  nd\ F=7/2,\ M=7$ has been computed.

For comparison against common alkali species, the Alkali.ne Rydberg calculator (ARC) package~\cite{SIBALIC2017319,ROBERTSON2021107814} is used to calculate the $C_6$ coefficients of Rb and Cs via second-order perturbation theory [i.e., solid blue and green curves in Fig. \ref{FigureRydberg}(d)]. In light of the fact such a perturbative calculation is only valid in the long-range limit (i.e., van der Waals regime), the ARC package is also utilized to extract the van der Waals radius through diagonalization of the pair-state interaction Hamiltonian. The critical $n$ value, above which the van der Waals radius exceeds the typical interatomic distance of 3 $\mathrm{\mu m}$ is shown as dotted line on the $C_6$ coefficient curve. It is evident from Fig. \ref{FigureRydberg}(d) that the $ns+ns$ $C_6$ coefficients for $^3$He are significantly smaller than those of $^2S_{1/2}$ states of Rb and Cs. This difference derives primarily from the fact that the s-p quantum defect splitting is close to half of an odd integer for the alkali atoms, meaning that the sum of the nearby alkali $np$ and $(n+1)p$ levels are nearly degenerate with the sum of the energies of the target $ns+ns$ levels. This near degeneracy leads to an unusually small denominator in the second-order perturbation expansion, far smaller than what occurs for the $ns\ F=3/2$ series of $^3$He, thus resulting in a larger $C_6$ for the alkali metal atoms.

Hence, to achieve a typical interaction strength of $\sim0.1$ GHz at 3 $\mathrm{\mu m}$ separation [i.e., $C_6\sim100$ GHz$\cdot\mu\mathrm{m}^6$, black dashed line of Fig. \ref{FigureRydberg}(d)], $ns\ F=3/2\ +\  ns\ F=3/2$ pair-state of \He\ would require a larger principal quantum number ($n\sim70$) than that of alkali atoms. The transition dipole matric element to couple to a Rydberg state from a low-energy state scales as $n^{-3/2}$, such that the transition strength for $n=70$ is still $\sim70$\% that of $n=55$. We also note that although DC Stark shifts from external electric fields scale as $n^7$, they are also dependent on the splittings between nearby opposite parity states, and thus the abnormally small $C_6$ coefficients for the $ns\ F=3/2$ series of $^3$He are expected to also correspond to an abnormally small DC polarizability. Moreover, it should be pointed out that, given their smaller quantum defects, the stretched pair-state $nd\ F=7/2\ +\  nd\ F=7/2$ exhibits much greater $C_6$ coefficients than their $ns+ns$ counterpart (even larger than Rb and Cs), hence requiring a much smaller $n$. We leave the assessment of critical $n$ value for this $nd+nd$ pair-state to future study.

\subsection{Coherent inter-tweezer tunneling \label{sec:tunneling}}

\begin{figure}[t!]
    \includegraphics[width=0.45\textwidth]{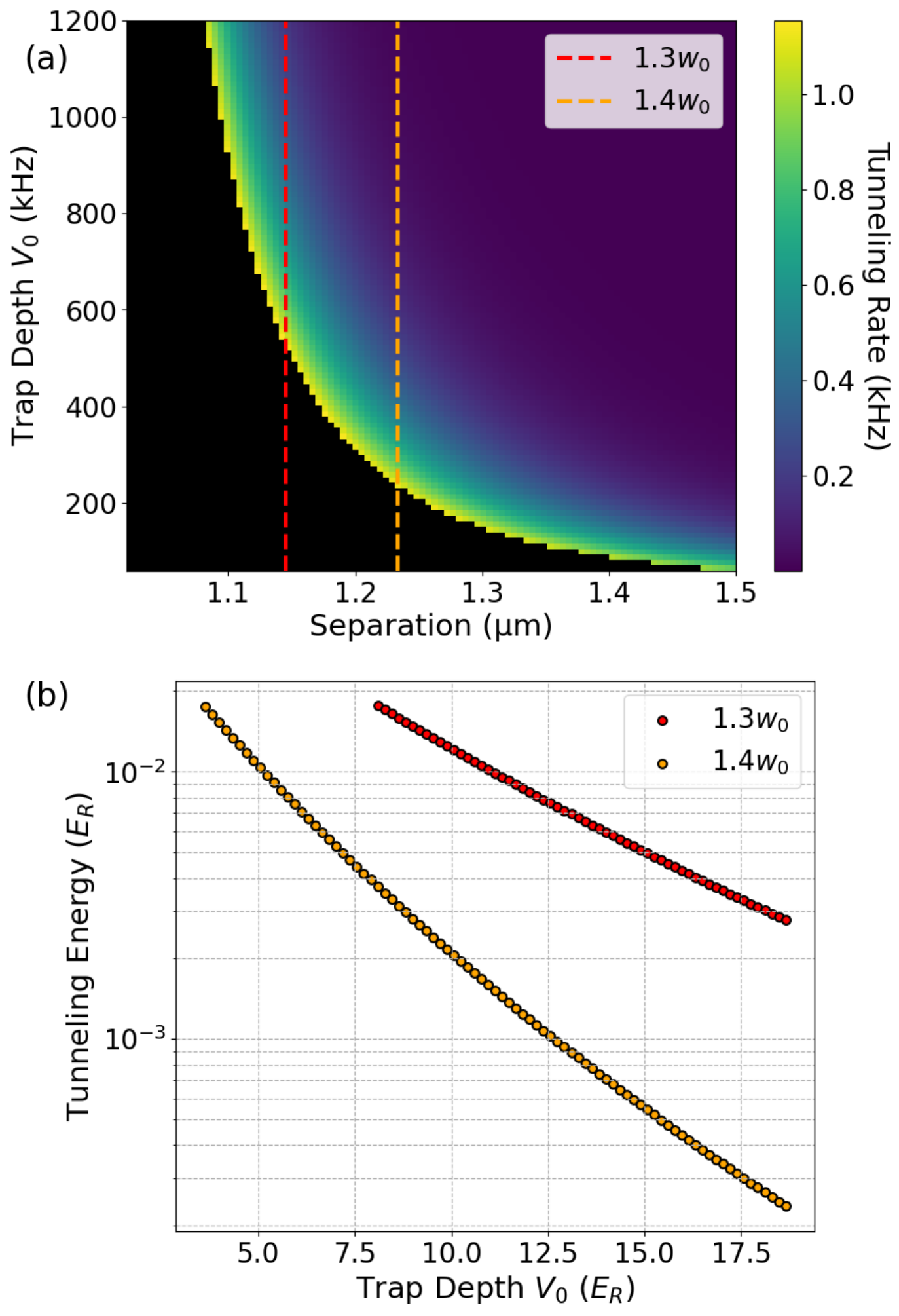}
    \caption{
        \textbf{Inter-tweezer coherent tunneling.}
        (a) Tunneling rate of a single \( ^3\mathrm{He} \) atom confined in a 3D optical double-well potential formed by 1013~nm tweezers, plotted as a function of trap depth \( V_0 \) and inter-tweezer separation. 
        The color scale represents the tunneling rate extracted from the energy splitting between the ground and first excited states of the full Hamiltonian. 
        Dashed vertical lines indicate separations of \( 1.3w_0 \) and \( 1.4w_0 \), corresponding to the one-dimensional cuts shown in panel~(b). 
        The black area is outside the region of interest, where two tweezers have merged sufficiently so as to not support separate bound states. In this area the barrier height is lower than the first excited state of the double well potential.
        (b) Tunneling energy, expressed in recoil units \( E_R \), as a function of \( V_0 \) for these separations, showing exponential suppression with increasing confinement. 
        The results are obtained from static 3D eigenvalue solutions of the full Hamiltonian.
        }
        \label{Figure8tunnel}
\end{figure}

The building blocks of tunnel-coupled tweezers, to a large extent, have already been demonstrated with fermionic Lithium-6 atoms in optical tweezers~\cite{Murmann2015, Spar2022, Yan2022}.
However, in these experiments, cooling of the atom into the optical trap was done by relatively slow evaporation from a bulk degenerate Fermi gas, taking $\sim$10 seconds to prepare; the fast, direct laser-cooling to the 3D motional ground state of lithium has not been demonstrated to date, due in large part to lithium's poorly resolved hyperfine structure. State preparation with \He ~could overcome these challenges, as explained in the previous sections.
Furthermore, the factor-of-two lighter mass contributes to faster absolute tunneling rates proportional to the mass ratio (see Fig.~\ref{Figure8tunnel}), leading to faster gates that are more robust to experimental decoherence and certain types of disorder.

We model a single \( ^3\mathrm{He} \) atom confined in a three-dimensional double-well optical potential formed by two blue-detuned tweezers at a wavelength of 
\( \lambda = 1013~\mathrm{nm} \). 
Alternatively, red-detuned tweezers of 1150\,nm could be used instead.
The tweezers are focused with a numerical aperture of \( \mathrm{NA} = 0.7 \), giving a diffraction-limited waist of $   w_0 = 0.61 \frac{\lambda}{\mathrm{NA}} \approx 0.88~\mu\mathrm{m}$.
The dipole potential of each tweezer is modeled as
\begin{equation}
\begin{aligned}
    V_{\rm dip}(\mathbf{r}) &= -\frac{\alpha(\omega)}{2 \epsilon_0 c} \frac{2 P}{\pi w_0^2 \left( 1 + \frac{z^2}{z_R^2} \right)} 
    \exp\Bigg[-\frac{2 \rho^2}{w_0^2 \left( 1 + \frac{z^2}{z_R^2} \right)} \Bigg],\\
    \rho &= \sqrt{x^2+y^2}, \quad z_R = \frac{\pi w_0^2 }{\lambda}
\end{aligned}
\end{equation}
with \{$ \rho,z$\} the radial and axial coordinates, $P$ the power, $z_R$ the Rayleigh range, and $w_0$ the waist. 
For a blue-detuned trap, the dynamic polarizability $\alpha$ is negative, and the atoms are repelled from regions of high intensity, requiring the generation of an ``anti-tweezer'' potential—an inverted intensity profile relative to the standard red-detuned case. 
Experimentally, this is implemented by inverting the SLM hologram pattern such that low-intensity regions of the projected light correspond to potential wells. 
Visually, the raw hologram resembles a dark spot (low intensity) on a bright background, which, after inversion, forms a stable optical tweezer for a blue-detuned atom. 
The total double-well potential is constructed from two such beams separated by a distance \( d \), and the corresponding single-particle Hamiltonian is
\begin{equation}
    \hat{H} = -\frac{\hbar^2}{2 m} \nabla^2 + V_{\rm dip}\left(\mathbf{r}-\frac{d}{2}\hat{x}\right)+V_{\rm dip}\left(\mathbf{r}+\frac{d}{2}\hat{x}\right)
\end{equation}
The lowest two eigenenergies, \( E_0 \) and \( E_1 \), are obtained numerically, and the coherent tunneling rate is given by
\begin{equation}
    J = \frac{\Delta E}{2} = \frac{E_1 - E_0}{2}.
\end{equation}
By computing for the 3D 
eigenstates of the non-separable Gaussian tweezer potential with the finite difference discretization method using a box size of $5.6d\times4.2w_0\times4.6z_R$ where the energy levels converge sufficiently, (this can alternatively be computed by the discrete variable representation method as well~\cite{Wall2015,Wei2024}) we estimate absolute tunneling rates on the $\sim1$ kHz-scale could be achieved with \He\,in tweezers spaced by 1.2\,$\mu$m (see Fig.~\ref{Figure8tunnel}), compared to $\sim300\,$Hz in the Princeton $^6$Li experiment~\cite{Spar2022}.
We note that for tweezer separations below approximately 1.3\,$w_0$, the tweezer shape's deviation from a Gaussian potential becomes pronounced and is not captured in our model, so the separation of 1.2\,$\mu$m is chosen as exemplary.
Similar results would be obtained with red-detuned tweezers at \( 1150~\mathrm{nm} \), where the slightly larger waist produces only minor changes in tunneling rates. 

Blue-detuned operation provides several advantages: atoms are confined at intensity minima, which suppresses photon scattering and preserves coherent delocalization, as a single photon recoil from the trap is already sufficient to excite the motional wavefunction from the 3D ground state.
Furthermore, in ``anti-tweezer'' configurations, intensity fluctuations mainly modulate the barrier rather than the trap bottom, enhancing tunneling stability. 
To understand this, note that in the harmonic limit of a Gaussian tweezer, the energy offset $\Delta$ between two adjacent blue-detuned tweezers would be the difference of the ``zero-point energy" $\sim\hbar\omega/2$ of the tweezers, which scales as the square root of the tweezer intensity.
This is in contrast to red-detuned tweezers, where $\Delta$ is directly proportional to the tweezer intensity difference.
Uncontrolled variations in $\Delta$ of order the coherent tunneling rate $J$ are the largest source of disorder in a tunnel-coupled red-detuned tweezer array~\cite{Spar2022}, leading to undesired localization at the single-particle level.
For \He, a $h\times300\,$kHz blue-detuned trap could support $h\times1$\,kHz tunneling rates, requiring intensity homogeneity between adjacent sites at the $J/\hbar\omega_z{=}1$\,kHz$/20$\,kHz level, or 5\%, which is achievable with SLM algorithms.
By contrast, a red-detuned trap of the same depth and tunneling rate requires stabilizing the trap depth to within the tunneling at the 1\,kHz/300\,kHz = 0.3\% level, a far more demanding requirement for SLMs or multi-AOD tones.
State-of-the-art SLMs can achieve kHz-scale hologram refresh rates, which is sufficiently fast compared to the kHz-scale tunneling dynamics considered here~\cite{Knottnerus2025, Lin2025}. Moreover, digital micromirror devices (DMDs) have been used to generate anti-tweezers~\cite{Trisnadi2022} and state-of-the-art DMDs have refresh rates of 32 kHz~\cite{Ayoub2021}.

\begin{figure*}[t!]
    \centering
    \includegraphics{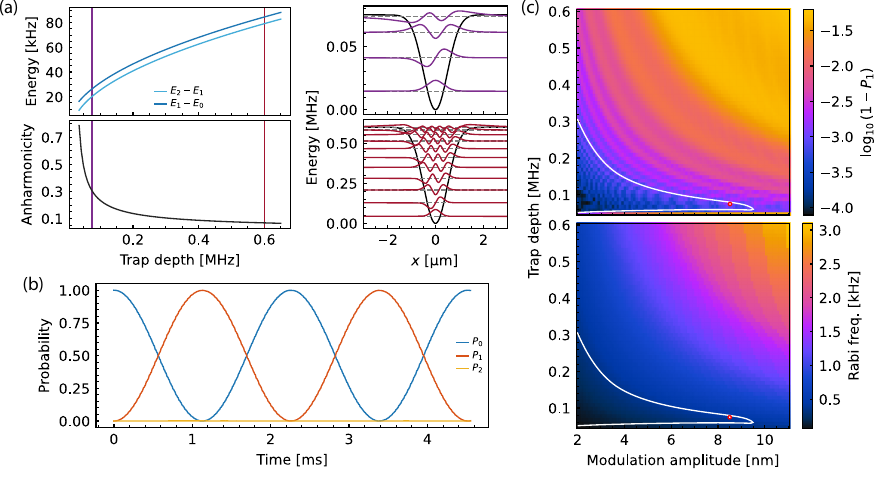}
    \caption{
        \textbf{Motional qubit control via trap position modulation.} (a) The energy intervals between the three lowest-energy motional modes as a function of overall trap depth, and the corresponding anharmonicity as defined in the main text. The eigenstate plots show all bound motional states at depths of $75\,\text{kHz}$ (purple, upper) and $600\,\text{kHz}$ (red, lower). We propose to operate at $75\,\text{kHz}$, where the anharmonicity is $\approx 30\%$, which is well beyond that of typical transmon qubits ($\approx 5-10\%$). (b) Modulating the trap position at frequency $\approx f_{01}$ drives $\sigma_x$ rotations between the $\ket{0}$ and $\ket{1}$ states and does not couple to the $\ket{2}$ state. (c) The approximate $\pi$-pulse fidelity and Rabi frequency plotted as functions of the modulation amplitude and overall trap depth, with the approximate region where fidelities of $> 0.999$--defined as the probability that an initial $\ket{0}$ state is taken to $\ket{1}$--are accessible outlined in white in both plots. The red points show the conditions used for the dynamics shown in (b), which are $8.5\,\text{nm}$ modulation amplitude and $75\,\text{kHz}$ overall trap depth, at which the $\pi$-pulse fidelity is roughly $0.999$.
        \label{Figure8}
    }
\end{figure*}

Finally, blue-detuned operation provides a similar suppression in undesired phase shifts during atom shuttling operations. 
During tweezer motion, changes in optical path experienced by the trap light as well as imperfections and nonlinearities in adaptive optical devices can lead to fluctuating intensities experienced by the atom, which introduces phase errors in the relative fermion wavefunctions for subsequent two-atom operations.
These phase errors can also be partially mitigated by spin-echo protocols.

\subsection{Encoding and manipulating motional qubits}

Atomic motion is a nascent and exciting tool in the atom array toolbox~\cite{Grochowski2023,Scholl2023b,Bohnmann2025}. In the NISQ era~\cite{Preskill2018}, resource efficiency is of vital importance to the pursuit of quantum advantage. Specifically, the overhead associated with quantum error correction remains largely prohibitive. Accordingly, efforts to incorporate ancilla qubits -- other ancilla degrees of freedom -- into data atoms have attracted enormous attention~\cite{Chen2022,Lis2023,Ma2023,Jia2024,Li2025}. Taking it further, efforts to directly encode robust qubits using many states within a single atom or molecule rather than a system of many atoms are very active~\cite{Gottesman2001,Albert2020}. Largely inspired by circuit QED, harmonic degrees of freedom have emerged as a promising platform for these goals~\cite{Crane2024,Liu2024a,Liu2024b} due to their relative ease of control, large Hilbert space, and long coherence time owed to their immunity from environmental noise such as magnetic fields. Indeed, coherence times for motional qubits in optical tweezers have been demonstrated to exceed $T_2^*\approx100$ ms with known technical limitations that can be straightforwardly removed~\cite{Scholl2023b}. 

We propose to encode quantum information in the motional states of \He. Our initial goal is to encode and manipulate a qubit in the lowest two motional states $|n=0\rangle$ and $|n=1\rangle$. Although we view this paradigm as a step towards the control over larger motional subspaces for qudit or logical qubit encoding, we emphasize that a motional qubit already opens many novel capabilities. For instance, a motional qubit could be used as an auxiliary memory qubit that is completely decoupled from the atomic spin qubit. Indeed, one- and two-qubit gates can be performed in the spin sector that are completely decoupled from the motional sector. SWAP operations can be used to convert or ``transduce" between the spin and motion sectors. A generic spin qubit state can be transduced into the motional qubit and be held there as memory while subsequent information processing can be done on the spin qubit. In principle, all three motional directions can be uniquely employed as memory. In fact, the Raman sideband control over all three directions is a necessary ingredient for efficient initialization of the 3D motional ground state.

\He~is well suited for exploiting the anharmonicity of optical tweezer traps due to its high trap frequency. A typical tweezer depth for cooling and imaging heavy alkalis or alkaline earth atoms is 10 MHz (0.5 mK). However, the trap frequency of \He ~in such a trap could be $\approx400$ kHz, meaning that only $\approx20$ bound states are supported. The anharmonicity can be greatly enhanced by further reducing the trap depth. Another option is to utilize optical potentials that are intrinsically anharmonic. Blue-detuned ``bottle-beam" traps are widely used for trapping Rydberg atoms and offer a quartic radial trapping potential~\cite{Isenhower2009,Barredo2020}. For heavy alkalis and alkaline earths, the lowest several trap states are too close together to be resolveable or useful for encoding information. However, the light mass of \He ~offers a unique opportunity to exploit such anharmonic traps.

The most straightforward approach is to use the anharmonicity of a shallow tweezer. Figure~\ref{Figure8}(a) shows the anharmonicity, defined as $(E_1-E_0)/(E_2-E_1)-1$ where $E_i$ is the energy of the $i$th level, as a function of trap depth. We assume a tweezer $1/e^2$ waist radius of $1\,\text{$\mu$m}$. The bound states for trap depths of $\approx$75 kHz and $\approx$600 kHz are also shown in Fig.~\ref{Figure8}(a). We see that the anharmonicity approaches $\approx$30\% for a trap depth of $\approx$75 kHz. In this regime, the lowest two levels can be uniquely controlled and addressed without coupling to the higher levels. We note that the effect of gravity is negligible in either the axial or radial direction over this full range of trap depths, owing to the small mass of \He. For comparison to superconducting circuits, the typical anharmonicity in transmon qubits is $\approx5-10$\%~\cite{Krantz2019}. Accordingly, we anticipate that pulse sequences designed to manipulate superconducting qubits will enhance our toolbox as well. These include the tools that have been used to encode qudits with dimension up to $d=12$~\cite{Blok2021,Wang2025,Champion2025,Li2025SCQ}. 

Since the $|n=0\rangle$ and $|1\rangle$ states have opposite parity, an antisymmetric perturbation is required to couple them, in direct analogy to electric dipole transitions. Figure~\ref{Figure8}(b) shows how a side-to-side modulation of the trap location at the qubit frequency $f_{01}$ drives high-fidelity Rabi oscillations between $|0\rangle$ and $|1\rangle$ without coupling to $|2\rangle$. The amplitude of the modulation determines the Rabi frequency, and we find that $\Omega/2\pi\approx1$ kHz (corresponding to $8.5$ nm modulation for a trap depth of $75$ KHz) can be used without significantly coupling to $|2\rangle$. These modulations drive $\sigma_x$ qubit rotations, but trap perturbations can also be used to perform $\sigma_z$ qubit rotations. A Z-gate can be performed simply by changing the qubit energy splitting in a well-controlled and adiabatic manner such that the frequency deviation times the time spent under that perturbed setting gives the desired phase shift. In this case, the qubit energy splitting can readily be changed by changing the trap depth.

\section{Application to fermionic quantum simulation and computing}

Understanding fermionic systems is a fundamental goal in physics and chemistry and could unlock solutions to important problems in quantum chemistry, condensed matter physics, and high-energy physics.
Fermionic problems -- both dynamical and in equilibrium -- pose enormous challenges to tackle on classical computers due to the notorious ``fermion sign problem.'' 
A major effort in AMO physics toward this goal over the past decades has been to use cold atom-based \textit{analog} quantum simulators, especially to implement the Fermi-Hubbard model, which has proved incredibly fruitful with the discovery of quantum phase transitions, strongly correlated phases, and exotic new quasiparticles~\cite{esslinger2010fermi, daley2022practical}.
The latest iterations of such analog optical lattice simulators offer a degree of programmability or reconfigurability in the lattice potential, allowing the exploration of a wider range of Hubbard models~\cite{bakr2025microscopy}.

In parallel, \textit{digital} quantum computers have made breakneck strides in the race toward universal gate-based quantum computing.
Atom-based computers rely on the flexibility, programmability, and motion of Rydberg-based qubits held in optical tweezer arrays.
Digital methods offer finer programmability, the possibility of error correction, and access to new observables.
However, encoding fermions with qubits is expensive due to fermionic exchange statistics~\cite{Gonzalez2023}, particularly for fermions with long-range interactions.
One can naturally encode a two-level system in the presence or absence of a fermionic particle, much like a two-level spin system of a qubit, but the fermionic anticommutations relations have an additional negative sign in natural operations on the wavefunctions, which usually requires mapping fermionic operators into Pauli strings, such as with Jordan-Wigner~\cite{jordan1928paulische} or Bravyi-Kitaev~\cite{Bravyi2002} transformations [see Fig.~\ref{Figure9}(a)]. 
While early experiments with qubit-based fermionic encodings have recently come online~\cite{barends2015digital, google2020hartree, cochran2025visualizing, nigmatullin2025experimental,
evered2025probing}, using native fermionic operations in a computer based on fermionic atoms~\cite{Gonzalez2023} presents an effective route that removes this challenging overhead [Figs.~\ref{Figure9}(b)-(d)].

\subsection{Experimental design for fermion quantum computing}

\begin{figure}[t!]
    \centering
    \includegraphics[width=0.48\textwidth]{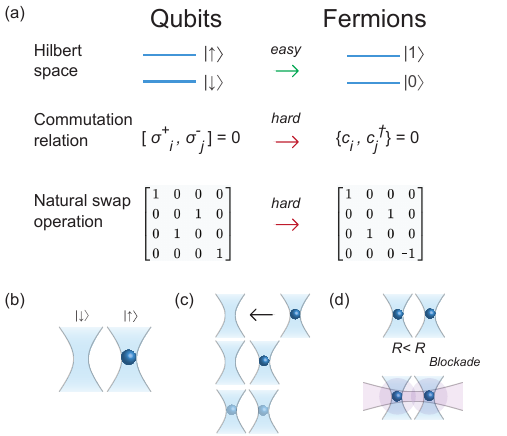}
    \caption{
        \textbf{Fermionic computing with neutral atoms. (a)} Logical qubits and logical fermions are both two-level systems, but the latter operators anti-commute while the former commute. \textbf{(b)} In neutral atoms, logical fermions are encoded in the presence or absence of a particle, shown confined in blue tweezer. \textbf{(c)} Fermionic SWAP gates are performed by allowing tunneling between two tweezers, and \textbf{(d)} CZ gates by Rydberg blockade. 
        \label{Figure9}
    }
\end{figure}

In this subsection, we present an experimental route to creating a quantum information processor using natively fermionic qubits, making use of two-qubit fermionic SWAP operations induced by tunneling.
We assume that the atoms are in the motional ground state, as described in Sec. II.

\begin{figure*}
    \centering
    \includegraphics[width=\linewidth]{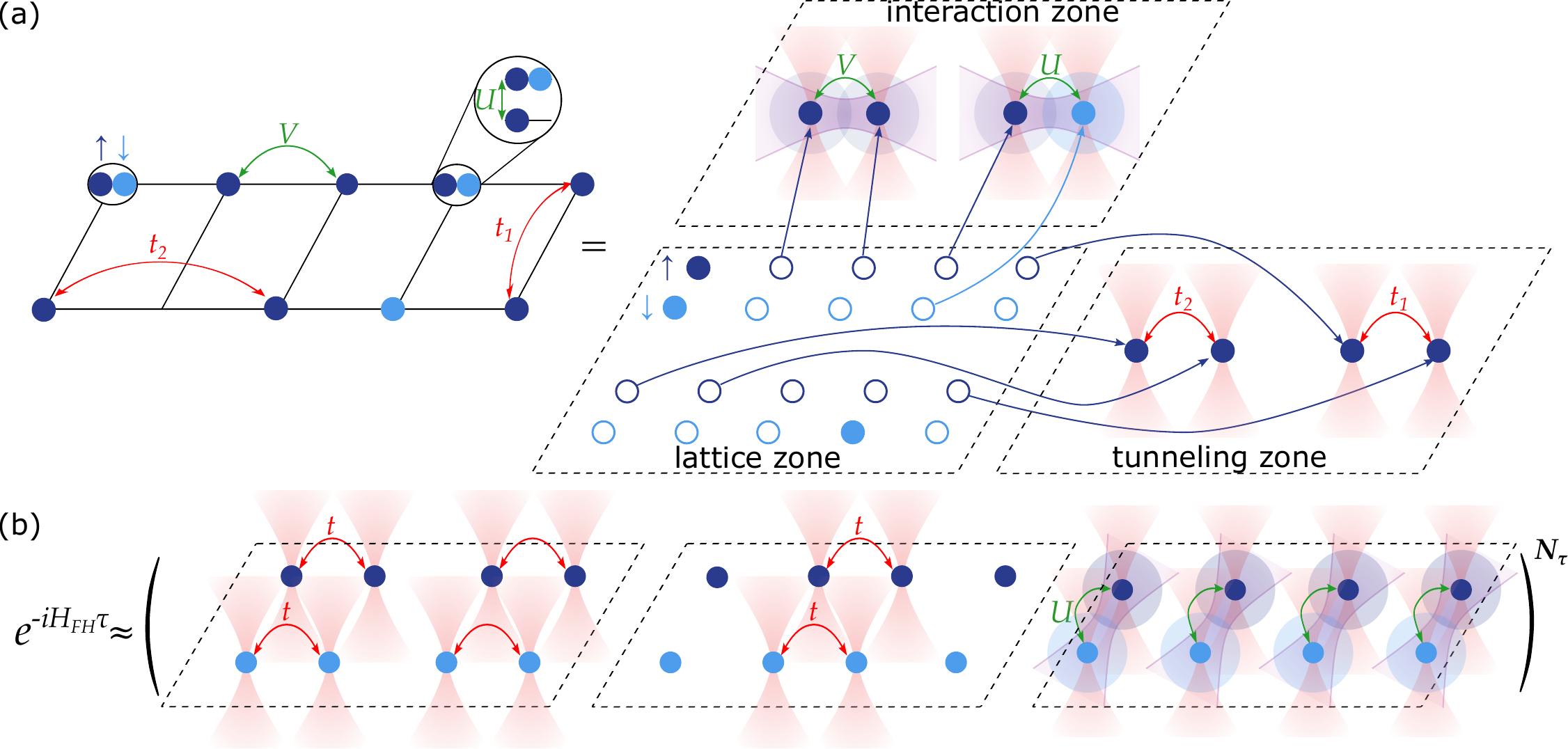}
    \caption{\textbf{Design of the fermionic digital quantum simulator}. (a) The spinful fermion dynamics on a physical lattice. The ``hardcore'' \He~fermions are held in a lattice zone, while the tunneling gate $U^{\rm tun}_{i j}$ and Rydberg interaction gate $U^{\rm int}_{i j}$ between arbitrary pair of sites $i$ and $j$ are realized by tweezer shuttles moving atoms back and forth (labeled by the arrows) from sites $i$ and $j$ to separate gate zones (the dashed regions). With sufficient spatial isolation, both gate zones apply multiple gates simultaneously. The spatial sites (labeled by dark blue and sky blue circles) encode spin (or orbital) degrees of freedom, and $U^{\rm tun}_{i j}$ and $U^{\rm int}_{i j}$ gates dynamically independently realize the tunneling and density-density interaction, either spin-, orbital- and geometry-specific. (b) To simulate the Trotterized quantum dynamics, the digital simulator rearranges atoms and applies gates in the gate zone to time-evolve each term of the target Hamiltonian. The dynamics of a standard Fermi-Hubbard model in a $4$-site chain is sketched step by step, with two rearrangements to simulate nearest-neighbor tunneling terms and one rearrangement to simulate the on-site Hubbard interaction between spin up and down. Then, such Trotter step repeats $N_\tau$ times to realize a time-$\tau$ quantum dynamics.}
    \label{fig:Qsim}
\end{figure*}

We will start with spinless fermions, for which the `qubit' is encoded by fermionic occupation (0 or 1) in a tweezer site. 

Equipped with the reconfigurable tweezer arrays, the \He-based quantum architecture natively dynamically applies the tunneling gates:
\begin{equation}
    U^{\rm tun}_{i j}(\vec{\theta}) = \exp\left\{-i\left[ \frac{\theta_1}{2}(e^{-i\theta_2} c_i^\dagger c^{\phantom\dagger}_j + \mathrm{h.c.}) +\frac{\theta_3}{2}(n_i-n_j)\right]\right\}
\end{equation}
and the Rydberg interaction gates:
\begin{equation}
    U^{\rm int}_{i j} (\theta) = \exp\left\{-i\theta n_i n_j\right\}
\end{equation}
on any pair of tweezer sites $i$ and $j$. To avoid cross-talk in the gate operations, shuttle tweezers are used to move target atoms into and out of designated gate zones, as shown in Fig.~\ref{fig:Qsim}(a). In the tunneling zone, an isolated tunneling gate is realized by bringing a pair of tweezer sites adjacent to each other. The tweezer tunneling and relative on-site potential are controlled by changing the trapping laser detuning~\cite{schuckert2025fault} and tweezer spacing, or even merging two tweezers~\cite{Gonzalez2023}.
Rydberg-mediated two-qubit gates (see Sec.~III B) are applied in the interaction zone. With sufficient spatial separation, both zones can operate multiple gates simultaneously, offering parallel operations on this platform.

This design has demonstrated the essential elements for digital quantum computing. The single-fermion operations, $\mathrm{Z}$, and the two-fermion operations -- $\rm fSWAP$ gate and $\rm CZ$ gate -- are $U^{\rm tun}_{i j}$ and $U^{\rm int}_{i j}$ gates at special parameter sets
\begin{align}
    \mathrm{Z}_{i} \equiv\blank e^{i\frac{\pi}{4}n_i} = U^{\rm tun}_{ii}\left(\frac{\pi}{4},0,0\right),\\
    \mathrm{fSWAP}_{ij} =\blank U^{\rm tun}_{i j}(-\pi,0,0)\mathrm{Z}_{i}\mathrm{Z}_{j},\\
    \mathrm{CZ}_{ij} =\blank U^{\rm int}_{i j} \left(\frac{\pi}{2}\right).
\end{align}
In fact, $U^{\rm tun}_{i j}$ and $U^{\rm int}_{i j}$ gates on arbitrary $i$ and $j$ form a universal gate set for all particle-number-conserving unitaries \cite{oszmaniecUniversalExtensionsRestricted2017,BK2002}. We note that a subset of our fermionic gate set based on intra-site collisions has been demonstrated with $^6$Li in optical lattices, reaching fidelities of over 99.7\%~\cite{bojovic2025high}. 

The universality enables the platform to perform general-purpose quantum operations, with one important direction being the simulation of spinful fermions, even using only spinless fermions. This is important because directly encoding the spin degree of freedom in the \He~spin state causes difficulties implementing on-site $U_{ii}$ interactions due to atom-atom loss processes. As indicated in Fig.~\ref{fig:Qsim}(a) with distinct colors, two sets of tweezer sites can be used to encode spins. A similar design extends the simulation to fermions with pseudo-spin degrees of freedom, such as orbital.

Exciting first applications of the \He-atom platform include simulating the quantum dynamics by Trotterization, and preparing the ground states of the FH-type Hamiltonians by the variational, adiabatic, or other quantum algorithms \cite{Gonzalez2023,Li2023fermionic,Preiss2024,tabaresProgrammingOpticalLatticeFermiHubbard2025,mcardleQuantumComputationalChemistry2020,peruzzo2014variational,haahQuantumAlgorithmSimulating2023,childsNearlyOptimalLattice2019}. In the next subsections, we identify two use cases for a natively fermionic digital computer, focusing on two major applications that extend far beyond the capabilities of existing analog quantum simulation platforms: simulating fermionic lattice models relevant to condensed matter physics, and molecular energy calculations in quantum chemistry. We will also see how the capabilities to simulate the (pseudo-)spin degrees of freedom will play an important role.

\subsection{Quantum simulation of strongly correlated many-body physics}

As concrete examples of the utility of the \He-based digital fermionic quantum simulation platform, we highlight its capacity to explore fermionic quantum many-body physics in the strongly correlated regime of two- and three-dimensional materials, where rich, unsolved physics, such as high-$T_c$ superconductivity (SC), quantum critical metals with non-Fermi-liquid behaviors, and interacting topological phases of matter, reside. 

The spin-$\frac{1}{2}$ Fermi-Hubbard (FH) model is:
\begin{equation}
    H_{\rm FH} = -t\sum_{\braket{ij},\sigma}\left (c^\dagger_{i\sigma} c^{\phantom\dagger}_{j\sigma} + h.c.\right ) + U\sum_{i} n_{i\uparrow} n_{i\downarrow} - \mu \sum_i n_i,
    \label{eq:FH}
\end{equation}
where $\braket{ij}$ refers to the nearest-neighbor sites and $U$ is the onsite interaction,
fails to capture many experimental observations of real materials, even qualitatively. It also has a number of pathological -- fine-tuned -- features that are not enjoyed by real materials, such as the coincidence of integer filling (and associated Mottness), nesting, and a van Hove singularity.
Extensions with kinetic frustration and multiple orbitals, which account for the physics mechanism behind ubiquitously observed phenomena, are of great research interest; yet, they remain largely inaccessible to both classical computational methods and analog quantum simulators.
The state-of-the-art classical computational tools face significant challenges in studying the landscape of phases and dynamical properties in strongly correlated FH-type models, especially when metallic degrees of freedom are prevalent and cannot be explained by the Fermi-liquid description, or when the geometry is complex and high-dimensional.
Meanwhile, analog neutral-atom simulators are limited by their tunability, including fixed tunneling geometries and limited control of physics beyond a single band.

On the other hand, the \He-based fermionic quantum simulator may provide critical insights that surpass those of existing analog quantum simulators and classical algorithms. Perhaps the most direct avenue, but already a fertile one, is to use the \He~platform to programmably simulate quantum dynamics via Trotterization, which repeats $N_\tau$ Trotter steps of a sequence of short-time tunnelings and density-density interactions:
\begin{equation}
    e^{-iH\tau}\approx \left(\prod_{ij} e^{-i\delta\tau  t_{ij} \left(c_i^\dagger c^{\phantom\dagger}_j + \mathrm{h.c.}\right)} \prod_{ij} e^{-i\delta\tau V_{ij}n_i n_j} \cdots\right)^{N_\tau},
\end{equation}
natively realized by $U^{\rm tun}_{i j}$ and $U^{\rm int}_{i j}$, where $H=\sum_{ij} (t_{ij} c^\dagger_i c^{\phantom \dagger}_j + \text{h.c.} +  V_{ij} n_i n_j)$ is the simulation target, $\tau$ is the dynamics time and $\delta\tau=\tau/N_\tau \ll \tau$ is the Trotter time slice. In Fig.~\ref{fig:Qsim}(b), we show how the dynamics of a single-band spinful FH model can be simulated on a $4$-site chain via Trotterization by illustrating a rearrangement sequence in one Trotter step, by encoding the spin degree of freedom in trap location to avoid multiple occupancies. Similar to what is shown in this case, for all the short-range interacting models we propose to simulate in this section, the total gate count scales linearly with the system size $L$, while the number of rearrangements is constant in each Trotter step.
Since $i$ and $j$ could be arbitrary for spatial, spin, and orbital degrees of freedom, this platform readily simulates FH-type models with extended-range tunneling and density-density interaction terms, as well as high-dimensional geometries, as we will show in the three examples for the rest of this section.

\subsubsection{Study geometric frustration for high-$T_c$ superconductivity}

The extensive study of the single-band original FH model \cite{leeDopingMottInsulator2006} on the 2D square lattice reflects the massive effort to reveal the unconventional fermion pairing mechanism in high-$T_c$ superconductivity, as well as the motion of holes or other dopants more broadly, as reflected in pseudogap and strange metal phenomena. While a variety of experimentally pertinent phenomena appear in the Hubbard model, including the robust antiferromagnetic spin order at half-filling and charge order such as stripes for special dopings~\cite{bourgundFormationIndividualStripes2025,simonscollaborationonAbsenceSuperconductivityPure2020,huangStripeOrderPerspective2018,maiIntertwinedSpinCharge2022,poilblancChargedSolitonsHartreeFock1989,whiteGroundStatesDoped1997,whiteDensityMatrixRenormalization1998,wietekStripesAntiferromagnetismPseudogap2021,xuStripesSpindensityWaves2022,zaanenChargedMagneticDomain1989,zhengStripeOrderUnderdoped2017}, it does not appear to host a superconducting phase, at least in any appreciable regime comparable to that of the real materials~\cite{simonscollaborationonAbsenceSuperconductivityPure2020,eberleinSuperconductivityTwodimensionalt1t2Hubbard2014}. 
The mismatch between the phase diagram of the FH model and that of realistic materials has inspired studies to include key features of these materials. One direction is to include the next-nearest-neighbor (NNN) tunneling ($t'$ term) to form the $t$-$t'$ model:
\begin{equation}
    H_{t\text{-}t'} = H_{\rm FH} - t'\sum_{\braket{\braket{ij}},\sigma}\left (c^\dagger_{i\sigma} c^{\phantom\dagger}_{j\sigma} + h.c.\right ),
    \label{eq:t1t2}
\end{equation}
where $\braket{\braket{ij}}$ labels all NNN site pairs $i$ and $j$. The $t'$ term kinetically frustrates the spin orders and pushes the ordering more favorably towards $d$-wave pair-forming~\cite{eberleinSuperconductivityTwodimensionalt1t2Hubbard2014,zhangFrustrationInducedSuperconductivityt1t22025}. 
It also breaks the particle-hole symmetry in the original FH model in the bipartite lattices, enabling the study of the discrepancy between electron- and hole-doped regimes lacking in the original FH model~\cite{zhangFrustrationInducedSuperconductivityt1t22025}.

While classical methods suggested the existence of the superconducting phase at $T=0$ in the $t$-$t'$ model~\cite{xuCoexistenceSuperconductivityPartially2024,jiangGroundstatePhaseDiagram2021,whiteCompetitionStripesPairing1999a}, there are plenty of questions unsolved: the stability of the potential $d$-wave superconductivity to thermal fluctuation -- the dome-shaped $T_c$ as a function of doping density has yet to be confirmed -- and the pairing mechanism like the $k$-space pairing gap. The appearance of and mechanisms related to phenomena such as the pseudogap and strange metal also remain poorly understood in this model. Computing dynamical quantities, such as conductivity and inelastic magnetic structure factors, would help understand the coexistence and structure of superconductivity, spin, and charge orders, and their sensitivity to doping and $t'$, is important yet challenging for classical methods.
Analog fermionic quantum simulators have limited access to the rich physics introduced by NNN tunneling. The NNN tunneling in the neutral atom quantum simulation experiments is usually weak -- in fact, for typically used separable lattices, it vanishes due to the structure of the Wannier functions, while for many more general lattices, the NNN tunneling tends to be small when confined to the Hubbard regime~\cite{Wei2024}. Analog quantum simulators may engineer substantial, tunable anisotropic $t'$ tunneling in a few tunable yet highly restricted geometries, such as the square-triangle crossover~\cite{xuFrustrationDopinginducedMagnetism2023}. While triangular lattices possible host SC phases \cite{chenUnconventionalSuperconductivityTriangular2013,xuPairDensityWave2019,zhuDopedMottInsulators2022}, they fail to capture the $C_4$ crystal symmetry of most high-$T_c$ superconductors~\cite{chmaissemNeutronPowderDiffraction1993,titovaCrystalElectronicStructure2016,chenIronbasedHighTransition2014,liuElectronicMagneticStructure2020,wangStructureResponsibleSuperconducting2024}. 

The \He-based digital architecture, on the other hand, offers programmable and individually and arbitrarily tunable $t$ and $t'$ tunnelings in arbitrary lattice geometries -- even with lattice defects or in three dimensions, as may be important for real materials -- via the tunneling gates shown in Fig.~\ref{fig:Qsim}(a), making it a powerful dynamical and single-site-resolvable tool. The tweezer platform also allows greater flexibility and gate efficiency than proposals based on optical superlattices~\cite{tabaresProgrammingOpticalLatticeFermiHubbard2025}. The relevant superconducting critical temperature $T_c\sim 0.1t'\sim0.03 t$ observed in real materials \cite{pavariniBandStructureTrendHoleDoped2001} is ``cold'' but not implausible in near-future experiments. For example, a recent large-scale neutral-atom Hubbard simulator has already reported temperatures as low as $T\simeq 0.05 t$ \cite{xuNeutralatomHubbardQuantum2025}; on the tweezer array side, in addition to dynamical techniques that can be transferred from optical lattices, entropy engineering such as post-selection on low-defect shots~\cite{Yan2022} also provides a route for state-of-the-art experiments to reach temperatures close to the target regime.

\subsubsection{Engineer multi-orbital many-body physics}

Faithful low-energy descriptions of realistic materials, including cuprate \cite{damascelliAngleresolvedPhotoemissionStudies2003} and multi-layer nickelate high-$T_c$ superconductors \cite{nomuraSuperconductivityInfinitelayerNickelates2022,sunSignaturesSuperconductivity802023,zhangHightemperatureSuperconductivityZero2024,liSuperconductivityInfinitelayerNickelate2019,liaoElectronCorrelationsSuperconductivity2023}, iron pnictides \cite{sobotaAngleresolvedPhotoemissionStudies2021}, and heavy-fermion compounds \cite{checkelskyFlatBandsStrange2024a,poseyTwodimensionalHeavyFermions2024}, require active orbital degrees of freedom beyond the single-band Hubbard model. These can be straightforwardly implemented in the \He~architecture as well. A broad class of examples is captured by the multi-band FH model:
\begin{align}
    H_{\rm mFH} 
    = \blank-\sum_{ij,mm',\sigma}t^{(mm')}_{ij}\left (c^\dagger_{im\sigma} c^{\phantom\dagger}_{jm'\sigma} + \mathrm{h.c.}\right ) \nonumber\\
    \blank\hspace{0.3in}+ \sum_{ij,mm'}U^{(mm')}_{ij} n_{im} n_{jm'} - \mu \sum_{im} n_{im},
    \label{eq:mbFH}
\end{align}
where $m$, $m'$ are the band indices, and $t$ and $U$ are generalized to be extended-range and band-dependent. 

Understanding multi-band metallic systems is challenging, and the simplest examples offer low-hanging fruit for quantum simulation. For example, in the Kondo lattice systems below the Kondo temperature \cite{gleisEmergentPropertiesPeriodic2024}, the Doniach competition between the Kondo coupling and the RKKY coupling leads to the peculiar non-Fermi-liquid behaviors and the loss of quasi-particle weight at the fan region right above the quantum critical point (QCP)~\cite{gleisEmergentPropertiesPeriodic2024,huQuantumCriticalMetals2024,hartnollColloquiumPlanckianDissipation2022}.
The nature~\cite{kuglerOrbitalSelectiveMottPhase2022,huQuantumCriticalMetals2024} and the behaviors of this Kondo destruction QCP are a topic of disagreement, given the challenges of classical calculations.
In addition to the two-band FH model, the periodic Anderson model (PAM) \cite{gleisEmergentPropertiesPeriodic2024} is also proposed to capture the quantum critical physics of an itinerant conduction band locally hybridized with a lattice of strongly correlated $f$ orbitals:
\begin{align}\label{eq:PAM}
H_{\rm PAM}
=\blank
\sum_{\bm{k},\sigma}\epsilon_kc_{\bm{k}\sigma}^\dagger c^{\phantom\dagger}_{\bm{k}\sigma} +V\sum_{i,\sigma}\!\left(c_{i\sigma}^\dagger f^{\phantom\dagger}_{i\sigma}+{\rm h.c.}\right) \nonumber\\
\blank\hspace{0.3in} +\epsilon_f\sum_{i,\sigma} n^f_{i\sigma}+U\sum_i n^f_{i\uparrow}n^f_{i\downarrow},
\end{align}
where $c$ ($f$) denotes the conduction (localized) orbital and $n^f_{i\sigma}=f_{i\sigma}^\dagger f^{\phantom\dagger}_{i\sigma}$.

Another realm where multi-orbital models have attracted broad attention is \moire~materials. \Moire~materials have a tunable, narrow bandwidth and carrier filling~\cite{xieFractionalChernInsulators2021}, suited for exploring both persisting and emerging puzzles. One such puzzle is the pairing mechanism of observed superconductivity \cite{caoUnconventionalSuperconductivityMagicangle2018,luSuperconductorsOrbitalMagnets2019,ohEvidenceUnconventionalSuperconductivity2021,kimEvidenceUnconventionalSuperconductivity2022,xieInteractingElectronsTwisted2022}. The SC $T_c$ relative to the bandwidth in twisted bilayer graphene (TBG) is extremely high, even compared to high-$T_c$ superconductors \cite{caoUnconventionalSuperconductivityMagicangle2018}, suggesting an unconventional pairing mechanism. However, irreconcilable experimental signatures contradict such a claim, calling for more studies towards a conclusive understanding~\cite{nuckollsMicroscopicPerspectiveMoire2024}. 
Another puzzle lies in the correlated insulating phases. At integer \moire-band fillings, the correlated insulating phase \cite{caoSuperlatticeInducedInsulatingStates2016,caoUnconventionalSuperconductivityMagicangle2018,yankowitzTuningSuperconductivityTwisted2019,luSuperconductorsOrbitalMagnets2019,caoStrangeMetalMagicAngle2020} is believed to be associated with the iso-spin band degeneracy lifting \cite{nuckollsMicroscopicPerspectiveMoire2024}, yet its coexistence with the adjacent dome-shaped SC phase is not understood. At fractional fillings, the observed insulating generalized Wigner crystal (GWC) is likely tied with the competing fractional Chern insulator (FCI)~\cite{xieFractionalChernInsulators2021,caiSignaturesFractionalQuantum2023a,parkObservationFractionallyQuantized2023,xuObservationIntegerFractional2023,xuInterplayTopologyCorrelations2025}. Studying the possible symmetry-breaking GWC melting into topological FCI would be interesting.
Notably, the \moire~physics is intrinsically connected to the multi-band-ness: the insulating GWC cannot be modeled by the single-band Mottness, and more essentially, any non-trivial topological property cannot exist in a single-band system, as captured by the effective models \cite{bistritzerMoireBandsTwisted2011,koshinoMaximallyLocalizedWannier2018a,khalafMagicAngleHierarchy2019,calugaruTwistedSymmetricTrilayer2021a}.

Accessing the rich multi-band physics described by models such as Eqs.~\ref{eq:mbFH} and~\ref{eq:PAM} poses a significant challenge for analog quantum simulators. Alkaline-earth atom optical lattices simulate high-spin models with tunable spin-symmetric~\cite{taieObservationAntiferromagneticCorrelations2022,ibarraManybodyPhysicsUltracold2024} or non-symmetric~\cite{mongkolkiattichaiQuantumGasMicroscopy2025} Hubbard interaction strengths, but are unable to tune orbital-selective tunnelings and inter-orbital hybridizations, given the insensitivity of hyperfine levels to the laser detuning, and generally do not have independent control of inter-orbital interactions, having only a single control parameter (magnetic field) to control all of them through a Feshbach resonance. 
Protocols using motional states suffer from multi-particle loss and lack orbital-selective tunability~\cite{Wei2024}. 
Optical superlattice proposal \cite{schlomerLocalControlMixed2024} and experiment \cite{gallCompetingMagneticOrders2021} are promising bilayer quantum simulators, but they also lack local tunability and geometric flexibility.

On the other hand, the reconfigurable \He-based digital platform with locally controllable tunnelings, interactions, and orbital-selective fillings is well-suited for multi-orbital simulations. Equations~\ref{eq:mbFH} and~\ref{eq:PAM} can be immediately implemented with Trotterization in exactly the same fashion as the single orbital case, by encoding different orbitals in different spatial sites of the tweezers.
For Kondo lattice models, a simulation target is to reach the Kondo onset temperature $T_0$, which is at a scale comparable to the renormalized superconducting transition temperature $T_c$. 
Measuring non-Fermi-liquid signatures at this temperature range would clarify the range of the quantum critical fan. The scalable quantum platform offers $k$-space resolvability for long-wavelength physics in this metallic system, which classical dynamical methods struggle to match. This resolvability would be the key to understanding the QCP and the Fermi-surface reconstruction.
The \He-based platform would also deepen the understanding of the \moire~physics beyond the implications from the self-consistent mean-field theories, whose accuracy is reliable only in special limits~\cite{peottaSuperfluidityTopologicallyNontrivial2015}. Detecting $k$-space pairing gap and \moire~band filling helps understand the underlying SC pairing mechanism in TBG and its robustness; the dynamical tuning and imaging over the GWC-FCI transition can probe the unexplored interacting topological phase transition.

\subsubsection{Resolve single sites in 3D strongly correlated physics}

One long-time research subject in strongly correlated systems is the role of dimensionality. In real three-dimensional (3D) materials, anisotropy can introduce an interplay between physics in reduced dimensions and physics in 3D. Understanding this is challenging: numerical methods face even greater challenges in 3D, and analog fermionic neutral atom simulators lack single-site resolution beyond 2D due to fundamental limitations in quantum gas microscopy.

The digital \He-based platform instead offers flexible control over the lattice connectivity, making dynamical, single-site-resolved simulation of FH-type models on arbitrary lattice geometries, including 3D lattices, possible.
This vastly opens the range of dimension-related questions in strongly correlated systems to quantum simulation, a few examples of which follow.

A first example of such a direction is to understand the role of anisotropy in models of the cuprate superconductors. While the superconducting mechanism is widely believed to be of a 2D nature, the effect of inter-layer tunneling is not fully clear. The $c$-axis charge transport \cite{kleinerIntrinsicJosephsonEffects1992} suggests a weak Josephson-type coupling in the superconducting phase; however, relevant microscopic and phenomenological models fail to explain the strong dependence of the transition $T_c$ on the layer thickness \cite{leggettAspectsTwodimensionalityCuprates}. Even though Berezinskii-Kosterlitz-Thouless (BKT) physics explains a suppressed $T_c$ in the 2D limit for some cuprate and transition metal dichalcogenides (TMD) superconductors \cite{saitoHighlyCrystalline2D2016a,dawsonApproachingTwoDimensionalSuperconductivity2020}, the experiment-observed thickness dependence is orders of magnitude stronger than the theory. Moreover, experiments on monolayer Bi-2212 \cite{jiangHighTcSuperconductivityUltrathin2014,yuHightemperatureSuperconductivityMonolayer2019} show no significant change in $T_c$, and most strikingly, enhanced $T_c$ is reported for the monolayer FeSe/STO interface \cite{Wang_2012}. A comprehensive study on the microscopic inter-layer coupling is necessary to gain a solid understanding of the relationship between SC pairing and inter-layer coupling. Another question is about the quantum criticality of 3D Kondo lattice metals. Beyond all points above, the sharply different windows of the strange-metal behaviors in the 3D heavy-fermion metals \cite{gegenwartQuantumCriticalityHeavyfermion2008,liuObservationAntiferromagneticQuantum2020,hartnollColloquiumPlanckianDissipation2022} and the layered cuprates \cite{phillipsStrangerMetals2022,checkelskyFlatBandsStrange2024a} as well as 2D FH models \cite{huangStrangeMetallicityDoped2019} could also be studied by realizing the microscopic models with the proposed digital quantum simulator. 

\subsection{Molecular energy calculations with spinless fermions}

Another concrete example of the utility of a spinless fermionic simulator is finding ground states of systems of many interacting fermions, a challenging problem for chemistry and materials science. A variational quantum eigensolver (VQE)~\cite{peruzzo2014variational} can be implemented with spinless \He ~fermions in tweezers, with the goal being to prepare an individual eigenstate (typically the ground state) in molecular Hamiltonians.  
At its core, the VQE minimizes a cost function determined by average energy as the optimal circuit converges to the ground eigenstate as the output. The target molecular Hamiltonian is~\cite{Gonzalez2023}
\begin{equation}
    \mathcal{H} = \sum_{ij} t_{ij}^{(1)} c_i^{\dagger} c_j + \sum_{ijkl} t_{ijkl}^{(2)} c_i^{\dagger} c_j^{\dagger} c_k c_l
    \label{eq:vqe}
\end{equation}
with complex coupling parameters $t_{ij}^{(1)}$ and $t_{ijkl}^{(2)} $.
(Other formulations of the molecular Hamiltonian are also possible, see Ref.~\cite{Li2023fermionic}).
The simulation starts with a choice of a trial wavefunction $|\Psi(\vec{\phi})\rangle = U(\vec{\phi})|\Psi_0\rangle$, where $U$ is a unitary operator with parameters $\vec{\phi}$.
One designs a circuit to output the average energy $\langle \Psi(\vec{\phi})|\mathcal{H}|\Psi(\vec{\phi})\rangle$, and by minimizing this quantity with respect to the parameters $\vec{\phi}$, one obtains the ground state energy for an optimal set of parameters: $E_{\rm ground} = \langle \Psi(\vec{\phi_{\rm g}}) | \mathcal{H} | \Psi(\vec{\phi_{\rm g}})\rangle$.
One layer of the fermion circuit for a 1D chain could, for example, consists of first the tunneling gates $U^{\rm tun}$ across nearest-neighbor bonds, followed by the interaction gate $U^{\rm int}$ across the same bonds~\cite{Li2023fermionic,Gonzalez2023} (with order $N$ two-qubit operations). 

A resource estimation of fermionic implementations of VQE, compared to qubit implementations~\cite{Li2023fermionic},
shows that fermionic circuit constructions are superior to qubit circuit constructions even in 1D chains (by a factor of $\sim$2-3 in two-body gate count, and fewer iterations with shallower circuits), with the advantage growing with higher-dimensional systems or for larger interaction strengths. 

\section{Concluding discussion}
We have presented a comprehensive architectural blueprint for the use of \He ~atoms in programmable optical tweezer arrays. This included a concrete analysis of atomic structure considerations as well as Rydberg-mediated interactions. We have shown that inter-tweezer hopping of \He ~atoms can be $\gtrsim3\times$ faster than previous demonstrations with lithium-6. We have also demonstrated a new toolbox for encoding and manipulating qubits directly in the tweezer trap potential, uniquely enabled by the light mass of \He. Finally, we have provided several examples of new opportunities for fermionic quantum computation that leverage the transport and inter-tweezer hopping of \He ~atom arrays.

We also anticipate several compelling opportunities that combine all three of the unique advantages of \He ~described above: fermionic inter-tweezer hopping, bosonic intra-tweezer motional state encoding, and ultrafast tweezer-based atomic transport. Broadly, these opportunities pertain to physical and chemical systems with itinerant fermionic modes coupled to a bosonic bath -- a paradigm that is ubiquitous in nature. For example, lattice gauge theories~\cite{Wilson1974,Bazavov2010} (LGTs) provide the foundation for our nonperturbative understanding of strong interactions, including hadron spectroscopy~\cite{Bulava:2022ovd}, 
the equation of state of the quark-gluon plasma~\cite{Borsanyi:2013bia,HotQCD:2014kol}, 
and the nuclear potential~\cite{Beane:2003da}. They also describe condensed matter phenomena including deconfined criticality and the fractional quantum Hall effect~\cite{Karthik:2015sgq}. 
However, the simulation of LGTs and nuclear matter in particular is plagued by several challenges~\cite{Bauer2023}. In this context, our \He ~architecture offers a method to (1) utilize the harmonic oscillator mode of tweezer-trapped atoms for direct and dense encoding of gauge bosons~\cite{Grochowski2023,Scholl2023b,Crane2024,Liu2024a,Liu2024b,Bohnmann2025}, and (2) utilize the intrinsic fermionic quantum statistics of our atoms for direct implementation of fermionic hopping operations that sidesteps the costly Jordan-Wigner transformation of fermions to spins~\cite{Jordan1928,Bravyi2002}. 

Countless phenomena in condensed matter physics, materials science, and quantum chemistry share this paradigm. These include phonon-mediated electron pairing in materials such as the BCS theory of superconductivity~\cite{Bardeen1957}, the interplay of electrical conductivity and magnetism with thermal conductivity~\cite{Gu2018}, optical conductivity and plasmonic behavior in photon-dressed materials~\cite{Yu2019}, and chemistry outside of the Born-Oppenheimer approximation in which nuclear motion during the electron dynamics cannot be neglected~\cite{McArdle2020}. Our resource-efficient hardware implementation of dense boson encoding and direct fermionic hopping at the hardware level will enable complementary studies across this wide spectrum of physical and chemical manifestations. Moreover, we foresee an intimate connection between the direct encoding of bosonic modes and bosonic error correction as a resource-efficient strategy for fault-tolerant quantum computation~\cite{Liu2024a,Bohnmann2025}. To this end, we will anticipate methods to interpret phononic/gauge modes in a physical model as superpositions of code words for the, e.g., `grid states' of the Gottesman-Kitaev-Preskill (GKP) logical qubit~\cite{Gottesman2001}.

Finally, we highlight the opportunities for metastable helium atom arrays for precision measurements of fundamental physics. As shown in Fig.~\ref{Figure2}(a), He* offers a narrow ``clock" transition from the 1s2s $^3S_1$ ``metastable ground state" to the 1s2s $^1S_0$ ``clock state." The lifetime of this state is $\tau=20$ ms, limited by decay to the absolute ground state. The 8 Hz natural linewidth of this ``clock" transition offers an excellent system for testing fundamental physics. Namely, since nuclear physics calculations for atoms as light as helium are tractable, isotope shifts of the clock transition for \He ~and $^4$He* offer a direct probe of the nuclear size and its relation to the ``proton charge radius puzzle"~\cite{vanRooij2011,LiMuli2025}. Although remarkable precision measurements have been performed of this isotope shift, they have only been done in bulk gases (a Bose-Einstein condensate of $^4$He* and a degenerate Fermi gas of \He). Crucially, collisional shifts and the Doppler shifts due to degeneracy pressure in the \He ~Fermi gas limit the linewidth of the clock transition to $\approx10$ kHz~\cite{vanRooij2011,Rengelink2018}, which is roughly three orders of magnitude above the natural linewidth. Following the development of ``atomic array optical clocks" of alkaline earth(-like) atoms~\cite{Norcia2019,Madjarov2019}, we anticipate that metastable helium atom arrays offer an incredible opportunity for precision measurements. In fact, the isotope shifts are large enough that dual-isotope~\cite{Sheng2022,Nakamura2024} metastable helium atom arrays are quite realistic. The ``clock-magic" wavelength that has been used for these bulk studies is $\approx$320 nm, which is blue detuned. Following from the blue-detuned ``anti-tweezer" trapping proposed above and demonstrated with cesium~\cite{Trisnadi2022}, we believe that such as system is straightforward, and the atoms could be transferred back and forth~\cite{Young2020} between tweezer arrays at the wavelengths proposed above (1013 or 1150 nm) for loading, imaging, and cooling.  

\begin{acknowledgments}
We thank Peter Mueller, Michael Bishof, Connor Holland, Lawrence Cheuk, Jon Hood, Cheng Chin, Fang Xie, and the Covey Lab for stimulating discussions. We thank Connor Holland for feedback on the manuscript. J.P.C acknowledges funding from the DOE Early Career Award from the Office of Nuclear Physics Quantum Horizons Program (Award No. DESC0025655) and the Army Research Office (Award No. W911NF-25-1-0214).  K.R.A.H. and H.-T.W. acknowledge support from the National Science
Foundation (DGE-2346014), the Department of Energy (DE-SC0024301), the W. M. Keck Foundation
(Grant No. 995764), and the Office of Naval Research (N00014-20-1-2695). Z.Y.~acknowledges funding from the Neubauer Family Assistant Professors Program and the David and Lucile Packard Foundation (2024-77404).  C.H.G. and J.D.P. have received partial support from the Department of Energy, Office of Science, Basic Energy Sciences, Award No. DE-SC0018251.
\end{acknowledgments}
\appendix
\section{\He~polarizability}
\label{Appendix, He polarizability}
Following the convention commonly found in literature \cite{Steck2007}, the ac Stark shift can be expressed as
\begin{align}
    \begin{split}
        &\Delta E(F,m_F;\omega) = -\alpha^{(0)}(F;\omega)\left|E_0^{(+)}\right|^2 \\
        &- \alpha^{(1)}(F;\omega){\left(i\mathbf{E}_0^{(-)}\times\mathbf{E}_0^{(+)}\right)}_z\frac{m_F}{F}\\
        &-\alpha^{(2)}(F;\omega)\frac{\left(3\left|E_{0z}^{(+)}\right|^2 - \left|E_0^{(+)}\right|^2\right)}{2}\left(\frac{3m_F^2-F(F+1)}{F(2F-1)}\right)
    \end{split}
\label{AC Starck Shift}
\end{align}
where the scalar, vector, and tensor polarizabilities are
\begin{align}
    \begin{split}
        \alpha^{(0)}(F;\omega) =& \sum_{F'}\frac{2\omega_{F'F}{\left|\langle F||\mathbf{d}||F'\rangle\right|}^2}{3\hbar(\omega_{F'F}^2-\omega^2)}\\
        \alpha^{(1)}(F;\omega) =& \sum_{F'}(-1)^{F+F'+1}\sqrt{\frac{6F(2F+1)}{F+1}}\\
        \times&\left\{
        \begin{array}{ccc}
        1 & 1 & 1\\
        F & F & F'
        \end{array}
        \right\}
        \frac{\omega{\left|\langle F||\mathbf{d}||F'\rangle\right|}^2}{\hbar(\omega_{F'F}^2-\omega^2)}\\
        \alpha^{(2)}(F;\omega) =& \sum_{F'}(-1)^{F+F'}\sqrt{\frac{40F(2F+1)(2F-1)}{3(F+1)(2F+3)}}\\
        \times&\left\{
        \begin{array}{ccc}
        1 & 1 & 2\\
        F & F & F'
        \end{array}
        \right\}
        \frac{\omega_{F'F}{\left|\langle F||\mathbf{d}||F'\rangle\right|}^2}{\hbar(\omega_{F'F}^2-\omega^2)}
    \end{split}
\label{polarizability_1}
\end{align}
where $\omega_{F'F}$ is the energy difference between states $|F'\rangle$ and $|F\rangle$.

Assuming the tweezer is linearly polarized (i.e., $E_{0z}^{(+)} = \cos\theta E_0^{(+)}$) with electric field parallel to the bias magnetic field, the vector polarizability term would automatically vanish; hence, Eq. (\ref{AC Starck Shift}) would become
\begin{equation}
\Delta E(F,m_F;\omega) = -\left|E_0^{(+)}\right|^2\alpha(\omega)
\end{equation}
where $\alpha(\omega)$ is the total polarizability
\begin{equation}
    \alpha(\omega) = \alpha^{(0)} + \alpha^{(2)}\left(\frac{2\cos^2\theta-1}{2}\right)\left[\frac{3m_F^2-F(F+1)}{F(2F-1)}\right]
\end{equation}
where $\theta=0$ is the angle between the tweezer polarization and the quantization axis determined by the applied magnetic field.

Using Wigner-Eckart theorem, the hyperfine reduced electric dipole matrix element could be decomposed
\begin{align}
    \begin{split}
        \langle F||\mathbf{d}||F'\rangle =& \langle J\ I\ F||\mathbf{d}||J'\ I\ F'\rangle\\
        =& \langle J||\mathbf{d}||J'\rangle(-1)^{F'+J+1+I}\sqrt{(2F'+1)(2J+1)}\\
        \times&\left\{
        \begin{array}{ccc}
        J & J' & 1\\
        F' & F & I
        \end{array}
        \right\}
    \end{split}
\label{reduced matrix element}
\end{align}
Substituting Eq.~(\ref{reduced matrix element}) into Eq.~(\ref{polarizability_1}), the scalar, vector, and tensor polarizabilities become
\begin{align}
    \begin{split}
        &\alpha^{(0)}(F;\omega) = \sum_{F'}\frac{2\omega_{F'F}{\left|\langle J||\mathbf{d}||J'\rangle\right|}^2}{3\hbar(\omega_{F'F}^2-\omega^2)}\\
        &\times
        (2F'+1)(2J+1)
        \left\{
        \begin{array}{ccc}
        J & J' & 1\\
        F' & F & I
        \end{array}
        \right\}^2\\
        &\alpha^{(1)}(F;\omega) = \sum_{F'}(-1)^{F+F'+1}\sqrt{\frac{6F(2F+1)}{F+1}}
        \left\{
        \begin{array}{ccc}
        1 & 1 & 1\\
        F & F & F'
        \end{array}
        \right\}\\
        &\times
        \frac{\omega{\left|\langle J||\mathbf{d}||J'\rangle\right|}^2}{\hbar(\omega_{F'F}^2-\omega^2)}
        (2F'+1)(2J+1)
        \left\{
        \begin{array}{ccc}
        J & J' & 1\\
        F' & F & I
        \end{array}
        \right\}^2\\
        &\alpha^{(2)}(F;\omega) = \sum_{F'}(-1)^{F+F'}\sqrt{\frac{40F(2F+1)(2F-1)}{3(F+1)(2F+3)}}
        \left\{
        \begin{array}{ccc}
        1 & 1 & 2\\
        F & F & F'
        \end{array}
        \right\}\\
        &\times
        \frac{\omega_{F'F}{\left|\langle J||\mathbf{d}||J'\rangle\right|}^2}{\hbar(\omega_{F'F}^2-\omega^2)}
        (2F'+1)(2J+1)
        \left\{
        \begin{array}{ccc}
        J & J' & 1\\
        F' & F & I
        \end{array}
        \right\}^2
    \end{split}
\label{polarizability_2}
\end{align}
In the limit of large detunings compared to hyperfine splittings, $\omega_{F'F}\approx\omega_{J'J}$; therefore, further adjustments have to be made to above polarizability equation, Eq.~(\ref{polarizability_2}).

\noindent
In light of the orthogonality relation of 6-j symbols
\begin{equation}
    \sum_j(2j+1)(2j''+1)
    \left\{
    \begin{array}{ccc}
    j_1 & j_2 & j\\
    j_3 & j_4 & j'
    \end{array}
    \right\}
    \left\{
    \begin{array}{ccc}
    j_1 & j_2 & j\\
    j_3 & j_4 & j''
    \end{array}
    \right\}
    =
    \delta_{j'j''}
\end{equation}
the scalar polarizability becomes
\begin{equation}
    \alpha^{(0)}(F;\omega) = \sum_{J'}\frac{2\omega_{J'J}{\left|\langle J||\mathbf{d}||J'\rangle\right|}^2}{3\hbar(\omega_{J'J}^2-\omega^2)}
\label{polarizability_3}
\end{equation}
In light of the Biedenharn-Elliott sum rule
\begin{align}
    \begin{split}
        &\left\{
        \begin{array}{ccc}
        j_1 & j_2 & j_{12}\\
        j_3 & j_{123} & j_{23}
        \end{array}
        \right\}
        \left\{
        \begin{array}{ccc}
        j_{23} & j_1 & j_{123}\\
        j_4 & j & j_{14}
        \end{array}
        \right\}\\
        &=
        \sum_{j_{124}}(-1)^{j_1+j_2+j_3+j_4+j_{12}+j_{23}+j_{14}+j_{123}+j_{124}+j}
        (2j_{124}+1)\\
        &\times
        \left\{
        \begin{array}{ccc}
        j_3 & j_2 & j_{23}\\
        j_{14} & j & j_{124}
        \end{array}
        \right\}
        \left\{
        \begin{array}{ccc}
        j_2 & j_1 & j_{12}\\
        j_4 & j_{124} & j_{14}
        \end{array}
        \right\}
        \left\{
        \begin{array}{ccc}
        j_3 & j_{12} & j_{123}\\
        j_4 & j & j_{124}
        \end{array}
        \right\}
    \end{split}
\end{align}
the vector and tensor polarizabilities become
\begin{align}
    \begin{split}
        &\alpha^{(1)}(F;\omega) = \sum_{J'}(-1)^{-2J-J'-F-I}\sqrt{\frac{6F(2F+1)}{F+1}}\\
        &\times
        (2J+1)
        \frac{\omega{\left|\langle J||\mathbf{d}||J'\rangle\right|}^2}{\hbar(\omega_{J'J}^2-\omega^2)}
        \left\{
        \begin{array}{ccc}
        1 & 1 & 1\\
        J & J & J'
        \end{array}
        \right\}
        \left\{
        \begin{array}{ccc}
        J & J & 1\\
        F & F & I
        \end{array}
        \right\}\\
        &\alpha^{(2)}(F;\omega) = \sum_{J'}(-1)^{-2J-J'-F-I}\sqrt{\frac{40F(2F+1)(2F-1)}{3(F+1)(2F+3)}}\\
        &\times
        (2J+1)
        \frac{\omega_{J'J}{\left|\langle J||\mathbf{d}||J'\rangle\right|}^2}{\hbar(\omega_{J'J}^2-\omega^2)}
        \left\{
        \begin{array}{ccc}
        1 & 1 & 2\\
        J & J & J'
        \end{array}
        \right\}
        \left\{
        \begin{array}{ccc}
        J & J & 2\\
        F & F & I
        \end{array}
        \right\}
    \end{split}
\label{polarizability_4}
\end{align}
Given the existence of alternate convention for the Wigner-Eckart theorem
\begin{equation}
    (\alpha\ j||\mathbf{T}^{(k)}||\alpha'\ j') = (-1)^{2k}\sqrt{2j+1}\langle\alpha\ j||\mathbf{T}^{(k)}||\alpha'\ j'\rangle
\end{equation}
and the fact that in NIST ASD database \cite{NIST_ASD}, line strength is provided as
\begin{equation}
    S = {\left|(J||\mathbf{d}||J')\right|}^2
\end{equation}
The reduced electric dipole matrix element in Eq.~(\ref{polarizability_2}) would become
\begin{equation}
    {\left|\langle J||\mathbf{d}||J'\rangle\right|}^2 = \frac{{\left|(J||\mathbf{d}||J')\right|}^2}{2J+1} = \frac{S}{2J+1}
\end{equation}
Using the recorded values of energy levels and line strengths from the NIST ASD database \cite{NIST_ASD}, we calculated the polarizabilities of various states of interest, as shown in Fig.~\ref{Figure3}.

\section{$|\beta|$ ratio and Raman fidelity limit}
\label{Appendix, Beta Ratio and Raman Fidelity Limit}
Consider a typical setup of two-photon stimulated Raman transition ($\Lambda$-configuration), where two hyperfine ground states $|g_1\rangle$ and $|g_2\rangle$ with energies $E_g$ are coupled to multiple hyperfine excited states $\{|e_n\rangle\}_{n=1}^{n_\mathrm{max}}$ by two optical fields, where the coupling strengths are characterized by the Rabi frequencies $\Omega_{1n}$ and $\Omega_{2n}$. The states $|e_n\rangle$ are states in the excited $P_J$ manifold, with energies $E_n$ in increasing order.

For the case of red-detuning (i.e., $\Delta < 0$), the laser detuning $\Delta = \hbar\omega-(E_1-E_g)$ characterizes the energy mismatch between the laser frequency $\omega$ and the atomic transition between ground $|g_{1,2}\rangle$ and lowest excited hyperfine state $|e_1\rangle$. The states $|e_n\rangle$ are $\delta_n = E_n-E_1$ relative to the lowest excited hyperfine state, such that $\Delta-\delta_n$ indicates the amount of energy mismatch between the laser frequency and atomic transition between ground and any abitrary excited hyperfine state $|e_n\rangle$.

For the case of blue-detuning (i.e., $\Delta > 0$), the laser detuning $\Delta = \hbar\omega-(E_{n_\mathrm{max}}-E_g)$ characterizes the energy mismatch between the laser frequency $\omega$ and the atomic transition between ground $|g_{1,2}\rangle$ and highest excited hyperfine state $|e_{n_\mathrm{max}}\rangle$.
The states $|e_n\rangle$ are $\delta_n = E_n-E_{n_\mathrm{max}}$ relative to the highest excited hyperfine state, such that $\Delta-\delta_n$ indicates the amount of energy mismatch between the laser frequency and atomic transition between ground and any abitrary excited hyperfine state $|e_n\rangle$.

Under such a setup, the (effective) Rabi frequency, which characterizes the Raman coupling strength, can be expressed as
\begin{equation}
\Omega_R=\sum_n^{n_\mathrm{max}}\frac{\Omega_{1n}\Omega_{2n}}{4[\Delta-\delta_n(B)]}
\label{effective rabi frequency}
\end{equation}
while the inelastic-scattering rate from spontaneous emission of excited states is
\begin{equation}
\Gamma_\mathrm{ine} = \gamma\sum_{m}^2\sum_n^{n_\mathrm{max}}\frac{\Omega_{mn}^2}{4[(\Delta-\delta_n(B)]^2}
\label{inelastic scattering rate}
\end{equation}
where $\gamma$ is the natural decay rate of the excited states.

In Eqs. (\ref{effective rabi frequency}) and (\ref{inelastic scattering rate}), $\Omega_{ge}$ is the Rabi frequency of individual hyperfine transitions (between ground and excited states), given by
\begin{equation}
\Omega_{ge} = -\frac{\langle\mathrm{g}|\hat{\varepsilon}\cdot\mathbf{d}|\mathrm{e}\rangle}{\hbar}\sqrt{\frac{2I}{\epsilon_0c}}
\end{equation}
where $\langle\mathrm{g}|\hat{\varepsilon}\cdot\mathbf{d}|\mathrm{e}\rangle$ is the transition dipole element, $I$ is the laser intensity, $c$ is the speed of light in vacuum, $\epsilon_0$ is the permittivity of free space, and $\hat{\varepsilon}$ is the laser polarization.

In Eqs. (\ref{effective rabi frequency}) and (\ref{inelastic scattering rate}), $\delta_n(B)$ is the field-dependent relative energy of hyperfine states, given by
\begin{equation}
\delta_n(B) = \delta_n(0) + [E_n(B) - E_n(0)]
\end{equation}
where $\delta_n(0)$ is the zero-field relative energy, obtained from experimental measurements in literature, $E_n(B)$ is the field-dependent eigen-enegy one acquires as by-product after solving the Hamiltonian for its eigen-state to calculate $\langle\mathrm{g}|\hat{\varepsilon}\cdot\mathbf{d}|\mathrm{e}\rangle$.

Although the Zeeman shifts in hyperfine state energy experienced by most alkali-metal atoms (e.g., Na, K, Rb, Cs, etc.) under external magnetic field may be small compared to their large fine-structure splitting, such that these Zeeman shifts can be ignored; this is generally not the case for atomic species like $\mathrm{^6Li}$ and $\mathrm{^3He}$, whose fine-structure, hyperfine-structure, and Zeeman splittings are on similar orders of magnitude, thus requiring our use of field-dependent relative energy, $\delta_n(B)$.

As proposed by Wei and Mueller~\cite{Wei2013}, a key figure of merit for stimulated Raman transition is the ratio of the (effective) Rabi frequency to the inelastic-scattering rate
\begin{equation}
|\beta| = \left|\frac{\Omega_R}{\Gamma_\mathrm{ine}}\right|
\end{equation}
which can be utilized to quantify the fundamental fidelity limit for single-qubit gates
\begin{equation}
F \approx 1-\frac{1}{|\beta|}
\end{equation}
since the inverse of the (effective) Rabi frequency reflects the time it takes for an atom to oscillate between $|g_1\rangle$ and $|g_2\rangle$ ground states while the inverse of the inelastic-scattering rate reflects the time between photon absorption events that leads to error. It should be noted that this metric is intended to be a general estimate of fidelity, independent of the exact gate operation performed.

The transition dipole moment $\langle\mathrm{g}|\hat{\varepsilon}\cdot\mathbf{d}|\mathrm{e}\rangle$ is obtained via diagonalizing the spin-coupled Hamiltonian as outlined below. Note that the general and simplified Hamiltonian approach follows the procedure used by Wei and Mueller~\cite{Wei2013}.

\subsection{General Hamiltonian in $(m_L, m_S, m_I)$ basis for $\mathrm{^6Li}$}
For the case of $\mathrm{^6Li}$, consider the general spin-coupled Hamiltonian in the $(m_L, m_S, m_I)$ basis that describes the hyperfine structure of an atom under magnetic field:
\begin{align}
\begin{split}
H &= H_A+H_B\\
&= c_\mathrm{f}\frac{\mathbf{L\cdot S}}{\hbar^2} + c_\mathrm{hf1}\frac{\mathbf{L\cdot I}}{\hbar^2} + c_\mathrm{hf2}\frac{\mathbf{S\cdot I}}{\hbar^2}\\
&\quad+\frac{\mu_B}{\hbar}(g_L\mathbf{L} + g_S\mathbf{S} + g_I\mathbf{I})\mathbf{\cdot B}\\
\end{split}
\end{align}
where for $\mathrm{^6Li}$, the Land\'e g-factors for angular momentum are $g_L = 0.99999587$, $g_S = 2.0023193043737$, $g_I = -0.0004476540$; the fine-structure and hyperfine-structure coefficients for the $\mathrm{2^2S}$ state are $c_\mathrm{f} = 0\mathrm{\ and\ }c_\mathrm{hf1} = c_\mathrm{hf2} = 152.1368407\mathrm{\ MHz}$; the fine-structure and hyperfine-structure coefficients for  $\mathrm{2^2P}$ state are $c_\mathrm{f} = (10.053044/1.5)\mathrm{\ GHz}\mathrm{\ and\ }c_\mathrm{hf1} = 17.386\mathrm{\ MHz},\ c_\mathrm{hf2} = -1.155\mathrm{\ MHz}$~\cite{GehmMichaelEric2003Thesis}.

Once the Hamiltonian is diagonalized, its corresponding eigenstate can be expressed as
\begin{equation}
|LQ\rangle = \sum_{m_Lm_Sm_I}C^Q_{m_Lm_Sm_I}|Lm_Lm_Sm_I\rangle
\end{equation}
From the Wigner-Eckart Theorem
\begin{align}
\begin{split}
\langle &Lm_Lm_Sm_I|\hat{\varepsilon}\cdot\mathbf{d}|L'm_L'm_S'm_I'\rangle\\
&=\delta_{m_Sm_S'}\delta_{m_Im_I'}\langle Lm_L|\hat{\varepsilon}\cdot\mathbf{d}|L'm_L'\rangle\\
&=\delta_{m_Sm_S'}\delta_{m_Im_I'}W_{m'_Lqm_L}^{L'L}\langle L||\mathbf{d}||L'\rangle
\end{split}
\end{align}
where
\begin{align}
\begin{split}
W_{m'_Lqm_L}^{L'L} &= (-1)^{L'-1+m_L}\sqrt{2L+1}
\left(
\begin{array}{ccc}
L'&1&L\\
m_L'&q&-m_L
\end{array}
\right)\\
&= \langle L',m_L';1,q|L,m_L\rangle.
\end{split}
\end{align}
Therefore, the transition dipole moment $\langle\mathrm{g}|\hat{\varepsilon}\cdot\mathbf{d}|\mathrm{e}\rangle$ is expressed in terms of coefficients $C^Q_{m_Lm_Sm_I}$ and reduced dipole matrix element $\langle L||\mathbf{d}||L'\rangle$
\begin{align}
\begin{split}
&\langle\mathrm{g}|\hat{\varepsilon}\cdot\mathbf{d}|\mathrm{e}\rangle\\
&= \langle LQ|\hat{\varepsilon}\cdot\mathbf{d}|L'Q'\rangle\\
&= \sum_{\bar{m}_L\bar{m}'_L\bar{m}_S\bar{m}_I}C_{\bar{m}_L\bar{m}_S\bar{m}_I}^QC_{\bar{m}'_L\bar{m}_S\bar{m}_I}^{Q'}W_{\bar{m}'_Lq\bar{m}_L}^{L'L}\langle L||\mathbf{d}||L'\rangle.
\label{transition dipole moment}
\end{split}
\end{align}

\subsection{Simplified Hamiltonian in $(m_I, m_J)$ basis for $\mathrm{^{23}Na}$}
For the case of $\mathrm{^{23}Na}$, since it operates under the intermediate field regime, where $c_\mathrm{hf}<E_B<c_\mathrm{f}$, the general Hamiltonian simplifies to the following form with respect to the $(m_I, m_J)$ basis
\begin{align}
\begin{split}
H &= H_\mathrm{hfs}+H_B^\mathrm{(hfs)}\\
&= c_\mathrm{hf}\frac{\mathbf{I\cdot J}}{\hbar^2}+\frac{\mu_B}{\hbar}(g_J\mathbf{J}+g_I\mathbf{I})\mathbf{\cdot B}\\
\end{split}
\end{align}
where for $\mathrm{^{23}Na}$, the Land\'e g-factors for atomic angular momentum are $g_J \approx 1+\frac{J(J+1)+S(S+1)-L(L+1)}{2J(J+1)}$, $g_I = -0.00080461080$; the hyperfine-structure coefficient for  $\mathrm{3^2S_{1/2}}$ state is $c_\mathrm{hf} = 885.81306440\mathrm{\ MHz}$, the hyperfine-structure coefficient for  $\mathrm{3^2P_{1/2}}$ state is $c_\mathrm{hf} = 94.44\mathrm{\ MHz}$, the hyperfine-structure coefficient for  $\mathrm{3^2P_{3/2}}$ state is $c_\mathrm{hf} = 18.534\mathrm{\ MHz}$~\cite{SteckAlkaliDataforSodium}.

Once the Hamiltonian is diagonalized, its corresponding eigenstate can be expressed as
\begin{equation}
|L\tilde{Q}\rangle = \sum_{m_Im_J}C^{\tilde{Q}}_{m_Im_J}|Lm_Im_J\rangle.
\end{equation}
In the intermediate regime, $Q$ is decomposed into $J$ and $\tilde{Q}$
\begin{equation}
C^Q_{m_Lm_Sm_I} = \langle m_Lm_S|Jm_J\rangle C^{\tilde{Q}}_{m_Im_J}.
\label{decomposition 1}
\end{equation}
Substituting Eq. (\ref{decomposition 1}) into Eq. (\ref{transition dipole moment}), the transition dipole moment can be expressed as
\begin{widetext}
\begin{equation}
\langle\mathrm{g}|\hat{\varepsilon}\cdot\mathbf{d}|\mathrm{e}\rangle= \sum_{\bar{m}_L\bar{m}'_L\bar{m}_S\bar{m}_I\bar{m}_J\bar{m}'_J}\langle\bar{m}_L\bar{m}_S|J\bar{m}_J\rangle C_{\bar{m}_I\bar{m}_J}^{\tilde{Q}}\langle\bar{m}'_L\bar{m}_S|J\bar{m}'_J\rangle C_{\bar{m}_I\bar{m}'_J}^{\tilde{Q}'}W_{\bar{m}'_Lq\bar{m}_L}^{L'L}\langle L||\mathbf{d}||L'\rangle.
\label{transition dipole moment 2}
\end{equation}
\end{widetext}
When $B=0$,
\begin{equation}
C^{\tilde{Q}}_{m_Im_J} = C^{Fm_F}_{m_Im_J} = \langle m_Im_J|Fm_F\rangle
\label{decomposition 2}
\end{equation}
Substituting Eq. (\ref{decomposition 2}) into Eq. (\ref{transition dipole moment 2}), the transition dipole moment can be expressed as
\begin{widetext}
\begin{equation}
\langle\mathrm{g}|\hat{\varepsilon}\cdot\mathbf{d}|\mathrm{e}\rangle= \sum_{\bar{m}_L\bar{m}'_L\bar{m}_S\bar{m}_I\bar{m}_J\bar{m}'_J}\langle\bar{m}_L\bar{m}_S|J\bar{m}_J\rangle\langle\bar{m}_I\bar{m}_J|Fm_F\rangle\langle\bar{m}'_L\bar{m}_S|J\bar{m}'_J\rangle\langle\bar{m}_I\bar{m}'_J|Fm_F\rangle W_{\bar{m}'_Lq\bar{m}_L}^{L'L}\langle L||\mathbf{d}||L'\rangle.
\label{transition dipole moment 3}
\end{equation}
\end{widetext}

\subsection{Phenomenological Hamiltonian for spin mixing in $\mathrm{^3He}$}
For the case of $\mathrm{^3He}$, its $2^3S$ ground state could still be treated using the general Hamiltonian in $(m_L, m_S, m_I)$ basis, where the Land\'e g-factors for angular momentum are $g_L = 0.9998250$, $g_S = 2.0022432$, $g_I = 2.3174824\times10^{-3}$; the fine-structure and hyperfine-structure coefficients for $\mathrm{2^3S}$ state are $c_\mathrm{f} = 0\mathrm{\ and\ }c_\mathrm{hf1} = c_\mathrm{hf2} = (-6.739701177/1.5)\mathrm{\ GHz}$~\cite{PhysRevA.32.2712}.

Nevertheless, the $2^3P_J$ excited states of $\mathrm{^3He}$ require special treatment, since the fine and hyperfine mixing of $2^1P_J$ into $2^3P_J$ is critical in the proper description of $\mathrm{^3He}$ $2^3P_J$ state hyperfine Zeeman splitting structure.

As a matter of fact, without the appropriate off-diagonal elements of the Hamiltonian matrix coupling $2^1P_J$ states with $2^3P_J$ states, one could not even arrive at the correct hyperfine level structure at zero magnetic field. See for example, Fig. 1(a) of Zhao et al.~\cite{PhysRevLett.66.592} for illustration.

In their 1985 paper~\cite{PhysRevA.32.2615}, Hinds, Prestage, and Pichanick proposed a phenomenological Hamiltonian that takes this $2^1P_J$ and $2^3P_J$ mixing into account, which in turn gave the first consistent interpretation of experimental data. Hinds et al. used the following phenomenological Hamiltonian to describe the hyperfine interactions of $\mathrm{^3He}$
\begin{align}
\begin{split}
H_\mathrm{hfs} &= C\mathbf{I\cdot S}+C'\mathbf{I\cdot K}+D\mathbf{I\cdot L}\\
&\quad+2\sqrt{10}E\mathbf{I\cdot}\{\mathbf{S}C^{(2)}\}^{(1)}+2\sqrt{10}E'\mathbf{I\cdot}\{\mathbf{K}\tilde{C}^{(2)}\}^{(1)}
\end{split}
\end{align}
where $\mathbf{K}$ is the antisymmetric spin operator $\mathbf{s_1-s_2}$, $C^{(2)}$ and $\tilde{C}^{(2)}$ are the exchange symmetric and antisymmetric tensors $C_1^{(2)}\pm C_2^{(2)}$ where $C_i^{(2)} = (4\pi/5)^{1/2}Y_i^{(2)}(\mathbf{\theta,\phi})$.

Next, combined with the fine-structure interaction Hamiltonian, Hinds et al. proposed an effective Hamiltonian $H_\mathrm{eff}$, which they have expressed explicitly in terms of its matrix elements. See for example, Table II of Hinds et al.~\cite{PhysRevA.32.2615}.

The corresponding coefficients of this effective Hamiltonian are: the spin-symmetric hyperfine (contact) constant $C = -4283.890\mathrm{\ MHz}$, the ratio of spin-antisymmetric to spin-symmetric hyperfine (contact) interaction constant $C'/C = 1.004$, the dipole hyperfine interaction constant $D = -28.128\mathrm{\ MHz}$, and the ratio of the dipole interaction constant to the tensor interaction constant $D/E = -3.945$.

Hence, the actual Hamiltonian under magnetic field is
\begin{equation}
H = H_\mathrm{eff}+H_B
\end{equation}
where $H_B$ is same Zeeman Hamiltonian as that used in the general Hamiltonian approach.

\begin{table}[t]
\caption{\label{tab:table2}
\textbf{$|\beta|$ ratio comparison between $(\sigma^+,\sigma^+)$ and $(\pi,\pi)$ stimulated Raman transition of \He.}
}
\begin{ruledtabular}
\begin{tabular}{l|r|r|r|r}
$\mathrm{Polarization}$&$B\mathrm{\ (G)}$&$\Delta\mathrm{\ (GHz)}$&$|\beta|\qquad\quad$&$F\qquad\quad$\\
\colrule
$(\sigma^+,\sigma^+)$&$0$&$-9.835$&$1942.986$&$0.999485$\\
$(\sigma^+,\sigma^+)$&$0$&$29.054$&$1959.174$&$0.999490$\\
\colrule
$(\sigma^+,\sigma^+)$&$800$&$-9.384$&$2077.032$&$0.999519$\\
$(\sigma^+,\sigma^+)$&$800$&$26.351$&$2105.516$&$0.999525$\\
\colrule
$(\pi,\pi)$&$0$&$-9.384$&$2113.947$&$0.999527$\\
$(\pi,\pi)$&$0$&$22.598$&$1964.401$&$0.999491$\\
\colrule
$(\pi,\pi)$&$800$&$-9.535$&$2117.231$&$0.999528$\\
$(\pi,\pi)$&$800$&$24.099$&$2062.955$&$0.999515$\\
\end{tabular}
\end{ruledtabular}
\end{table}

\subsection{$|\beta|$ ratio comparison between $(\sigma^+,\sigma^+)$ and $(\pi,\pi)$ stimulated Raman transition of $\mathrm{^3He}$}
Increasing $\Delta$ from being red-detuned relative to $2^3P_2\ F=5/2$ hyperfine level to being blue-detuned relative to $2^3P_0\ F=1/2$ hyperfine level, we observe two local maxima of $|\beta|$ ratio with respect to detuning, one local maximum occurs when the laser is blue-detuned with respect to $2^3P_2\ F=3/2$ while still red-detuned with respect to $2^3P_0\ F=1/2$, and the other local maximum occurs when the laser is blue-detuned with respect to $2^3P_0\ F=1/2$.

As shown by Fig.~\ref{Comparison between (sigma^+,sigma^+) and (pi,pi)}(b) and Table~\ref{tab:table2}, for the $(\sigma^+,\sigma^+)$ stimulated Raman transition described in Fig.~\ref{Raman_Fidelity_Limit_Figure} of main text, the $|\beta|$ ratio is larger for $B=800\mathrm{\ G}$ compared to cases with both lower and higher magnetic fields; moreover, the blue-detuned maximum is larger than that of the red-detuned maximum (with respect to $2^3P_0\ F=1/2$), suggesting that $B=800\mathrm{\ G}$ and blue-detuned by 26.351 GHz is an optimal configuration for stimulated Raman transition to take place with high fidelity.

In addition, we also investigated the case of a $(\pi,\pi)$ stimulated Raman transition. As shown by Fig.~\ref{Comparison between (sigma^+,sigma^+) and (pi,pi)}(b) and Table~\ref{tab:table2}, this polarization scenario exhibits red-detuned and blue-detuned local maxima with $|\beta|$ similar to that of $(\sigma^+,\sigma^+)$, indicating that $(\pi,\pi)$ is also a viable polarization configuration to work with.

\begin{figure}[t]
    \centering
    \includegraphics[width=0.48\textwidth]{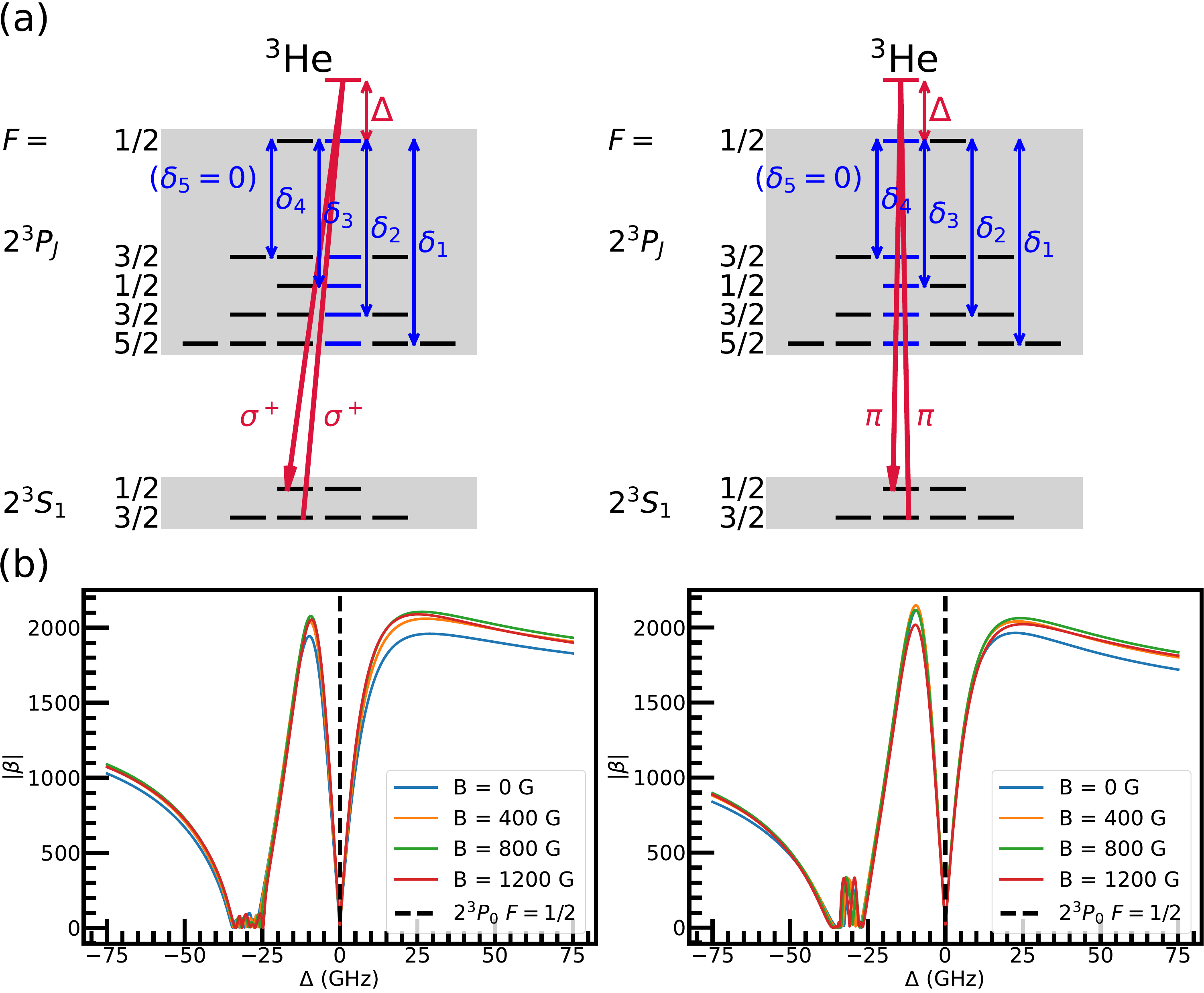}
    \caption{
        \textbf{$|\beta|$ ratio comparison between $(\sigma^+,\sigma^+)$ and $(\pi,\pi)$ stimulated Raman transition of \He.} (a) The hyperfine level structure of states involved in the stimulated Raman transition. (b) $|\beta|$ ratio as a function of detuning $\Delta$ for different magnetic field $B$.
        \label{Comparison between (sigma^+,sigma^+) and (pi,pi)}
    }
\end{figure}

\subsection{$|\beta|$ ratio comparison between $\mathrm{^{23}Na}$ and $\mathrm{^{171}Yb}$ at $B=0$}
For the zero-field scenario ($B=0$), the transition dipole moment $\langle\mathrm{g}|\hat{\varepsilon}\cdot\mathbf{d}|\mathrm{e}\rangle$ involved in the calculation of individual Rabi frequency $\Omega_{mn}$ can be expressed as Eq. (\ref{transition dipole moment 3}), which is essentially the product of a few Clebsch-Gordan coefficients with the reduced dipole matrix element $\langle L||\mathbf{d}||L'\rangle$. Therefore, diagonalizing the spin-coupled Hamiltonian is not even required for this case, as one could directly compute Eq. (\ref{transition dipole moment 3}), use the zero-field relative energy $\delta_n(0)$ and natural decay rate $\gamma$ from literature (\cite{SteckAlkaliDataforSodium} for $\mathrm{^{23}Na}$ and \cite{PhysRevA.107.063107, PhysRevA.86.051404} for $\mathrm{^{171}Yb}$) to calculate the $|\beta|$ ratio (note that $\langle L||\mathbf{d}||L'\rangle$ would cancel out between the numerator and denominator of $|\beta|$, thus $\langle L||\mathbf{d}||L'\rangle$ does not have to be known in order to perform the $|\beta|$ calculation). 

While calculating the $|\beta|$ ratio for $\mathrm{^{23}Na}$, suppose we only took the $3^2P_{1/2}$ excited fine-structure manifold into account [green box of Fig.~\ref{Comparison between Na-23 and Yb-171}(a)]; then, compared to the scenario where we considered both $3^2P_{1/2}$ and $3^2P_{3/2}$ excited fine-structure manifolds [orange box of Fig.~\ref{Comparison between Na-23 and Yb-171}(a)], we observe that the inclusion of the upper $3^2P_{3/2}$ manifold exerts a ``destructive interference'' on the lower $3^2P_{1/2}$ manifold such that instead of the monotonic increase in $|\beta|$ due to $3^2P_{1/2}$ manifold alone [green curve of Fig.~\ref{Comparison between Na-23 and Yb-171}(b)], the combined effect of $3^2P_{1/2}$ and $3^2P_{3/2}$ manifolds causes the $|\beta|$ ratio to level out and reach an asymptotic value at large detuning $\Delta$ [orange curve of Fig.~\ref{Comparison between Na-23 and Yb-171}(b)].

\begin{figure}[t]
    \centering
    \includegraphics[width=0.48\textwidth]{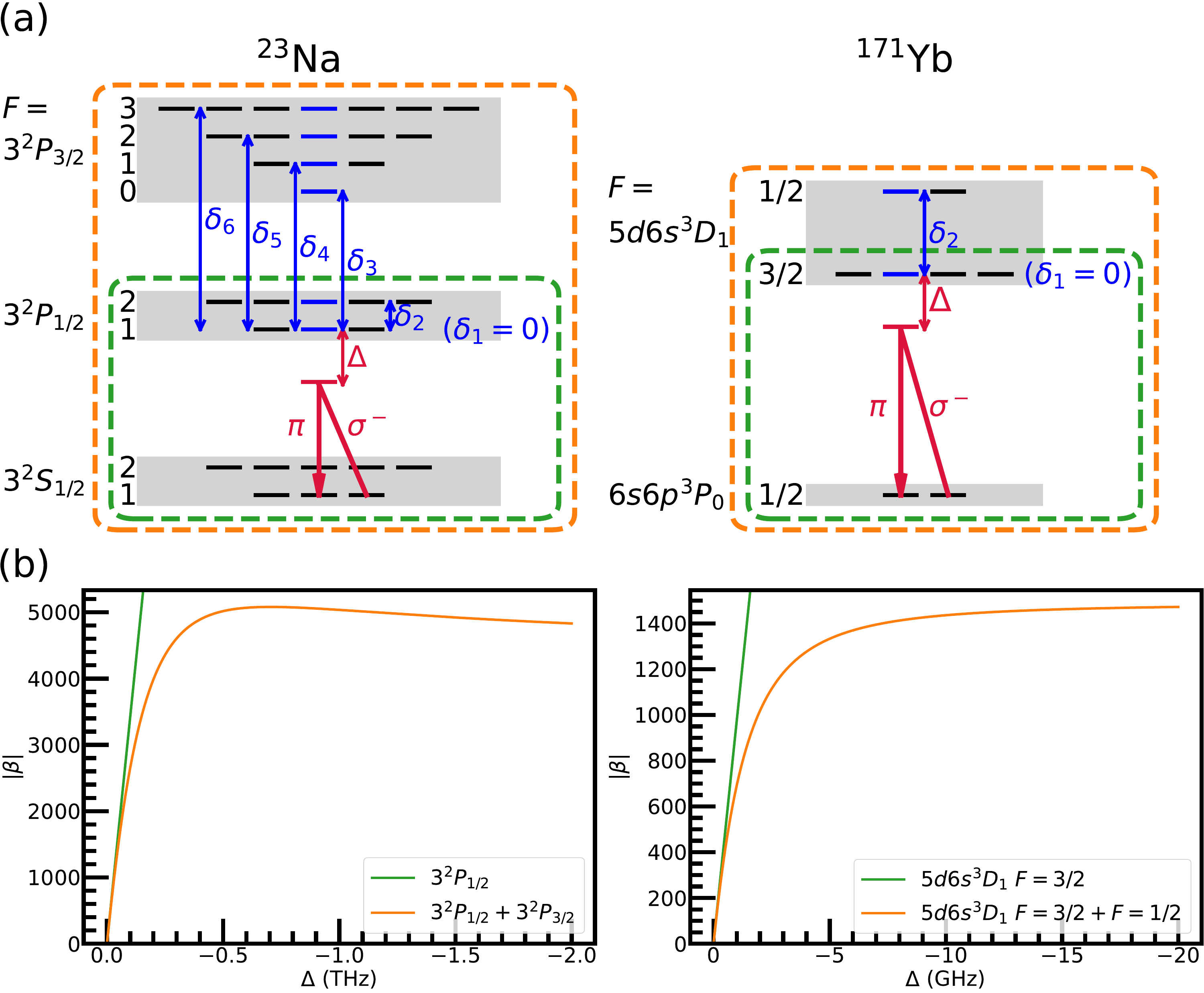}
    \caption{
        \textbf{$|\beta|$ ratio comparison between $^{23}$Na and $^{171}$Yb at $B=0$.} (a) The hyperfine level structure of states involved in the stimulated Raman transition. (b) $|\beta|$ ratio as a function of detuning $\Delta$.
        \label{Comparison between Na-23 and Yb-171}
    }
\end{figure}

One finding worth noting is that, compared to alkali-metal atoms like $\mathrm{^{23}Na}$, where the ``destructive interference'' generally takes place between fine-structure manifolds of different $J$, such ``destructive interference'' also occurs between hyperfine levels of different $F$ within a single fine-structure manifold for atom like $\mathrm{^{171}Yb}$. In this case, instead of the monotonic increase in $|\beta|$ due to $5d6s~ ^3D_1~ F=3/2$ level alone [green curve of Fig.~\ref{Comparison between Na-23 and Yb-171}(b)], the combined effect of $5d6s~ ^3D_1~ F=3/2$ and $5d6s~ ^3D_1~ F=1/2$ levels causes the $|\beta|$ ratio to level out and reach an asymptotic value at large detuning $\Delta$ [orange curve of Fig.~\ref{Comparison between Na-23 and Yb-171}(b)].

This observation could be explained with the completeness relation derived by Wei and Mueller~\cite{Wei2013}. In their 2013 paper, Wei and Mueller proposed that for a complete set of excited states $\{|e_n\rangle\}_{n=1}^{n_\mathrm{max}}$, we have the following relationship for stimulated Raman transition between a pair of ground states $|g_1\rangle$ and $|g_2\rangle$
\begin{equation}
\sum_n^{n_\mathrm{max}}\Omega_{1n}\Omega_{2n} = 0
\end{equation}
Since the $5d6s~^3D_1$ fine-structure state is the only excited state that the $\mathrm{^{171}Yb}$ ground state $6s6p~^3P_0$ is allowed to transition by selection rules, the above completeness relation would suggest that the product of individual Rabi frequencies for the $5d6s~^3D_1~ F=3/2$ and $5d6s~^3D_1~ F=1/2$ levels must have opposite signs (Table~\ref{tab:table3}), thus leading to the ``destructive interference'' observed in $|\beta|$ ratio calculation. However, for alkali-metal atoms like $\mathrm{^{23}Na}$, since both $D_1$ and $D_2$ transition are allowed by selection rule, the above completeness relation imposes a less strict constraint on the product of individual Rabi frequencies, such that opposite signs of hyperfine levels within the same fine-structure manifold are no longer required (Table~\ref{tab:table3}).

\begin{table}[t]
\caption{\label{tab:table3}
\textbf{Individual Rabi frequencies $\Omega_{mn}$ involved in calculation of $|\beta|$ ratio for $^{23}$Na and $^{171}$Yb at $B=0$}. $\Omega_{mn}$ are expressed as multiples of $\frac{\langle L||\mathbf{d}||L'\rangle}{\hbar}\sqrt{\frac{2I}{\epsilon_0c}}$ for the corresponding atomic species.}
\begin{ruledtabular}
\begin{tabular}{r|r|r}
&$\mathrm{^{23}Na}$&$\mathrm{^{171}Yb}$\\
\colrule
$\Omega_{11}$&$\sqrt{1/36}$&$\sqrt{1/9}$\\
$\Omega_{21}$&$0$&$-\sqrt{2/9}$\\
$\Omega_{12}$&$\sqrt{1/36}$&$\sqrt{2/9}$\\
$\Omega_{22}$&$-\sqrt{1/9}$&$\sqrt{1/9}$\\
\colrule
$\Omega_{13}$&$\sqrt{1/9}$&\\
$\Omega_{23}$&$\sqrt{1/9}$&\\
$\Omega_{14}$&$\sqrt{5/36}$&\\
$\Omega_{24}$&$0$&\\
$\Omega_{15}$&$\sqrt{1/36}$&\\
$\Omega_{25}$&$-\sqrt{1/9}$&\\
$\Omega_{16}$&$0$&\\
$\Omega_{26}$&$0$&\\
\end{tabular}
\end{ruledtabular}
\end{table}

\section{Multichannel Quantum Defect Theory}
\label{Appendix, FT-MQDT}
A few detailed studies of the Rydberg state spectroscopy have been carried out for $^3$He.  The great majority of the high resolution studies of Rydberg states has considered only the $np$-states that are excited by a single electric dipole photon out of the $1s2s\ ^3S_1$ metastable level~\cite{vassen1989high,DRAKE199493,ClausenMerkt2025prl,drake2025helium}.

For $^3$He the states of interest take the form $1snl$, where the angular momentum quantum numbers that matter are total angular momentum of the core $f_c$, which couples all of the core quantum numbers $s_c,l_c,j_c,I$, and the total angular momentum of the Rydberg electron $j_e$. Since the core electron is in an s state we can write $j_c=s_c=\frac{1}{2}$ and when coupled together with the nuclear spin $I=\frac{1}{2}$ we get two hyperfine core states, $f_c=0$ and $f_c=1$, the latter being the ground state, with a core hyperfine splitting $\Delta_{f_0-f_1}= 8.66564986577$ GHz~\cite{Schneider2022-gs}. If we take the degeneracy weighted average of the hyperfine thresholds to be the zero of our energy scale, the threshold energies would be $\{E_1=-2.16641246644,\ E_0=6.49923739933\}$ GHz.

The theory that describes the hyperfine couplings for atomic Rydberg states that converge to hyperfine-split ionization threshold energies has been well-developed since the 1980s~\cite{sun1988hyperfine}.  It follows the idea of the frame transformation combined with MQDT (FT-MQDT) which was introduced by Fano, Lu, Lee, and their collaborators and followers~\cite{Lu1970,Fano1970,Lee1973}.

The details of FT-MQDT for the present context are covered well in \cite{Robicheaux2018}, but we can summarize the basic idea as follows. The physics when the Rydberg electron is at short distances, which is especially relevant for Rydberg states having low values of the principle quantum number $n$, the atom is well described in $(LS)J$ coupling scheme.  In the case of the $1sns$ low-lying Rydberg states, the approximately good quantum numbers are $L=0$ and total spin 0 or 1, i.e. singlet and triplet states. However, as the Rydberg electron probes long distances from the ion core, i.e. for the states at high $n$, the approximately good quantum numbers of the $^3$He atom change. In FT-MQDT, the primary quantity that governs all of the Rydberg channel coupling information is the reaction matrix $\underline{K}$, which is approximately diagonal in the singlet-triplet representation at short range, $K_{SS'}=\delta_{SS'}\tan\pi\mu_S(\epsilon)$. The quantum defects vary slowly with energy $\epsilon$, and in some cases are well-approximated by energy-independent constants, but for the present calculations the energy dependence of the eigenquantum defects was taken into account as was done for the $np$ states by Ref.~\cite{ClausenMerkt2025prl} and with a less precise assumption of linear energy dependence in the earlier work by Ref.~\cite{vassen1989high}. 

Known values of the lower-lying $ns$ singlet and triplet quantum defects for the $n\leq 10$ states of $^3$He and $^4$He \cite{Morton2006-ke} were analyzed to extract an isotope shift correction to the quantum defect which could be applied to the excited states, in order to take advantage of the better-known states of $^4$He. After obtaining the quantum defects for $^3$He, they were fitted to a function of energy 
\begin{equation}
    \mu(\epsilon)=\mu_0+\mu_1\epsilon+\mu_2\epsilon^2+...
    \label{eq:qdft}
\end{equation}
where $\epsilon=\frac{Ry(^3He)}{\nu^2}$, fitting parameters used in this calculation can be found in Table ~\ref{tab:qdft}.

With the short-range $K$-matrix then known in the singlet-triplet representation as a function of energy, the first real step of FT-MQDT is applied, namely, to recouple that short-range $K$-matrix into a representation that is better described by the quantum numbers appropriate at long-range. This recoupling is a real orthogonal frame transformation matrix $\underline{U}$, given by Eq. \ref{umat}, 
\begin{widetext}
\begin{equation}
\label{umat}
   \underline{U}=\braket{[(Ij_c)f_c\ j_e]F|[I(SL)J]F}=
    [S,L,j_c,j_e,f_c,J]
   (-1)^{I+j_c+j_e+J}
\begin{Bmatrix}
I & j_c & f_c \\
j_e & F & J
\end{Bmatrix}
\begin{Bmatrix}
s_c & s_e & S \\
l_c & l_e & L \\
j_c & j_e & J
\end{Bmatrix},
\end{equation}
\end{widetext}
where the notation [a,b,...] = $\sqrt{(2a+1)(2b+1)...}$.  This is applied to the short-range K matrix, transforming it into the frame of our long-range good quantum numbers of core+electron $\underline{K}^{LR}\ =\underline{U}\ \underline{K}^{LSJ}\ \underline{U}^T$.

At this point, bound state energies are found by solving for the roots of the following equation:
\begin{equation}
    |\underline{K}^{LR}+\tan\pi\underline{\nu}|=0
\end{equation}
where $\underline{\nu}$ is a diagonal matrix consisting of the effective quantum numbers in each of the channels. The bound state energies can be displayed in the form of a Lu-Fano plot as seen in Fig. \ref{FigureRydberg}(b) where the two channels of the S F=$\frac{1}{2}$ series interact with each other.

For the work done here, special interest was taken in the $1sns \ ^3$S$_1$ F=$\frac{3}{2}$ series of $^3$He due to its isolation from nearby perturbers, but FT-MQDT calculations were done for all of the symmetries for the S, P, D, and F series in order to gain an accurate model of the long range interactions. 

\begin{table*}[!t]
\centering
\resizebox{\textwidth}{!}{
\begin{tabular}{ccccccccccccccc}
\toprule
$\mu_i$ & $^1S_0$ & $^3S_1$ & $^1P_1$ & $^3P_0$ & $^3P_1$ & $^3P_2$ & $^1D_2$ & $^3D_1$ & $^3D_2$ & $^3D_3$ & $^1F_3$ & $^3F_2$ & $^3F_3$ & $^3F_4$\\
\midrule
$\mu_0$ & 0.139716 & 0.296655 & -0.0121597 & 0.0683488 & 0.0683787 & 0.0683811 & 0.00211422 & 0.00288581 & 0.00289117 & 0.00289155 & 0.000440873 & 0.000445445 & 0.000449169 & 0.000447954 \\
$\mu_1$ & 0.054412 & 0.0754184 & 0.0136929 & -0.0385954 & -0.0385723 & -0.0385699 & -0.00722717 & -0.0137631 & -0.0137623 & -0.0137633 & -0.00472127 & -0.00482053 & -0.00479613 & -0.00482045 \\
$\mu_2$ & 0.323622 & 0.262775 & 0.337192 & 0.218744 & 0.218726 & 0.218737 & 0.158517 & 0.159786 & 0.159803 & 0.159795 & 0.27874 & 0.279593 & 0.279751 & 279605 \\
$\mu_3$ & -23.4026 & -20.269 & -26.8352 & -25.0712 & -25.0681 & -25.0708 & -10.9504 & -10.9372 & -10.9383 & -10.9382 & -26.575 & -26.5759 & -26.5768 & -26.5772 \\
$\mu_4$ & 727.44 & 598.537 & 876.553 & 795.634 & 795.521 & 795.618 & 254.678 & 254.559 & 254.592 & 254.593 & 863.541 & 863.522 & 863.502 & 863.565 \\
\bottomrule
\end{tabular}}
\caption{\label{tab:qdft}\textbf{Hyperfine fitting parameters from eq. \ref{eq:qdft} for energy in atomic units.} Note: the $p$ state fit parameters agree with those from \cite{ClausenMerkt2025prl},except that ours are less accurate. That paper was discovered only after the present calculations were completed.  
}
\end{table*}

\section{Rydberg Interactions}
\label{Appendix, Rydberg}
\begin{figure}[!t]
    \centering
    \includegraphics[width=0.4\textwidth]{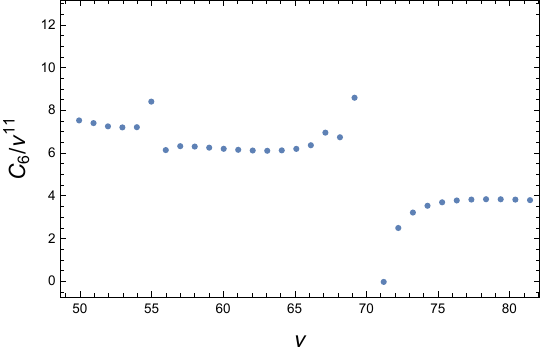}
    \caption{\textbf{ Scaled $C_6$ as a function of the effective principal quantum number $n^* \equiv \nu$ for one eigenstate of the ns F=1/2 equal state pair of $^3$He.} The $C_6$ coefficient for $n=70$ is above the bounds of the plot with a $C_6/\nu^{11}$ value of $\sim$ 37 a.u.  These scaled $C_6$ coefficients are in atomic units (a.u.); to convert to $\mathrm{GHz}\cdot\mu\mathrm{m}^6$ multiply by $1.44\times10^{-19}$.}
    \label{fig:SS12}
\end{figure}
Consider two interacting atoms, one in state $b_1$ and the other in state $b_2$, separated by some internuclear distance $R$. For a simple system like this, we can express the pair Hamiltonian as:
\begin{equation}
    H=h_1+h_2+H_\mathrm{int},
\end{equation}
where $h_1$ and $h_2$ are the single atom Hamiltonians and $H_\mathrm{int}$ is the interaction energy between the two atoms. For this calculation, it is of particular interest to work with Rydberg atoms, so we can simplify our system further by considering each atom as a frozen ionic $1s$ core plus one Rydberg electron. Then, $H_\mathrm{int}$ has the form:
\begin{equation}
    H_{int}=\frac{1}{R}+\frac{1}{|\textbf{r}_1-\textbf{r}_2-\textbf{R}|}-\frac{1}{|\textbf{r}_1-\textbf{R}|}-\frac{1}{|\textbf{r}_2-\textbf{R}|},
\end{equation}
where $\textbf{r}_1$ and $\textbf{r}_2$ are the position vectors for the Rydberg electrons with respect to their nuclei with $\textbf{R}$ defining our internuclear axis. A multipole expansion of the electron-electron Coulomb repulsion gives the dipole-dipole interaction potential, written here as a tensor operator:
\begin{equation}
    V_{dd}=-\frac{\sqrt{24\pi}}{R^3}\{[r_1^{(1)}\otimes r_2^{(1)}]^{(2)}\otimes Y^{(2)}(\hat{R})\}_0^{(0)},
\end{equation}
\\
where $r^{(k)}_{i\ q}=\sqrt{4\pi/(2k+1)}r_i^kY^{(k)}_q(\Omega_i)$. For brevity, we only consider here the case where the internuclear axis, $\textbf{R}$, is aligned with the quantization $z$-axis, and for that choice the total magnetic quantum number $M =m_1+m_2$ is conserved. The matrix element of this operator in a pair state basis, $\ket{b_1b_2}$, is characterized by the $C_3$ coefficient:
\begin{equation}
    \braket{b_1b_2|V_{dd}|b_1'b_2'}=\frac{C_{3;b_1b_2,b_1'b_2'}}{R^3}.
\end{equation}
The energies of the system have been studied both by numerical diagonalization and in perturbation theory. After the Hilbert space is restricted to states energetically close to any particular target states of the two interacting atoms, the Hamiltonian matrix can be set up to diagionalize at multiple values of the internuclear distance, $R$.  Those eigenenergies serve as potential energy curves for the dipole-dipole interaction and yield the corresponding $C_6$ and $C_3$ coefficients~\cite{Weber2017-uv}. For the present calculation, the Hilbert space was restricted to relevant states which also satisfy $|n_{target}-n'|\leq 2$.

The $C_6$ coefficient for this long-range interaction is defined by a second-order perturbative expansion, which involves a sum over all intermediate states. This expansion couples Rydberg pair states, $\ket{b_1b_2}$, of opposite parity as seen below:
\begin{equation}
C_{6;b_1b_2}=\sum_{b_1'b_2'}\frac{\braket{b_1b_2|V_{dd}|b_1'b_2'}\braket{b_1'b_2'|V_{dd}|b_1b_2}}{\bar{E}-E_{b_1'}-E_{b_2'}}
\end{equation}
where $\bar{E}=E_{b_1}+E_{b_2}$. The $C_6$ coefficient, which scales approximately like $\nu^{11}$, can change dramatically when the pair states that couple to the initial and final pairs are nearly degenerate, i.e. $\bar{E}\approx E_{b_1'}+E_{b_2'}$. An example of this wild variation due to near degeneracies can be seen in Fig.\ref{fig:SS12}.

\section{Imaging survival}
\label{Appendix, Imaging}

\begin{figure}[t]
    \centering
    \includegraphics[width=0.9\linewidth]{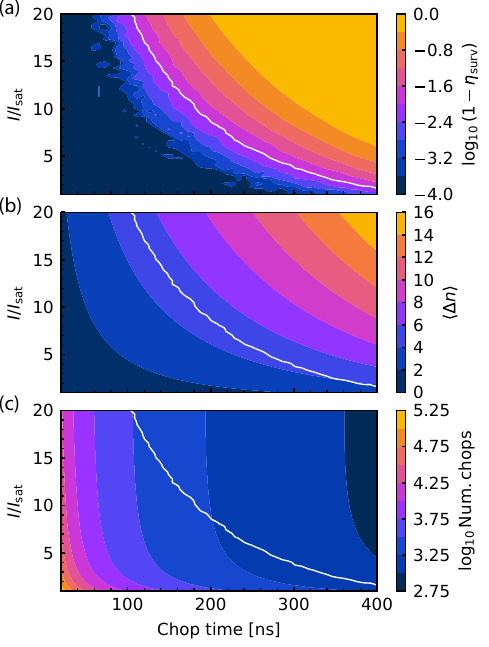}
    \caption{
        \textbf{Simulation of photon scatter.} We compute the (a) overall loss, (b) heating rates, and (c) number of chops required from our chopped readout scheme as functions of the laser intensity $I / I_\text{sat}$ and the chop time. We base our calculation of the overall loss on a targeted total of $1000$ photons scattered, assuming the sideband cooling step takes atoms back to their motional ground state after each chop. Heating is based on the average change in motional quantum number estimated using a harmonic trap approximation of a $h \times 20\,\text{MHz}$-deep tweezer and is conditioned on atom survival. The white level curves in each plot show where the overall loss crosses $10^{-2}$.
        \label{fig:s16}
    }
\end{figure}

We estimate the effects of imaging via Monte Carlo simulation of two-level atoms scattering into free space. Considering atoms initially in the ground state of a 1150-nm tweezer $1\,\text{$\mu$m}$ in waist radius, the trap is quenched off at $t = 0$, and the atom is allowed to move in free space while illuminated by two orthogonal lasers with wavevectors oriented in the plane perpendicular to the tweezer axis. Both lasers that have equal intensity are resonant to the 389-nm imaging transition, on which the atom scatters with radiation pattern following that for entirely $\sigma^\pm$-polarized light. After a given chop time passes, the tweezer is turned back on, and the atom is considered to have been recaptured if its total energy is less than the tweezer's depth, $h \times 20\,\text{MHz}$.

Based on the recapture rate $\eta$ and number of scattered photons $\phi$, we estimate the overall survival rate of atoms in an imaging sequence as $\eta_\text{surv} = \eta^{\phi_0 / \phi}$. $\phi_0 = 1000$ is a target total number of photons chosen as a rough requirement for acceptable imaging using a realistic photon collection device. We also estimate heating via the average change in motional quantum number in the radial direction as $\langle \Delta n \rangle = \langle E \rangle / \hbar \omega - 1 / 2$, where $\langle E \rangle$ is the average total energy of atoms in the tweezer at the end of the chop time, conditioned on recapture, and $\omega$ is the radial trap frequency. We plot the results in Fig.~\ref{fig:s16} as functions of the per-laser intensity $I / I_\text{sat}$ and chop time, and identify a sizable region where $\eta_\text{surv} > 0.99$ and $\langle \Delta n \rangle < 4$.

We note that an acousto-optic modulator on the laser beam that delivers the trapping light to the AOD or SLM can straightforwardly be used to modulate the trap with rise and fall times $<100$ ns. We therefore specifically highlight the case in the bottom right corner of the plots in Fig.~\ref{fig:s16}, where the chop time is $\approx300$ ns and the intensity is $I/I_\text{sat}\approx2$. The corresponding survival after scattering $1000$ photons is $\approx0.999$, as shown in Fig.~\ref{fig:s16}(a), and the heating per chop is $\langle\Delta n\rangle\approx2$, as shown in Fig.~\ref{fig:s16}(b). Figure~\ref{fig:s16}(c) shows the number of chops needed to scatter $1000$ photons, which is $\approx10^3$ under these conditions. For a chop time of $\approx300$ ns, this suggests that a total ``bright" time of $\approx300$ $\mu$s is required. Given that typical high-survival readout protocols for tweezer-trapped atoms take $\approx10$ ms, we could intersperse substantial cooling with this blue imaging while maintaining a typical readout time. For instance, the total probe time would be 10 ms if the cooling time is $\approx30\times$ longer than the ``bright" time on each chop, corresponding to $\approx10$ $\mu$s of cooling per chop period. Following the analysis of Raman sideband cooling with dressed state optical pumping in Fig.~\ref{Figure4}(d), this should be enough time to remove $\langle\Delta n\rangle\approx2$ of heating per chop.

In contrast to the far-detuned regime discussed in main text, for the on-resonant scenario of above chopped imaging protocol, population would start to accumulate in the $1s3p\ ^3P$ state, and the two-step ionization process (i.e, transition to $1s3p\ ^3P$ state, followed by photoionization) becomes the dominant factor behind the two-photon ionization loss. Since we are operating under a different regime, the simple scaling relationship $\alpha_\mathrm{2ph}\sim s^2/\Delta^2$ is no longer valid, and loss rate calculation must be performed via a different approach.

From Dinnen \textit{et al.}\cite{Dinneen:92}, the net loss rate due to photoionization is the product of photon flux $I/h\nu$, photoionization cross section $\sigma$, and excited state population $\rho_\mathrm{ee}$
\begin{equation}
R_\mathrm{loss} = \frac{I}{h\nu}\sigma\rho_\mathrm{ee}
\end{equation}
where $h$ is Planck constant and $\nu$ is frequency of light.
\newline
The excited state population is expressed as
\begin{equation}
\rho_\mathrm{ee} = \frac{s/2}{1+\left(2\Delta/\Gamma\right)^2+s}
\end{equation}
where $\Gamma$ is the transition linewidth and $s=I/I_\mathrm{sat}$ is the saturation parameter.

Using cross section of $1s3p\ ^3P$ state ($\sigma=3.627\times10^{-18}\mathrm{\ cm^2}$) from Chang \textit{et al.}\cite{Chang1995} and neglecting small wavelength dependence, the net loss rate due to photoionization is calculated to be $R_\mathrm{loss}\approx0.016\mathrm{\ s^{-1}}$ for detuning $\Delta=0\mathrm{\ MHz}$ and intensity $I=2I_\mathrm{sat}$, which is quite similar to the $\alpha_\mathrm{2ph}\approx0.023\mathrm{\ s^{-1}}$ estimate of the far-detuned scenario described in main text.

\bibliography{library_revision_2}

\end{document}